\documentclass[aip,superscriptaddress,amsmath,amssymb,reprint]{revtex4-1}
\usepackage{hyperref}

\usepackage{graphicx}
\usepackage{dcolumn}
\usepackage{bm}

\usepackage[utf8]{inputenc}
\usepackage[T1]{fontenc}
\usepackage{mathptmx}
\usepackage{etoolbox}

\usepackage{siunitx}
\usepackage{braket}
\usepackage[]{empheq}
\usepackage{booktabs}
\usepackage{multirow}
\usepackage{bbold}
\usepackage{xcolor}
\usepackage[normalem]{ulem}
\DeclareSIUnit\electronmass{\ensuremath{\mathit{m_e}}}
\DeclareSIUnit\hartree{\ensuremath{\mathit{E}_h}}
\DeclareSIUnit\bohr{\ensuremath{\mathit{a}_0}}

\usepackage{cleveref}

\definecolor{lightblue}{RGB}{68,114,196}
\definecolor{darkgreen}{RGB}{0,100,0}

\begin{document}

\title[High-dimensional quantum dynamics applied to a dendrimer building block]{Wavepacket and Reduced-Density Approaches for High-Dimensional Quantum Dynamics:
Application to the Nonlinear Spectroscopy of Asymmetrical Light-Harvesting Building Blocks}
\author{Joachim Galiana}
\altaffiliation{Current affiliation: Departamento de Química, Universidad Autónoma de Madrid, Madrid, Spain}
\affiliation{ICGM, Univ Montpellier, CNRS, ENSCM, Montpellier, France}
\author{Michèle Desouter-Lecomte}%
\affiliation{Institut de Chimie Physique, Université Paris-Saclay-CNRS, UMR8000, F-91400 Orsay, France
}%

\author{Benjamin Lasorne}
\email{benjamin.lasorne@umontpellier.fr}
\affiliation{%
ICGM, Univ Montpellier, CNRS, ENSCM, Montpellier, France
}%

\date{\today}

\begin{abstract}

Excitation-energy transfer (EET) and relaxation in an optically excited building block of poly(phenylene ethynylene) (PPE) dendrimers are simulated using wavepackets with the multilayer multiconfiguration time-dependent Hartree (ML-MCTDH) method and reduced-density matrices with the hierachical equations of motion (HEOM) approach. The dynamics of the ultrafast electronic funneling between the first two excited electronic states in the asymmetrically \textit{meta}-substituted PPE oligomer with two rings on one branch and three rings on the other side, with a shared ring in between, is treated with 93-dimensional \textit{ab initio} vibronic-coupling Hamiltonian (VCH) models, either linear or with bilinear and quadratic terms. 
The linear VCH model is also used to calibrate an open quantum system that falls in a computationally demanding non-perturbative non-Markovian regime.
The linear-response absorption and emission spectra are simulated with both the ML-MCTDH and HEOM methods. 
The latter is further used to explore the nonlinear regime towards two-dimensional (2D) spectroscopy. 
We illustrate how a minimal VCH model with the two main active bright states and the impulsive-pulse limit in third-order response theory may provide at lower cost polarization-sensitive time-resolved signals that monitor the early EET dynamics. 
We also confirm the essential role played by the high-frequency acetylenic and quinoidal vibrational modes.
\end{abstract}

\maketitle

\section{\label{sec:introduction}Introduction}

The photophysics and ultrafast funneling dynamics in treelike poly(phenylene ethynylene) (PPE) dendrimers have drawn great interest since their original synthesis in 1995.\cite{xu_design_1994}
They have been extensively studied, both experimentally\cite{devadoss_energy_1996,shortreed_directed_1997,melinger_optical_2002,swallen_dendrimer_1999,kopelman_spectroscopic_1997} and  theoretically,\cite{minami_frenkel-exciton_2000,chernyak_exciton_2000} for their astonishing light-harvesting properties. 
In the present work, we investigate such features in an asymmetrical building block of PPE dendrimers, for which we extract time-dependent spectral fingerprints of ultrafast electronic excitation-energy transfer (EET) from simulations with performant quantum-dynamics methods using an \textit{ab initio} calibrated vibronic coupling Hamiltonian (VCH) model in full dimensionality as regards in-plane motions.

\begin{figure*}[!ht]
  \includegraphics[width=0.9\textwidth]{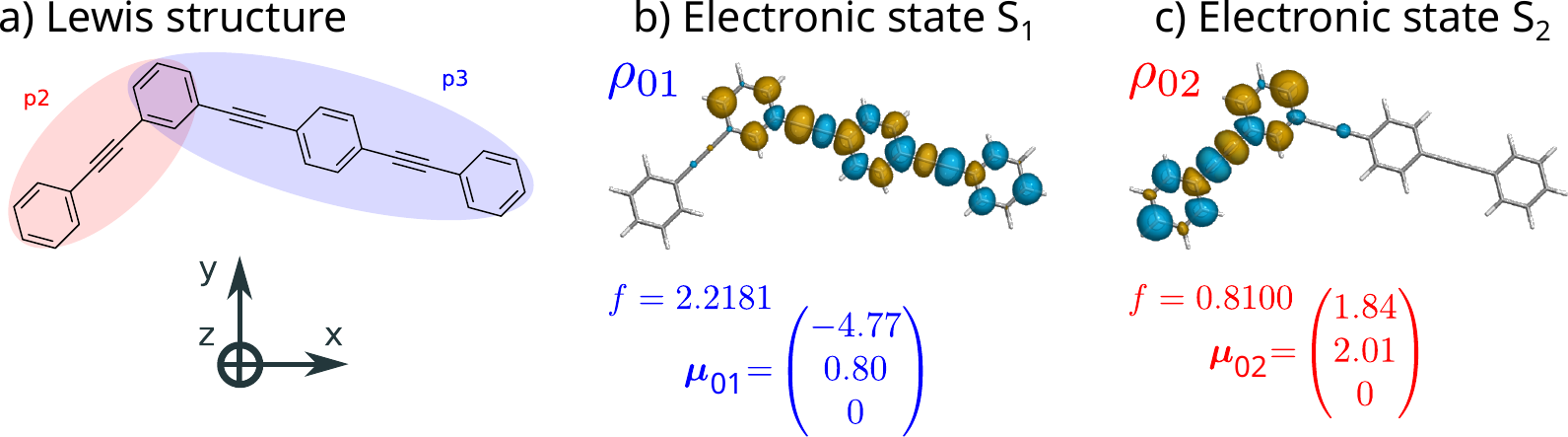}
  \caption{
    a) Lewis structure associated to the Franck-Condon geometry of the first asymmetrical \textit{meta}-substituted PPE oligomer, called m23 in the following.
    The Cartesian axes are also given.
    b) and c) Transition density, oscillator strentgh, and transition dipole moment (in atomic units) for each of the first two singlet electronic excited states at the Franck-Condon geometry.
  }
  \label{fig:m23_est}
\end{figure*}

In the present work, our system of interest for simulating such signals is the exhaustively studied asymmetrically \textit{meta}-substituted PPE oligomer, made of four phenyl or phenylene rings and three acetylenic bonds, called m23 in the following (\cref{fig:m23_est} a).
This building block of PPE dendrimers was shown to be an ideal model system for EET, having two distinct localized excitons as its first two singlet electronic excited states (see \cref{fig:m23_est}, b and c).\cite{fernandez-alberti_nonadiabatic_2009,huang_theoretical_2015,galiana_excitation_2024}
Here, we use both wavepacket and reduced-density approaches to estimate electronic population dynamics, linear- and nonlinear time-resolved spectroscopic signals, altogether to characterize ultrafast intramolecular EET in m23.

Alternative full-dimensional simulations of the photoinduced dynamics of m23 were originally exposed by Fernandez-Alberti and co-workers using the hybrid quantum-classical trajectory surface hopping (TSH) method \cite{fernandez-alberti_nonadiabatic_2009,malone_nexmd_2020} and the multiconfiguration Ehrenfest approach.\cite{fernandez-alberti_non-adiabatic_2016} Such simulations were also recently used to produce nonlinear spectroscopic signals to study exploitable spectral signatures of EET.\cite{hu_spectral_2021,zhang_what_2025}
The present work explores similar objectives but with completely different computational approaches.

We recently proposed a quantum wavepacket study -- based on the multiconfiguration time-dependent Hartree (MCTDH) method\cite{meyer_multi-configurational_1990,beck_multiconfiguration_2000} -- of the photoinduced dynamics of m23 with dimensionally reduced VCH models obtained from its \textit{ab initio} potential energy surfaces (PESs).\cite{galiana_excitation_2024}
Such a strategy was concomitantly followed by other authors to go toward higher-dimensional model, using a wavepacket tensor-train formalism.\cite{liu_ultrafast_2024}
Here, we revisit the photodynamics of m23 with both the multilayer MCTDH (ML-MCTDH)\cite{wang_multilayer_2003,manthe_multilayer_2008,wang_multilayer_2015} wavepacket and the hierarchical equations of motion (HEOM)\cite{Kubo1989,Tanimura2006,Tanimura2020,Yan2007,Yan2009,BaiShi2024}  reduced-density approaches, using a systematic parametrization procedure for generating VCH models adapted to two-exciton systems in high dimensionality.

Wavepacket methods take their historical roots in simulating energy-resolved molecular photodissociation or light-absorption cross-section spectra from time-dependent quantum dynamics, as triggered by the seminal work of Heller in the late 1970s that related the cross-section spectrum to the Fourier transform of a wavepacket correlation function.\cite{heller_timedependent_1975}
This is a linear-response approach based on first-order time-dependent  perturbation theory.

More recently, some pressing questions have arisen within the theoretical/computational community about what we actually simulate and how we should better relate to time-resolved spectroscopy experiments beyond linear response.
Indeed, since Mukamel's seminal works \cite{Mukamel1992,Mukamel1995} there has been a clear incentive for theoretical chemistry to address nonlinear and/or two-dimensional (2D) spectroscopy techniques, which are no longer exotic but rather the new standard.\cite{Mukamel2017,Cerullo2019,Collini2021,Krich2023,segatta_time-resolved_2024,Bittner2023}
In particular in the cases of photoinduced (energy or charge) transfers and chemical reactions, the relation between electronic populations and experimental observables has become central.

In addition, taking into account the role of the environment that surrounds the molecule (dissipation, solvent and thermal effects, \textit{etc}.) with the reduced-density operator formalism has become crucial again in this context.
This has led to many flavours of open quantum dynamics approaches, one of them -- used in the present work -- being known as the HEOM method, in which vibrations are considered as a continuous bath (with a statistical spectral density) coupled to the discrete electronic subsystem (quantum few-level reduced-density matrix).
We illustrate herein how a VCH model in a minimal electronic basis set and HEOM simulations in the strong system-bath coupling regime with approximate spectral densities may provide relevant information on spectroscopic signals monotoring the early EET in the m23 system as a complement to alternative semiclassical simulations.\cite{hu_spectral_2021,zhang_what_2025}

The paper is organized as follows.
In the second section, we describe our procedure for the realistic parametrization of both the discrete and continuous high-dimensional models that are used for ML-MCTDH and HEOM dynamics, respectively, and recall the underlying formalism for calculating various spectroscopic responses from simulations.
In the third section, we first provide and compare time-resolved electronic populations and steady-state spectra obtained with both approaches within linear response. We continue with an exposition of nonlinear and 2D spectra; in particular we discuss transient polarization-sensitive signals and show how they provide a fingerprint for EET.
Concluding remarks and outlook are gathered in the last section. 

\section{\label{sec:methods}Methods and Concepts}

\subsection{The models and their Hamiltonians}

\subsubsection{\label{sec:model}High-dimensional vibronic coupling Hamiltonians}

Within an \textit{ab initio} adiabatic description, the electronic system of the m23 molecule is reduced herein to its electronic ground state,  S\textsubscript{0}, and its first two bright singlet electronic excited states, S\textsubscript{1} and S\textsubscript{2}, which are nonadiabatically coupled and have been described in detail in previous works.\cite{fernandez-alberti_nonadiabatic_2009,huang_theoretical_2015,galiana_excitation_2024}

The minima of the S\textsubscript{0} (Franck-Condon point; FCP), S\textsubscript{1}, and S\textsubscript{2} states, and their frequencies, as well as the minimum-energy conical intersection (MECI) between S\textsubscript{1} and S\textsubscript{2}, and its branching-space vectors, have all been optimized and characterized previously with DFT and TD-DFT calculations (CAM-B3LYP/6-31+G\textsuperscript{*} level of theory)\cite{galiana_excitation_2024} using the \textsc{Gaussian} 16 (Revision A.03) quantum-chemistry software package.\cite{frisch_gaussian_2016}
In particular, it has been shown that S\textsubscript{1} is locally excited on the 3-ring branch p3 (at \qty{3.88}{\electronvolt}) while S\textsubscript{2} is locally excited on the shorter 2-ring branch p2 (at \qty{4.55}{\electronvolt}).
The corresponding transition densities are given in \cref{fig:m23_est}, b) and c), together with the electronic transition dipole moments and oscillator strengths.
Let us insist that we focus here on the two locally-excited (excitonic-like within this pseudo-dimer) bright states. 
Potential charge-transfer-like dark states, are weakly-coupled to the locally-excited ones (low differential overlap of $n\pi^*/\pi\pi^*$-type) and are expected not to affect the ultrafast dynamics of EET (see, \textit{e.g.}, our previous investigation in Ref. \cite{breuil_bright--dark--bright_2023}).

In order to run our simulations, we have parametrized a diabatic high-dimensional VCH model for the ground state S\textsubscript{0} and the two coupled PESs of the singlet electronic excited states S\textsubscript{1} and S\textsubscript{2} of m23 with all the $N=93$ in-plane (A') normal modes of vibration.
Such a type of model refers to as a `diabatization by Ansatz'.
At the FCP, the two excited diabatic states that form the interacting basis set, D\textsubscript{1} and D\textsubscript{2}, coincide with the adiabatic ones, S\textsubscript{1} and S\textsubscript{2} by construction.
All the in-plane modes are involved \textit{a priori} in both the intra-state energy and inter-state coupling gradients for the first two electronic excited states (of same symmetry A').
Such a selection includes in particular the dominant quinoidal and acetylenic modes, previously identified as being essential for EET in the PPE dendrimers, \cite{ho_diabatic_2019,fernandez-alberti_nonadiabatic_2009,fernandez-alberti_unidirectional_2010,galiana_unusual_2023,galiana_excitation_2024} here augmented with extra low-frequency modes and C--H vibrations.

Note that out-of-plane modes are soft and anharmonic distortions.
In addition, the bright-excited-state bonding pattern is cumulenic (double-double-double) and much more rigid than the alternate single-triple-single bonding pattern, so that the molecules that have absorbed light are likely to stay planar in the early dynamics after excitation. 
In this situation, all relevant gradients and branching-space vectors are A' (in-plane). 
Hence, the out-of-plane motions are expected to play a dynamical role over the timescale of the overall rotations ($> \qty{1}{\pico\second}$), and thus be treated as essentially inducing statistical spectral broadening as regards the ultrafast vibronic dynamics of the ideal planar system involved in the EET process.

Our most sophisticated model Hamiltonian is in-between the customary linear vibronic coupling (LVC) and the full quadratic vibronic coupling (QVC) models; namely, the off-diagonal bi-linear terms are set to zero here, but diagonal quadratic and bi-linear terms remain flexible (three approximate flavors of it will be discussed in what follows).
The model Hamiltonian matrix (centered at the FCP geometry) thus reads
\begin{equation}
\label{eq:m23_HDQD_LVC_model}
\begin{aligned}
\mathbf{H}(\mathbf{Q})
&=
\begin{pmatrix}
0 & 0 & 0 \\
0 & E^{(1)} & 0 \\
0 & 0 & E^{(2)}
\end{pmatrix}
+
\left(
\widehat{T}_{\text{nu}}
+
\sum_i\frac{1}{2}k^{(0)}_iQ_i^2
\right)
\mathbb{1}_{3}\\
&+
\sum_i
\begin{pmatrix}
0 & 0 & 0 \\
0 & \kappa_i^{(1)}Q_i & 0 \\
0 & 0 & \kappa_i^{(2)}Q_i
\end{pmatrix}
+
\sum_i
\begin{pmatrix}
0 & 0 & 0 \\
0 & 0 & h_i'Q_i \\
0 & h_i'Q_i & 0 \\
\end{pmatrix}\\
&+
\sum_i\sum_j
\begin{pmatrix}
0 & 0 & 0 \\
0 & \frac{1}{2}\gamma_{ij}^{(1)}Q_iQ_j & 0 \\
0 & 0 & \frac{1}{2}\gamma_{ij}^{(2)}Q_iQ_j
\end{pmatrix}
\quad,
\end{aligned}
\end{equation}
where $\widehat{T}_{\text{nu}}$ is a typical one-state normal-mode nuclear kinetic-energy operator and the second matrix (proportional to the identity matrix) is the harmonic reference, parametrized at the minimum of the electronic ground state (FCP) through the knowledge of the frequencies $\omega_i^{(0)^2}=k_i^{(0)}$ (assuming mass-weighted normal coordinates, $Q_i$).

The curvatures of the electronic excited states are allowed to be different together and with the electronic ground state (thus including Duschinsky effects) thanks to the diagonal quadratic terms $\gamma_{ii}^{(s)}$.
In other words, we model the PESs with a Taylor expansion to the second order for the diabatic potential energies (diagonal) and to the first order for the inter-state coupling (off-diagonal).

In the following, we shall refer to \cref{eq:m23_HDQD_LVC_model} as the LVC+$\gamma$ model.
From it, we define the pure LVC model, upon setting the last matrix to zero (quadratic corrections and bi-linear mode mixings).
The intermediate model with quadratic corrections obtained from the LVC+$\gamma$ model but restricted to $i=j$ for $\gamma_{ij}$ will be called LVC$+\gamma_{ii}$.

The LVC model will be used as the system-bath Hamiltonian associated to our open quantum system (see below). Within second quantization, it may be termed a Frenkel-Holstein excitonic Hamiltonian (in relation with the p2 and p3 locally excited states) coupled to a harmonic bosonic bath, or a spin-boson model as regards the two-level excited manifold at the FCP (D\textsubscript{1} and D\textsubscript{2}).

\subsubsection*{Local fitting procedure \textit{via} identification of energy derivatives}

Herein, we give details about the parametrization of the model Hamiltonian defined in \cref{eq:m23_HDQD_LVC_model}. It uses:
i) the vertical transition energies, gradients, and Hessians at the FCP;
ii) the position (with respect to FCP) of the MECI, $\Delta\mathbf{Q}_X$, and the branching-space vectors, $(\mathbf{g},\mathbf{h})$, at this geometry.

The strategy for ensuring that the diabatic and adiabatic states coincide at both the FCP and MECI geometries is similar to the one used in a previous study on a dimensionally-reduced model for m23.\cite{galiana_excitation_2024,gonon_generalized_2019}
The orthogonal pair of branching-space vectors, $(\mathbf{g},\mathbf{h})$, is obtained numerically from the Hessian of the squared energy difference at the MECI, and is rotated into a pair of new branching-space vectors, $(\mathbf{g}',\mathbf{h}')$, such that
\begin{equation}
\mathbf{h}'\cdot\Delta\mathbf{Q}_X=0 \quad,
\end{equation}
where the $\mathbf{h}'$-vector components define the off-diagonal parameters in \cref{eq:m23_HDQD_LVC_model}.
With the 93 selected normal modes, the parametrized rotation angle is $\theta=\qty{18.82}{\degree}$, which accounts for a moderate photochemical reaction-path curvature (second-order relaxation) from FCP to MECI.

We now focus on the parameters for the diabatic potential energies.
The diabatic energies at the FCP geometry, $E^{(s)}(\mathbf{Q}=\bm0)$, simply identify to the adiabatic vertical transition energies of S\textsubscript{1} and S\textsubscript{2}.
Similarly, the diabatic gradients can be made identical to the adiabatic gradients of S\textsubscript{1} and S\textsubscript{2} within a Hellmann-Feynman spirit for a local crude-adiabatic representation.

The generation of the curvatures and bi-linear cross-terms for the excited diabatic states is more involved.
The matrices $\bm{\gamma}^{(s)}$ (which account for intra-state mode mixing) cannot be directly identified to the vertical transition Hessians of the adiabatic states S\textsubscript{1} and S\textsubscript{2}.
Indeed, because we defined an inter-state coupling gradient, $\mathbf{h}'$, the definition of the diabatic Hessians (sums of the ground-state Hessian $\mathbf{K}_{\text{S}_0}$ and the $\bm{\gamma}^{(s)}$ matrices) must reflect the effect of the coupling when producing the adiabatic Hessians.
To ensure this, we define the diabatic Hessians as, according to a second-order Jahn-Teller-type formula,\cite{gonon_applicability_2017}
\begin{equation}
\begin{aligned}
\mathbf{K}_{\text{S}_0}+\bm{\gamma}^{(1)}&=\mathbf{K}_{\text{D}_1}=\mathbf{K}_{\text{S}_1}+2\frac{\mathbf{h}'\mathbf{h}'^{\text{T}}}{E_{\text{S}_2}-E_{\text{S}_1}} \quad,\\
\mathbf{K}_{\text{S}_0}+\bm{\gamma}^{(2)}&=\mathbf{K}_{\text{D}_2}=\mathbf{K}_{\text{S}_2}-2\frac{\mathbf{h}'\mathbf{h}'^{\text{T}}}{E_{\text{S}_2}-E_{\text{S}_1}} \quad.
\end{aligned}
\end{equation}
Such an effect is physically motivated but occurs here to have little consequences numerically, only because the FCP energy gap, $E_{\text{S}_2}-E_{\text{S}_1}$, is large enough (a more detailed analysis is provided in \textcolor{black}{SI, section SI-I and fig. SI-1}).

Let us recall that we have restricted ourselves to the 93 in-plane normal modes of the m23 molecule herein.
Overall, the most important values of the Hessians occur to be on-diagonal.
In other words, the normal modes of S\textsubscript{0} almost form an orthonormal eigenbasis for the Hessians of the excited states (almost diagonal matrices).
This means that there is little mode mixing, as proven by the small magnitude of the elements of the matrices $\bm{\gamma}^{(s)}$.
They are also shown in \textcolor{black}{SI, fig. SI-1}.
Such an analysis can be related to the evaluation of Duschinsky matrices, only with the fact that, here, the excited states are computed at the same geometry as the ground-state equilibrium geometry.

However, let us examine, for the mode-mixing matrices, $\bm{\gamma}^{(s)}$, some finer features as regards the frequencies of the normal modes.
There are greater values (hence greater mode mixing) for the group of quinoidal and acetylenic modes (from \qtyrange{1600}{2400}{\per\centi\meter}), both among them and with the other modes.
To some extent, this is also true for the triangular modes (from \qtyrange{1000}{1600}{\per\centi\meter}), with significant mode-mixing among them (see  \textcolor{black}{SI, fig. SI-1}).

The previous analysis can serve two different (but not unrelated) purposes:
i) finding mode-combination strategies for optimally taking into account the correlation between the strongly mixing modes;
ii) establishing relevant system-bath partitions wrt. active modes (as regards spectroscopy and EET) and spectator modes.
Such considerations are expected to help in forming a rationale for designing an optimal tree for ML-MCTDH simulations under high-rank-tensor compression approaches.

\subsubsection*{Validity and limitations of the local fitting procedure}

The electronic energies at the minimum of 
S\textsubscript{0} (FCP) are set with no ambiguity from the onset.
Now, in order to evaluate the validity of our VCH approximate descriptions, we should compare the energies of the \textit{ab initio} S\textsubscript{1} and S\textsubscript{2} adiabatic states at some critical points (obtained wih routine optimization procedures) to the model eigenvalues.
The electronic energies at the minima of S\textsubscript{1}, S\textsubscript{2}, and at the MECI are given in \cref{tab:m23_HDQD_critical_points}.

\begin{table*}[!ht]
\begin{center}
\caption{
Energies in \unit{\electronvolt} of the first two adiabatic and diabatic excited states at the critical points in the \textit{ab initio} PESs (at the CAM-B3LYP/6-31+G\textsuperscript{*} level of theory) and in the LVC model (8-dimensional, from previous work\cite{galiana_excitation_2024}), and LVC+$\gamma$ model (present work; 93-dimensional).
}
\label{tab:m23_HDQD_critical_points}
\small{
\begin{tabular}{l|ccc|ccc|ccc}
  \toprule
  Model & \multicolumn{3}{c}{Full-dimensional (\emph{ab initio})\textsuperscript{a}} & \multicolumn{3}{c}{8-dimensional (LVC)} & \multicolumn{3}{c}{93-dimensional (LVC + $\gamma$)} \\
  Critical Point & MinS\textsubscript{1} & MinS\textsubscript{2} & MECI & MinS\textsubscript{1} & MinS\textsubscript{2} & MECI & MinS\textsubscript{1} & MinS\textsubscript{2} & MECI\textsuperscript{b}\\
  \midrule
  $E(\text{S}_1)$      & 3.61 & 3.99 & 4.30 & 3.67 & 4.00 & 4.40 & 3.62 & 4.13 & -- \\
  $E(\text{S}_2)$      & 4.62 & 4.17 & 4.30 & 4.59 & 4.23 & 4.40 & 4.64 & 4.13 & -- \\
  $\Delta E(\text{S}_1-\text{S}_2)$ & 1.01 & 0.18 & $<5\text{e}^{-4}$ & 1.08 & 0.23 & $<5\text{e}^{-3}$ & 1.02 & $<5\text{e}^{-3}$ & -- \\
  $E(\text{D}_1)$      &   -- &   -- &   -- & 3.67 & 4.01 & 4.40 & 3.64 & 4.13 & -- \\
  $E(\text{D}_2)$      &   -- &   -- &   -- & 4.59 & 4.22 & 4.40 & 4.62 & 4.13 & -- \\
  \bottomrule
\end{tabular}
}
\end{center}
\textsuperscript{a}\footnotesize{Diabatic energies are not determined directly from \emph{ab initio} calculations but are post-processed for the VCH models.} \\
\textsuperscript{b}\footnotesize{For the 93-dimensional VCH model, the effective MECI occurs to be peaked and thus merges with MinS\textsubscript{2}.}
\end{table*}

First, we can remark the excellent energy agreement as regards the S\textsubscript{1} minimum (MinS\textsubscript{1}) obtained within our 93-dimensional LVC + $\gamma$ model and the true S\textsubscript{1} minimum of the \textit{ab initio} PESs.
However, the S\textsubscript{2} minimum (MinS\textsubscript{2}) obtained within our 93-dimensional LVC + $\gamma$ model occurs to merge numerically with an S\textsubscript{1}/S\textsubscript{2} conical intersection, which is a perfectly acceptable situation when the MECI is peaked (the S\textsubscript{2} minimum is the MECI).
Since the \emph{ab initio} S\textsubscript{2} minimum (MinS\textsubscript{2}) exists, the \emph{ab initio} MECI should in fact be slopped (at least slightly).
Such small discrepancies are to be expected from a model that is imperfect as regards anharmonicity effects.
Yet, we expect that the semi-quantitative orders of magnitude shown here are good enough for our model to be relevant.

\subsubsection{Parametrization as an open quantum system}

An open quantum system results from the bipartite  partition of a complex system into an active sub-system treated by quantum mechanics coupled to an environment described by collective bath modes and statistical mechanics. \cite{Breuer2002,Weiss2012,Kuhn2011}
The corresponding generic Hamiltonian matrix of the full system is split into three contributions,
\begin{equation}
\mathbf{H}={\mathbf{H}_\text{S}}+{\mathbf{H}_\text{SB}}+{\mathbf{H}_\text{B}} \quad,
\end{equation}
where ${\mathbf{H}_\text{S}}$ is the Hamiltonian of the open biexcitonic (three-level) system, associated here to the partition of the electronic degrees of freedom from all the vibrational (bosonic) modes.
It is represented by the first diagonal matrix of \cref{eq:m23_HDQD_LVC_model}, with the energies at the FCP geometry of the ground state (set to zero) and of the two excited states, $E^{(1)}$ and $E^{(2)}$.

The environment is a collection of $N=93$ harmonic vibrational modes centered at the origin (FCP point) and associated to the ground-state frequencies. It is represented by ${\mathbf{H}_\text{B}}$ and given by the second matrix of \cref{eq:m23_HDQD_LVC_model}.

Such bosonic modes are separated into ${N}_\text{bath}$ baths that make the effective electronic energies and inter-state coupling fluctuate to first order wrt. the $\mathbf{H}_\text{S}$ reference. 
The system-bath coupling is given by the third and fourth matrices of the LVC model in \cref{eq:m23_HDQD_LVC_model}.
It can be recast as $\mathbf{H}_\text{SB}=\sum\nolimits_{n }^{{{N}_\text{bath}}}{{{\mathbf{S}}_{n }}}{{B}_{n }}$, where ${\mathbf{S}}_{n }$ are the matrices of the projection and transition operators wrt. the electronic states of the system. 
The corresponding excitonic Pauli-type system operators are $\hat{S}_{1}=\left| 1 \right\rangle \left\langle  1 \right|$, $\hat{S}_{2}=\left| 2 \right\rangle \left\langle  2 \right|$, and ${\hat{S}_{3}}=\left| 1 \right\rangle \left\langle  2 \right|+\left| 2 \right\rangle \left\langle  1 \right|$, respectively. 

The collective modes, ${B}_{n }$, are bosonic-type operators given here in coordinate representation for notational simplicity.
Note that they do not appear as such in the HEOM equations but only \textit{via} their statistical correlation functions, as described below.
Three collective modes ${B}_{n}$ are used here (${N}_\text{bath}=3$). 
The first two, ${B}_{1}=\sum\nolimits_{j}{\kappa _{j}^{(1)}}{{Q}_{j}}$ and ${B}_{2}=\sum\nolimits_{j}{\kappa _{j}^{(2)}}{{Q}_{j}}$, are defined from the gradients of the PESs at the reference FCP geometry. 
They are called tuning collective modes since they make the electronic energy gap vary to first order.
The third collective mode, ${B}_{3}=\sum\nolimits_{j}{h_{j}^{'}}{{Q}_{j}}$, is the coupling mode, which induces a first-order electronic inter-state coupling.
 
The evolution of the system reduced-density matrix, $\rho _\mathrm{S}(t)=\mathrm{tr}_\mathrm{B}\left[\rho(t)\right]$, obtained upon tracing out the bath degrees of freedom, is given by a non-Markovian master equation.
When the bath is harmonic, it follows Gaussian statistics and the main tool of the master equation is the matrix of the two-time correlation functions of the bath collective modes,
${{C}_{nn' }}(t)={{\left\langle {{B}_{n }}(t){{B}_{n' }}(0) \right\rangle }_\text{eq}}$,
where ${B}_{n }(t)$ is the time-evolved Heisenberg representation of the operator and ${{\left\langle \bullet  \right\rangle }_\text{eq}}$ denotes the average over a Boltzmann distribution at some given temperature $T$. We neglect here any correlation among the modes of the three baths and consider only diagonal terms ${C}_{nn }(t)={{\left\langle {{B}_{n }}(t){{B}_{n }}(0) \right\rangle }_\text{eq}}$,  with $n=1,2,3$, which will further be denoted ${{C}_{n }}(t)$ with a single index for the sake of simplicity.
The correlation functions may be computed with molecular dynamics in order to have an exhaustive description of the solvent.\cite{Mangaud2015,Geva2020,Dunnett2021}
In this work, it is obtained \textit{via} the spectral densities that encode the interaction of the baths at each frequency.

\begin{figure}[!ht]
    \centering
    \includegraphics[width=0.4\textwidth]{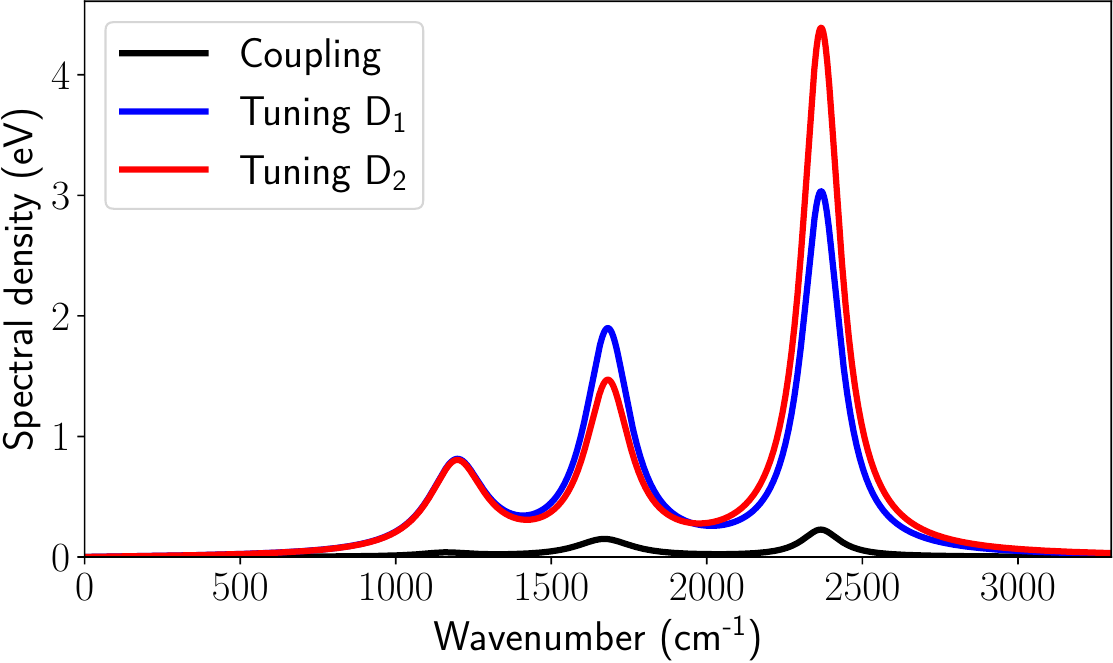}
    \caption{Spectral densities associated to the correlation functions for the three baths (tuning for D\textsubscript{1}, tuning for D\textsubscript{2}, and inter-state coupling between them.)}
    \label{fig:spectral_density}
\end{figure}

In a first step, discrete spectral densities involving the LVC Hamiltonian parameters only are given by
\begin{equation}
 {{J}_{n }}(\omega )=\frac{\pi }{2}\sum\nolimits_{k}{\frac{f_{j}^{(n )2}}{{{\omega }_{j}}}}\,\delta (\omega -{{\omega }_{j}}) \quad,
 \end{equation}
 where $f_{j}^{(n)}=\kappa _{j}^{(n)}$ for the tuning baths, $n =1,2$, and $f_{j}^{(3)}=h_{j}^{'}$ for the coupling bath.
They peak at all individual vibrational frequencies, ${\omega }_{j}$, and form a Dirac comb. In order to account qualitatively for the existence of some dissipative environment and decaying mechanisms, the Dirac delta distributions are broadened by a single spectral width, $\Gamma$, so as to be approximated as Lorentzian functions,
\begin{equation}
\delta (\omega -{{\omega }_{j}})\sim\frac{1}{\pi }\frac{\Gamma}{{{\left( \omega -{{\omega }_{j}} \right)}^{2}}+{{\Gamma }^{2}}} \quad,
\label{eq:delta}
\end{equation}
thus leading to a continuous spectral density.
This is a typical smoothing approach that has been used in another context (MCTDH dynamics) in order to systematically reduce the number of active modes.\cite{Burghardt2010,Burghardt2012}
In the absence of any further information on the dissipative environment, we adopted this heuristic procedure and we took $\Gamma =\qty{80}{\per\centi\meter}$, giving a spectral line enlargement of the same order of magnitude as in a previous simulation on the m22 dimer.\cite{galiana_unusual_2023} 
The corresponding spectral densities are shown in \cref{fig:spectral_density}.

Within this context, the correlation functions are given by the following Fourier transforms,
\begin{equation}
{{C}_{n }}(t-t')=\frac{\hbar}{\pi }\int_{-\infty }^{\infty }{\mathrm{d}\omega {{J}_{n }}(\omega ){{\tilde{n}}_{\beta }}(\omega )}{{e}^{-i\omega (t-t')}} \quad, 
\label{CFourier}
\end{equation}
where $\tilde{n}_{\beta }(\omega )=(1+\coth (\beta \hbar\omega /2))/2$ is the temperature-dependent Bose function with $\beta =1/{{k}_\mathrm{B}}T$, and  ${{k}_\mathrm{B}}$ is the Boltzmann constant.
We take here the standard room temperature, $T=298$ K.
Also, let us stress out here that the symbol $\omega$, as used in our various equations, duly refers to an actual angular frequency (an angle variation per unit of time).
However, for the sake of ease of comparison, we customarily provides its values in terms of its equivalent wavenumber (given in reciprocal centimeters), such is in \cref{fig:spectral_density}.

\subsection{Propagated quantities and time-evolution equations}
\subsubsection{\label{sec:mlmctdh}Quantum wavepacket propagation}

Using the aforementioned VCH models, quantum dynamics calculations were carried out using the ML-MCTDH method, as first implemented in the Heidelberg MCTDH package and further integrated to the \textsc{Quantics} suite of quantum-dynamics programs.\cite{meyer_multi-configurational_1990,beck_multiconfiguration_2000,worth_quantics_2020}

The ML-MCTDH wavefunction was defined by building a simple tree reflecting the different types of molecular vibrations among the 93 in-plane normal modes.
The separation of the branches in the first layer of the tree is based on the frequencies and reduced masses of the normal modes, sorted into four groups defined in \textcolor{black}{SI}.
The latter group reflects also the three-peak features of the spectral density shown in \cref{fig:spectral_density}.
The rest of the three was built in a simple iterative manner down to the last layer where the individual modes are combined two by two.

The number of single-particle functions (SPFs) for each node and combined modes was chosen so as to maintain a reasonable computational time ($<\qty{24}{\hour}$ of human time on a 16-core CPU for \qty{200}{\femto\second}) and small lowest natural weights ($< \num{1e-2}$) throughout the simulation.
A detailed representation of the ML-tree is given in \textcolor{black}{SI, fig. SI-3}.

\subsubsection{\label{sec:heom}Hierarchical equations of motion}

The HEOM method \cite{Kubo1989,Tanimura2006,Tanimura2020,Yan2007,Yan2009,Mangaud2023} for simulating the dissipative dynamics of an open quantum system relies on solving a non-Markovian master equation. It is exact for harmonic baths characterized by Gaussian statistics when the hierachy is  truncated soundly. The central point of the HEOM formalism is the expansion of the correlation functions for each bath, $C_{n}(t)$, and of their complex conjugates,  $\bar{C}_{n} (t)$, as sums of ordered contributions. The most popular expression, adopted here, is a weighted sum of $K_{n}$ rotating and decaying exponential functions,
\begin{equation}
{{C}_{n}}(t)=\sum\nolimits_{k=1}^{{{K}_{n}}}{\alpha _{k}^{(n)}}{{e}^{i\gamma _{k}^{(n)}t}} \quad,
\label{eq:Cdetexpan}
\end{equation}
with complex amplitudes$\alpha_k$ and phases $\gamma_{k}=\Omega_k+i\Gamma_k$. Every term corresponds to a so-called ``bath artificial decay mode'', with positive decay rate, $\Gamma_k$, and positive or negative rotating frequency, $\Omega_k$, associated to the absorption or emission energy between the system and the bath. The Fourier transform (\cref{CFourier}) provides analytical expressions of the $\alpha_k$, $\tilde{\alpha_k}$, and $\gamma_{k}$ coefficients \cite{Pomyalov2010,LeDe2024} when the spectral density is fitted by a sum of Tannor-Meier Lorentzian functions,\cite{Meier1999}
\begin{equation}
 J\left( \omega  \right)=\sum\limits_{l=1}^{{{n}_\text{lor}}}{\frac{{{p}_{l}}\omega }{\hbar^3\left[ {{\left( \omega +{{\Omega }_{l}} \right)}^{2}}+\Gamma _{l}^{2} \right]\left[ {{\left( \omega -{{\Omega }_{l}} \right)}^{2}}+\Gamma _{l}^{2} \right]}} \quad.
 \label{eq:Jexpansion}
\end{equation}
The expressions of the $\alpha_k$ and $\gamma_{k}$ parameters related to this Tannor-Meier method are summarized in SI, section SI-V.A.
For each bath, the spectral density was fitted here in practice by a sum of three functions, ${{n}_\text{lor}}=3$.
The values of the parameters of the Lorentzian functions are given in \textcolor{black}{SI, table SI-I}.

The HEOM formalism consists in a local-in-time system of coupled equations among auxiliary matrices, also called auxiliary density operators (ADOs) having the dimension of the system reduced-density matrix, here 3 $\times$ 3. The coupled equations involve only the parameters of the correlation function (\cref{eq:Cdetexpan}) and the system-bath operators. They are summarized in \textcolor{black}{SI, section SI-V.A}.
At room temperature, we neglect the poles coming from the Bose function (\cref{CFourier}) called the Matsubara frequencies. This approximation notably reduces the computational effort and nevertheless allows a satisfactory qualitative description. However, their consideration is crucial at very low temperature where their number becomes very large and requires different strategies to overcome the computational difficulties due to the prohibitive number of ADOs. \cite{Xu2022,LeDe2024}
Note that the fit by only some Tannor-Meier Lorentzian functions smooths out some structures and provides an average description. 
Different HEOM strategies could be
used such as the free-pole method \cite{Xu2022} or the discretization of the spectral density with
undamped modes.\cite{LiuShi2014} 
We have compared different approaches in a previous paper for a similar system (the symmetrical dimer parent, m22) involving only two baths. \cite{LeDe2024}
However, in the present case of m23, we use three baths and the computational cost should become prohibitive with a standard implementation of HEOM.
This is illustrated with the increase of the number of matrices with the number of modes in SI (SI-Table II).

The HEOM coupled equations were solved according to standard encoding within the adaptive Runge-Kutta method or within a tensor-train format, as described in refs. \cite{Mangaud2023,LeDe2024}.

\subsection{Simulation of spectroscopic signals}

We first present here the typical approaches based on  linear-response theory to obtain the steady-state absorption and emission spectra, using the ML-MCTDH or HEOM formalisms.
Then we summarize the main relations derived from the third-order optical response functions to simulate time-dependent spectroscopy with HEOM. \cite{IshizakiTanimura2005,Tanimura2006,Tanimura2012,Tanimura2020,Yan2010,ZhuYan2011,Shi2011,yan_theoretical_2021,Berkelbach2017}

\subsubsection{\label{sec:spectro}Steady-state absorption and emission spectroscopy}

The linear-response steady-state absorption and emission spectra obtained with quantum wavepacket dynamics, here within the ML-MCTDH formalism, are typically computed as the power spectra of autocorrelation functions.

Let us define without loss of generality a molecular wavepacket within a (1+2)-manifold of diabatic electronic states (the ground state and two excited states) as a set of three components,
\begin{equation}
    \Psi(\mathbf{Q},t) = 
    \left(\psi_0(\mathbf{Q},t)\quad \psi_1(\mathbf{Q},t)\quad \psi_2(\mathbf{Q},t)\right)^\text{T} \quad.
\end{equation}
For a given initial state, $\Psi(t=0)$, we define the autocorrelation function as
\begin{equation}
c(t)=
\Braket{\Psi(0)|\Psi(t)} \quad,
\end{equation}
where integration is performed over the nuclear degrees of freedom, $\boldsymbol{Q}$, according to standard Dirac's braket notations.

Unless otherwise specified, the so-called $t/2$-trick can be used (always valid for a real-valued initial wavepacket and a Hermitian Hamiltonian, according to time-reversibility and unitary propagation), 
\begin{equation}
c(t)=
\Braket{\Psi(-t/2)|\Psi(t/2)} \quad.
\end{equation}
From an infinite propagation time, we should obtain the power spectrum associated to the propagation of this initial wavepacket upon taking the Fourier transform of the autocorrelation function,
\begin{equation}
\label{eq:mctdh_power_spectrum}
    \sigma(\omega)\propto\int_{-\infty}^{\infty}c(t)e^{i\omega t}\mathrm{d}t \quad.
\end{equation}

Because, numerically speaking, the final propagation time, $t_{\rm{f}}$, is always finite, the autocorrelation function has to be multiplied by some periodic and smoothly decaying function, for example $g(t)=\cos^n(\pi t/2t_{\rm{f}})$, in order to make the Fourier-transformed signal well-behaved and avoid the so-called Gibbs phenomenon over the finite time window ($n=1$ in this work).
An extra artificial dissipative broadening was also applied upon convoluting the spectra with a Lorentzian, hence multiplying the autocorrelation function with a damping function $\exp(-t/\tau)$, with an \textit{ad hoc} damping time $\tau=\qty{60}{\femto\second}$ corresponding to a half-width at half-maximum (HWHM) of $\qty{88}{\per\centi\meter}$, so as to mimick the numerical spectral widths effectively obtained from HEOM linear-response simulations of the same spectra (related to, but not to be confused with the Lorentzian HWHM parameter, $\Gamma=\qty{80}{\per\centi\meter}$).

Such a time-to-energy procedure largely depends on how we define the initial state.
For the absorption spectrum, we typically assume a sudden excitation of the ground-state vibrational wavefunction from the electronic ground state to each of the two electronic excited states.
For the emission spectrum, the relaxed vibronic eigenstate within the excited state manifold must be computed beforehand.
In a strongly-coupled and high-symmetrical situation such as in our previous work on the m22 dimer,\cite{galiana_unusual_2023} we could have considered each component of the superposition of the two vibrational contributions for each diabatic state and perhaps play with various mixtures.
However, for m23, the electronic population ratio in the relaxed vibronic state is so close to one in the lowest diabatic state that we considered as a good approximation that we could reduce it to its single component in D\textsubscript{1} $\approx$ S\textsubscript{1}. 

The customary state-to-state power spectra given in \cref{eq:mctdh_power_spectrum} are independent of the transition dipole moments.
They can be further summed with relative  weights that reflect both transition dipole strengths so as to produce a total spectrum.
This is here one of the main differences in computing linear spectra for the two approaches (wavepackets or reduced-density matrices), which is related to the way that the excitation is modelled. 
Within the HEOM description, the polarization of the electric field is included from the onset, so that all the electronic excited states having nonzero transition dipole moments are absorbing. 
Within the (ML-)MCTDH description, the polarization is not directly taken into account, and one has to decide from the onset which of the two electronic excited state is absorbing.

While such simulations based on wavepacket quantum dynamics and Fourier transforms are now routinely used for the simulation of steady-state spectroscopy, their extensions to nonlinear spectroscopies -- requiring a third response formalism as in here -- are becoming pressing matters.
For instance, a recent and striking example of such an application based on the MCTDH formalism can be found in ref. \cite{segatta_time-resolved_2024}.

Now, as regards HEOM, the absorption and emission spectra within linear response are written as follows (involving a ultimate over-the-system trace), 
\begin{equation}
{{\sigma }^{\text{abs}}}(\omega )=\operatorname{Re}\left\{ \mathrm{FT}\left[ \mathrm{tr}_\mathrm{S}\left( \rho _{{{\mu }_{+}}}^{\dagger }(t){{\rho}_{{{\mu }_{+}}}}(0) \right) \right] \right\} \quad,
\label{eq:sigmaabs}
\end{equation}
\begin{equation}
{{\sigma }^{\text{em}}}(\omega )=\operatorname{Re}\left\{ \mathrm{FT}\left[ \mathrm{tr}_\mathrm{S}\left( \rho _{{{\mu }_{-}}}^{\dagger }(t){{\rho}_{{{\mu }_{-}}}}(\text{eq}) \right) \right] \right\} \quad,
\label{eq:sigmaem}
\end{equation}
where the prefactor proportional to $\omega$ or ${{\omega }^{3}}$ for absorption and emission, respectively, has been set equal to one for simplicity, and $\mathrm{FT}$ designates the Fourier transform.

The density matrices involved in them are built from the excitonic lowering and raising (transition-dipole) operators,
\begin{equation}
 \mu _{-}^{(p)}=\sum\nolimits_{j=1}^{2}{{{\mu }_{0j,p}}}\left| 0 \right\rangle \left\langle  j \right| \quad,
 \label{eq:mumoins}
\end{equation}
and
\begin{equation}
 \mu _{+}^{(p)}=\sum\nolimits_{j=1}^{2}{{{\mu }_{0j,p}}}\left| j \right\rangle \left\langle  0 \right| \quad,
\label{eq:muplus}
\end{equation}
where $p=x$ and $p=y$ denote the electric-field polarization directions along the $x$- or $y$-axes within the molecular plane (see \cref{fig:m23_est}). 

For the absorption, ${{\rho}_{{{\mu }_{+}}}}(0)=\sum\nolimits_{p}{\mu _{+}^{(p)}}{{\rho}_\mathrm{S}}(0)$ and all the ADOs are set to zero initially.
For the emission, the lowering operator acts on the system density matrix ${{\rho}_{{{\mu }_{-}}}}(\text{eq})=\sum\nolimits_{p}{\mu _{-}^{(p)}}{{\rho}_\mathrm{S}}(\text{eq})$ and on all the ADOs obtained when the equilibrium vibronic state within the excited electronic manifold -- denoted $\text{eq}$ -- has been reached after relaxation.
We consider different possibilities for the polarization: either parallel to the unit vectors $e_x$ or $e_y$ (see \cref{fig:m23_est}), or also (for completion, since the definition of the two axes is quite arbitrary), along another oblique direction, along the unit vector $(e_x+e_y)/\sqrt{2}$.
The first two cases will be referred to as XX and YY and lead to an initial condition with a superposed electronic state, ${{\mu }_{01,x}}\left| {\text{D}_{1}} \right\rangle +{{\mu }_{02,x}}\left| {\text{D}_{2}} \right\rangle $ or ${{\mu }_{01,y}}\left| {\text{D}_{1}} \right\rangle +{{\mu }_{02,y}}\left| {\text{D}_{2}} \right\rangle $. The oblique polarization, further denoted (X+Y)(X+Y), corresponds to the preparation of the superposed state $({{\mu }_{01,x}}+{{\mu }_{01,y}})\left| {\text{D}_{1}} \right\rangle +({{\mu }_{02,x}}+{{\mu }_{02,y}})\left| {\text{D}_{2}} \right\rangle $ (up to renormalization).

\subsubsection{\label{sec:timedptspectro}Nonlinear spectroscopy}

We now expose the formal ingredients for the simulation of time-frequency resolved or two-dimensional (2D) spectra (in particular within the HEOM formalism) when assuming delta-like pump-probe laser pulses.
In particular, we show how this impulsive approximation allows us to simulate significant polarization-dependent spectroscopic signals that provide fingerprints of the EET. 
We first recall the general formalism of multi-time correlation functions, which are involved in nonlinear 2D spectroscopy.\cite{Mukamel1995}

\begin{figure}[!ht]
    \centering
    \includegraphics[width=0.4\textwidth]{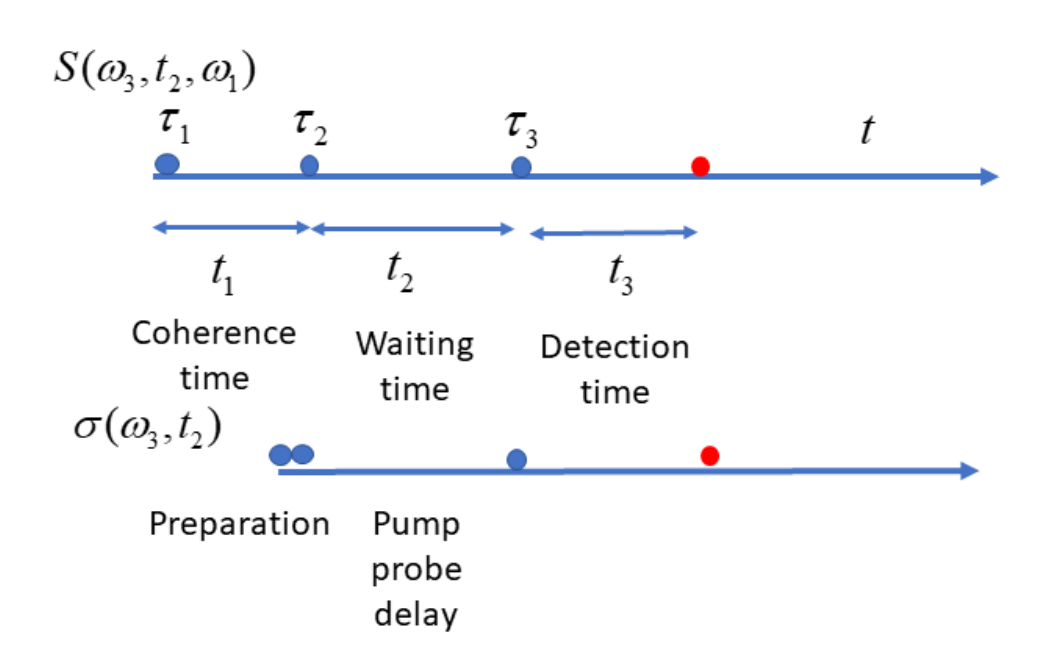}
    \caption{Scheme of principle of the three time intervals involved in the computation of 2D photon echo spectra involving Fourier transforms with respect to $t_1$ and $t_3$ or of the ESE (\cref{eq:ESE}), GSB (\cref{eq:GSB}), or TA (\cref{eq:TA}) spectra, for which $t_1 = 0$ and the Fourier transform is over $t_3$.}
    \label{fig:schema_spnonlin}
\end{figure}

The third-order optical response function  ${{R}^{(3)}}({{t}_{3}},{{t}_{2}},{{t}_{1}})$ that describes the system response to the light-induced perturbations is abundantly documented.\cite{Mukamel1992,Mukamel1995,Khalil2003,Mukamel2009,Valkunas2012,Domcke2015,Mukamel2017,Zhang2019,Collini2021,Krich2023,segatta_time-resolved_2024,Bittner2023}
The simulation of the 2D photon echo spectra involves the interaction between the system and three laser pulses at times $\tau_1$, $\tau_2$, and $\tau_3$ separated by delay times $t_1$ and $t_2$, generating a third-order polarization after a time interval $t_3$ as schematized in \cref{fig:schema_spnonlin}.

The first pulse prepares an optical coherence between the excited states and the ground state so that the field-free evolution during ${{t}_{1}}$ is customarily called the coherence time. 
The second pulse creates a population in the excited manifold or in the ground state and the next free evolution time ${{t}_{2}}$ is the waiting or population time.
The third pulse probes the system after this delay ${{t}_{2}}$ before a detection after the ${{t}_{3}}$ time interval.
The third-order polarisation is then obtained by convolution with the laser pulses centered at $\tau_1$, $\tau_2$, and $\tau_3$, respectively (see \cref{fig:schema_spnonlin}). 
They are each characterized by their polarization,
wave vector ${{k}_{n}}$, frequency, envelope, and initial phase. 
Spectra derived from the third-order reponse ${{R}^{(3)}}({{t}_{3}},{{t}_{2}},{{t}_{1}})$ without convolution with laser pulses are thus ideal spectra that would involve interactions with pulses described by delta distributions $\delta (t)$.

The response function
\begin{equation}
 {{R}^{(3)}}({{t}_{3}},{{t}_{2}},{{t}_{1}})={{R}_\text{rp}}({{t}_{3}},{{t}_{2}},{{t}_{1}})+{{R}_\text{nr}}({{t}_{3}},{{t}_{2}},{{t}_{1}}) \quad,
 \label{eq:Rsum_rp-nr}
\end{equation}
 may be split into two contributions, the rephasing one, ${R}_\text{rp}$ (with detected wave vector $k=-{{k}_{1}}+{{k}_{2}}+{{k}_{3}}$), and the nonrephasing one, ${R}_\text{nr}$ (with $k={{k}_{1}}-{{k}_{2}}+{{k}_{3}}$). The 2D spectrum for a given waiting time ${{t}_{2}}$ is obtained by two Fourier transforms,
\begin{align}
  & S({{\omega }_{3}},{{t}_{2}},{{\omega }_{1}}) \nonumber  \\ & =\operatorname{Re}\int_{0}^{\infty }{\mathrm{d}{{t}_{1}}}\int_{0}^{\infty }{\mathrm{d}{{t}_{3}}{{e}^{i{{\omega }_{1}}{{t}_{1}}+i{{\omega }_{3}}{{t}_{3}}}}{{R}_\text{nr}}({{t}_{3}},{{t}_{2}},{{t}_{1}})} \nonumber  \\ 
 & +\operatorname{Re}\int_{0}^{\infty }{\mathrm{d}{{t}_{1}}}\int_{0}^{\infty }{\mathrm{d}{{t}_{3}}{{e}^{-i{{\omega }_{1}}{{t}_{1}}+i{{\omega }_{3}}{{t}_{3}}}}{{R}_\text{rp}}({{t}_{3}},{{t}_{2}},{{t}_{1}})} \quad. \nonumber  \\ 
 \label{eq:S2D}
\end{align}

The integral over the $\omega_1$ frequency 
\begin{equation}
 \sigma ({{\omega }_{3}},{{t}_{2}})=\int_{0}^{\infty }{S({{\omega }_{3}},{{t}_{2}},{{\omega }_{1}})}\mathrm{d}{{\omega }_{1}} \quad,
\end{equation}
is the transient absorption (TA). 
Within the field of time-resolved spectroscopy, TA is usually defined as the differential absorbance within a pump-probe context: namely, $\Delta A=A(\mathrm{w/~pump}) - A(\mathrm{w/o~pump})$ (aka. differential optical density).  Hereby, we investigate TA from the three-pulse photon-echo 2D polarization upon assuming that the first two pulses overlap.\cite{Mukamel1995,Zhang2019} TA bears some resemblance with the so-called transient grating (TG) signal because the third-order polarization in TA is identical to that in TG, only the measurements are different in TA and TG.\cite{Fleming1996} The time ${{t}_{2}}$ takes the meaning of a delay between the pump and a probe. For the sake of convenience, the subscripts are removed for $t_2$ and $\omega_3$ and the time-dependent spectrum in \cref{eq:sigma} is denoted $\sigma ^\text{TA}(\omega ,t)$. 
Note that specific polarizations may be used for the pump and the probe for polarization sensitive signals. They will be specified when necessary. 
By using \cref{eq:S2D} with overlapping first two pulses, the computation of ${{\sigma }^\text{TA}}(\omega ,t)$ involves the responses for ${{t}_{1}}=0$,
\begin{equation}
 \sigma^\text{TA} ({{\omega }},{{t}})=\operatorname{Re}\int_{0}^{\infty }{\mathrm{d}{{t}_{3}}\left[ {{R}_\text{nr}}({{t}_{3}},{{t}},0)+{{R}_\text{rp}}({{t}_{3}},{{t}},0) \right]}{{e}^{i{{\omega }}{{t}_{3}}}} \quad. 
 \label{eq:sigma}
\end{equation}

The expressions of the responses for HEOM simulations are detailed in \textcolor{black}{SI, section SI-VI.A}. 
In principle, ${{\sigma }^\text{TA}}(\omega ,t)$ is the sum of three contributions related to different absorption or emission processes : ${\sigma }^{\text{ESE}}(\omega ,t)$ describes the excited-state stimulated emission (ESE), ${\sigma }^{\text{GSB}}(\omega ,t)$ corresponds to the ground-state bleaching (GSB), and ${\sigma }^{\text{ESA}}(\omega ,t)$ gives the excited-state absorption (ESA). 
The ESA process is not considered here because the electronic basis set is truncated to the ground and the two D\textsubscript{1} and D\textsubscript{2} excited states.
In our simulations, we consider
\begin{equation}
 {{\sigma }^{\text{TA}}}(\omega ,t)={{\sigma }^{\text{ESE}}}(\omega ,t)+{{\sigma }^{\text{GSB}}}(\omega ,t) \quad.   
\label{eq:TA}
\end{equation}
We refer to \textcolor{black}{SI, section SI-VI.A} for more details and we give here only the final operational expressions for the specific response functions involved in ESE and GSB. 
We have 
\begin{equation}
 {{\sigma }^{\text{ESE}}}({{\omega }},{{t}})=\operatorname{Re}\int_{0}^{\infty }{\mathrm{d}{{t}_{3}}{{e}^{i{{\omega }}{{t}_{3}}}}{{R}^{\text{ESE}}}({{t}_{3}},{{t}},0)} \quad,
 \label{eq:ESE}
\end{equation}
and
\begin{equation}
 {{\sigma }^{\text{GSB}}}({{\omega }},{{t}})=\operatorname{Re}\int_{0}^{\infty }{\mathrm{d}{{t}_{3}}{{e}^{i{{\omega }}{{t}_{3}}}}{{R}^{\text{GSB}}}({{t}_{3}},{{t}},0)} \quad,
 \label{eq:GSB}
\end{equation}
with
\begin{equation}
{{R}^{\text{ESE}}}({{t}_{3}},{{t}_{2}},{{t}_{1}})=\left\langle {\bar{\mu}}_{-}G({{t}_{3}})\left\{ G({{t}_{2}})\left[ G({{t}_{1}})\left( {\bar{\mu }_{+}}{\bar{\rho}_{0}} \right){\bar{\mu }_{-}} \right]{\bar{\mu }_{+}} \right\} \right\rangle \quad,  
\label{eq:R1}
\end{equation}
and
\begin{equation}
 {{R}^{\text{GSB}}}({{t}_{3}},{{t}_{2}},{{t}_{1}})=\left\langle {\bar{\mu }_{-}}G({{t}_{3}})\left\{ {\bar{\mu }_{+}}G({{t}_{2}})\left[ G({{t}_{1}})\left( {\bar{\rho}_{0}}{\bar{\mu }_{-}} \right){\bar{\mu }_{+}} \right] \right\} \right\rangle \quad. 
 \label{eq:R3}
\end{equation}
There, $\bar{\rho}_{0}$ represents a vector of matrices formed by the system density matrix and all the ADOs.
At the initial time, all the ADOs are nil and the system is in the ground electronic state. 
$G(t)$ is the HEOM propagator driving the ensemble of HEOM matrices during a time $t_1$, $t_2$, or $t_3$.
$\bar{\mu }={{I}_{N}}\otimes \mu $ where $I$ is the identity matrix, $N$ is the number of HEOM matrices, and $\mu$ is the dipole transition matrix of the system. 
This means that this $\mu$ matrix must be applied to the system density matrix and to every ADO. \cite{Tanimura2006}
The relevant polarization must be specified for each application of the lowering or raising transition operators.
The notation $\left\langle \bullet  \right\rangle $ means here the final trace over the system.
The trace over the bath is implicitly done by the HEOM propagator. 
The order of the right or left applications of the transition matrix after the propagation during the specific time intervals are explained in SI, section SI-IV.A.
When ${{t}_{1}}=0$, one has $G({{t}_{1}=0})=1$, and the right part of the bracket describes the initial state that is assumed to be prepared by an impulsive delta-like pulse. 
For ESE, this initial state, ${{\mu }_{+}}{{\rho}_{0}}{{\mu }_{-}}$, is the system reduced-density matrix of a superposition of the excited states weighted by the transition dipoles. 
For GSB, from \cref{eq:R3} the initial state is the ground state populated with the norm of the previous superposition. 
However, each initial state can be renormalized for convenience.

The GSB contribution does not depend on $t_2$ in the particular case of the time-dependent transient absorption spectra when computed by using the response for $t_1 = 0$, making a grid only along $t_2$ and $t_3$ and assuming impulsive laser pulses. 
Such a situation is discussed in the authoritative book by Mukamel (chapter 11).\cite{Mukamel1995}
In the HEOM case, the system is in the ground state and the bath is at equilibrium (with nil ADOS). 
Whatever the value of $t_2$, the total system remains the same as before the excitation and the following correlation function does not change. 
Within a wavepacket approach, the system is either described by an eigenstate or by a Boltzmann mixture of eigenstates, and only a phase is varying with $t_2$ without any effect on the spectrum, as expected. 
Note that, for our 2D-spectroscopy simulations, the responses (rephasing and non-rephasing) are computed for grids along $t_1$ and $t_3$ and the GSB is not taken as a constant.
Thus, $\sigma^{\text{GSB}}$ corresponds to the absorption spectrum of the linear response up to an arbitrary factor, which diseappears  when normalizing.
Similarly, the ESE contribution evolves towards the stationary emission spectrum of the linear response when the delay $t$ ($t_2$) becomes sufficiently long.
Using the ${R}^{\text{ESE}}({t}_{3},{t},0)$ response involves a specific initial superposed state but the new equilibrated thermal state for long $t_2$ is in principle independent on the initial state. Both the absorption and emission spectra of the linear response are limiting cases of the time-dependent approach.

Integrated signals over the $\omega$ ($\omega_3$) frequency contain quite significant fingerprints of the dynamics within the excited manifold.
In addition, they are easy to compute via the response functions since one has
\begin{align}
  & {{S}^{\text{TA}}}(t)=\int_{0}^{\infty }{d\omega \,({{\sigma }^{\text{ESE}}}(\omega ,t)+{{\sigma }^{\text{GSB}}}(\omega ,t))}  \nonumber \\ 
 & ={{S}^{\text{ESE}}}(t)+{{S}^{\text{GSB}}}(t)\nonumber \\
 & ={{R}^{\text{ESE}}}(0,t,0)+{{R}^{\text{GSB}}}(0,t,0) \quad. \label{eq:signal(t)}
\end{align}
$S^\text{ESE}(t)$ gives the surface area of the emission spectrum. 
As ${R}^{\text{GSB}}(0,t,0)$ does not depend on the delay $t$, it is just a constant and $S^\text{ESE}(t)$ is the main component. 
When the main active excited states are populated by different transition dipole components, polarization-sensitive detection of these integrated signals  may provide different types of information about the energy transfer. 
As illustrated below, the diagonal cases, ${S}^{\text{TA}}_{\text{XX}}(t)$ or ${S}^{\text{TA}}_{\text{YY}}(t)$, are linked to the decay of the electronic coherence in a purely dephasing case or of the the population or both. 
An off-diagonal case, ${S}^{\text{TA}}_{\text{YX}}(t)$ or ${S}^{\text{TA}}_{\text{XY}}(t)$, probes the increase of population in the final state. 
Finally, by combining detection signals with parallel or perpendicular polarizations, one gets insight into the rotation of the polarization by the anisotropy decay signal, 
\begin{equation}
r(t)=\frac{S_{\parallel }^{\text{TA}}(t)-S_{\bot }^{\text{TA}}(t)}{\Delta {{S}_{\text{iso}}}(t)} \quad,
\label{eq:raniso}
\end{equation}
where $\Delta {{S}_{\text{iso}}}(t)$ is the isotropic absorption difference, 
\begin{equation}
\Delta {{S}_{\text{iso}}}(t)=S_{\parallel }^{\text{TA}}(t)+2S_{\bot }^{\text{TA}}(t)  \quad.
\label{eq:deltaiso}
\end{equation}
Such a polarization-sensitive analysis has been proposed in the literature to decipher the different electronic or vibrational coherences during the relaxation.\cite{Bai2015,yan_theoretical_2021}

\section{\label{sec:results}Results and discussion}

We present herein population dynamics and spectroscopic linear and non-linear signals revealing EET with ML-MCTDH simulations using the full vibronic model and with HEOM using spectral densities parameterized by the Tannor-Meier method.
For the particular system investigated here, the computational effort happens to increase quite heavily as regards the convergence threshold, especially when generating multidimensional time grids for non-linear spectroscopy.
Our strategy here is to consider decent enough models for ensuring relevant semi-quantitative results at reasonable computational time. 
The first approximation concerns the spectral density. 
When the full density is retained with the three peaks for each bath, the model is denoted "3L model" for three Lorentzian functions per bath. 
It is expected that the most important modes are those at high frequencies corresponding to the acetylenic vibrations and to the quinoidal distortions.
By cutting off the low frequency peak in each bath, we define the "2L model" focusing on the two peaks around \num{1600} and \qty{2300}{\per\centi\meter}. 
The second approximation is relative to the HEOM level. 
Convergence tests are presented in SI, sections SI-V.C and SI-VI.E. 
We further discuss there the computational effort related to the increase of the number of matrices with the level and with Matsubara terms. 
We also illustrate that using a near-converged level may nevertheless provide relevant semi-quantitative information.
Such a strategy appears as being a good balance for exploring such a complex system with three baths that are strongly coupled to the electronic system.

\subsection{\label{sec:eet}EET dynamics}

\begin{figure*}[!ht]
    \centering
    \includegraphics[height=0.23\textheight]{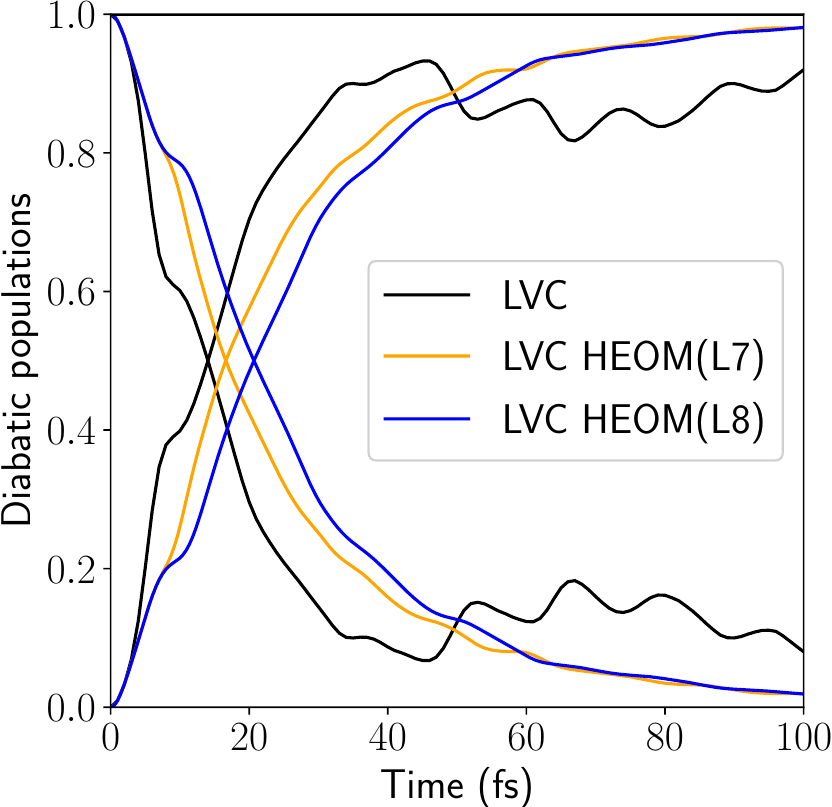}
    \includegraphics[height=0.23\textheight]{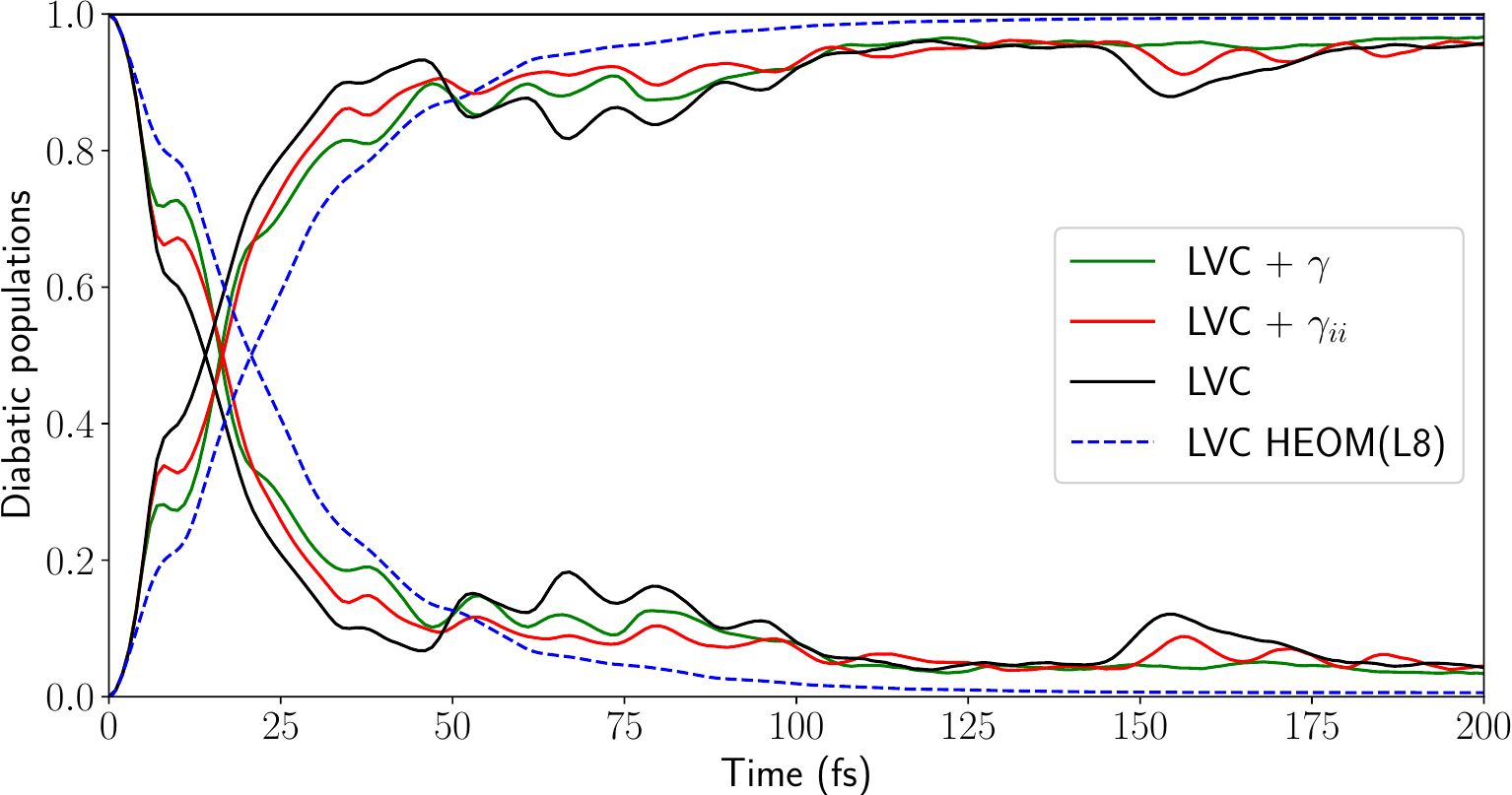}
    \caption{Diabatic populations for different models and propagation methods for the simulation of EET.
    Left: comparison of the use of the ML-MCTDH and HEOM (hierarchy levels L7 and L8) formalisms with the LVC model.
    Right: comparison of the use of the LVC, LVC$+\gamma_{ii}$, and LVC$+\gamma$ models within the ML-MCTDH formalism, up to longer timescales.
    }
    \label{fig:comparison_MLMCTDH_HEOM}
\end{figure*}

We first probe the characteristics of intramolecular EET within the m23 molecule by applying a sudden excitation from the ground state to the second electronic excited state.
In practice, this is achieved upon initializing the system with the ground-state equilibrium wavepacket or reduced-density and placing it at $t=0$ in the diabatic state localized on the p2 branch (D\textsubscript{2}, so-called donor state).
From this initial state, the system is propagated within the excited-state manifold, allowing transfer to the diabatic state localized on the p3 branch (D\textsubscript{1}, so-called acceptor state).
The electronic population transfer is shown in \cref{fig:comparison_MLMCTDH_HEOM} (left panel) for both ML-MCTDH and HEOM dynamics for the full 3L model. A comparison with the 2L model is given in SI, figures SI-fig.8 and SI-fig.10.

Population inversion occurs after around \qty{20}{\femto\second} in all types of simulations.
Comparing the time evolution in the case of the LVC model only, we see that the population transfer calculated within the HEOM formalism is slightly slower than the one computed with a discrete definition of the modes (ML-MCTDH).
Note that it is not surprising to keep observing sustained oscillations in populations calculated with a wavepacket method, since we really are simulating a finite, isolated, and closed system microcanonically (the broadening is phenomenological and only introduced after, at the Fourier-transform stage, to get a spectrum with realistic peak widths), while HEOM is a dissipative and canonical method per construction and is expected to provide full monotonic decay eventually.
We also note that the more advanced case with eight hierarchy levels (L8) yields slightly delayed transfer dynamics compared to L7.
This seems to be correlated with accounting for the small timelag due to the inflection feature around \qty{10}{\femto\second}, also observed in ML-MCTDH results. 

The ML-MCTDH propagation has also been carried out with the more advanced LVC$+\gamma_{ii}$ and LVC+$\gamma$ models, which include second-order ``intra-state couplings''.
The results are shown in \cref{fig:comparison_MLMCTDH_HEOM} (right panel) for longer timescales (up to \qty{200}{\femto\second}).
The three models with the same propagation method (ML-MCTDH and similar ML-tree) exhibit small differences.
The LVC model yields faster population transfer than the models including different curvatures (quadratic intra-state couplings) and mode mixing (bi-linear intra-state couplings).
This, again, goes with a smaller timelag associated to a less pronounced inflection feature around \qty{10}{\femto\second}.
After population inversion, the LVC model exhibits less stability once the acceptor state is populated (see around \qty{75}{\femto\second} and \qty{150}{\femto\second} for instance).
The more we account for intra-state couplings (quadratic, LVC$+\gamma_{ii}$ and all bi-linear, LVC$+\gamma$), the more monotonic are the transfer to and the relaxation in the acceptor state.

\begin{figure}[!ht]
    \centering
    \includegraphics[height=0.18\textheight]{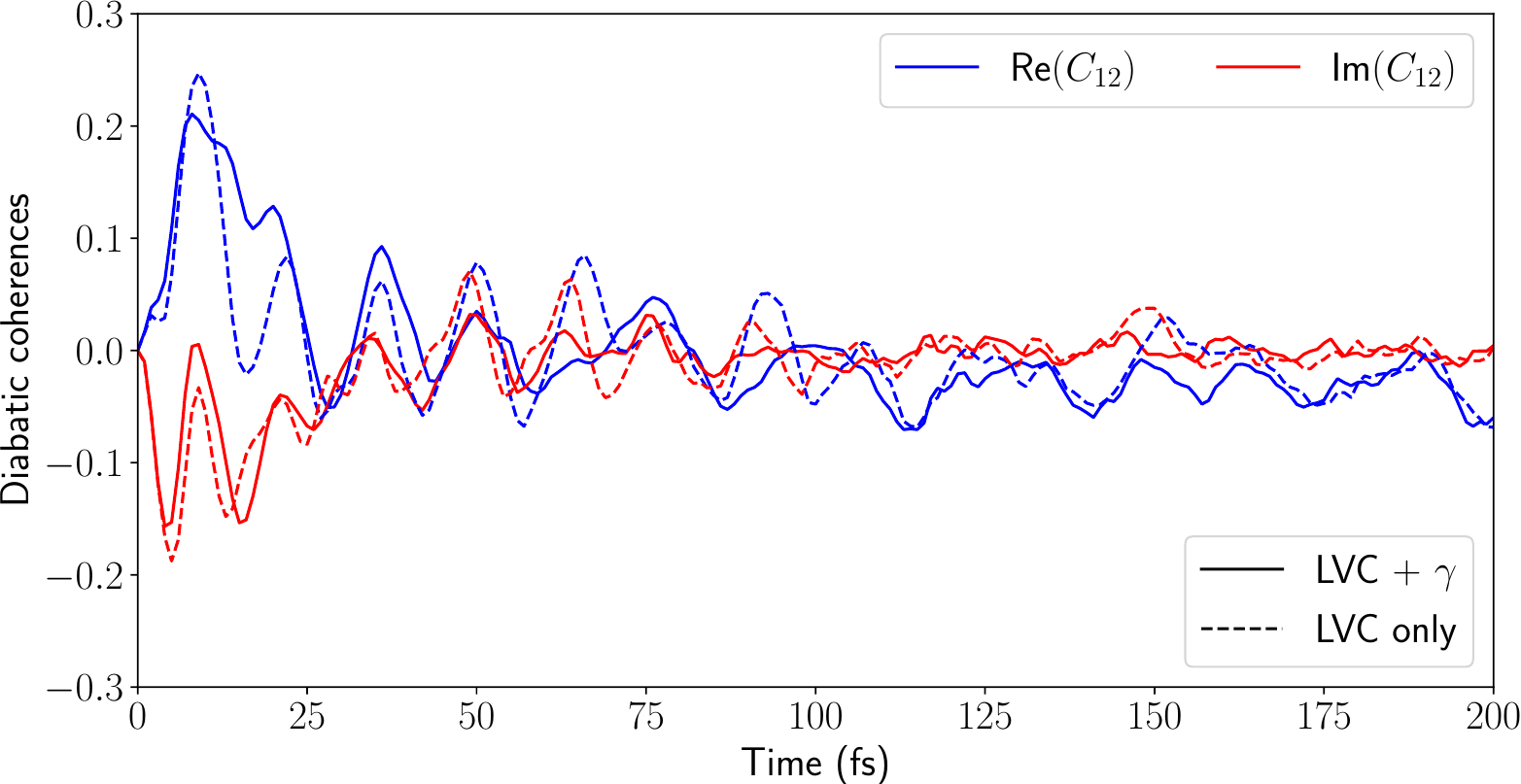}
    \caption{Real and imaginary parts (blue and red lines, respectively) of the diabatic coherence for the LVC$+\gamma$ and LVC only models, with plain and dashed lines, respectively.}
    \label{fig:diacoh}
\end{figure}

The mode-mixing parameters slow down the population transfer dynamics but are eventually required to better relax toward the equilibrium geometry of the diabatic potential energy surfaces.
This is consistent with a slightly stronger damping of the oscillatory features in the diabatic coherence, as shown in \cref{fig:diacoh}.
This means that the mode-mixing parameters somewhat enhance ``internal dissipation'' for the ML-MCTDH calculations, which is included from the onset in an \textit{ad hoc} manner within the HEOM approach.

Again, we stress here that our estimation of the mode-mixing parameters is done thanks to the energy derivatives wrt. a large number of degrees of freedom.
A more involved discussion of the importance of mode-mixing parameters, in particular as regards reducing dimensionality, is made in \textcolor{black}{SI, section SI-I}.

\subsection{\label{sec:linear}Steady-state spectroscopy}

\begin{figure*}[ht!]
    \centering
    \includegraphics[width=0.31\textwidth]{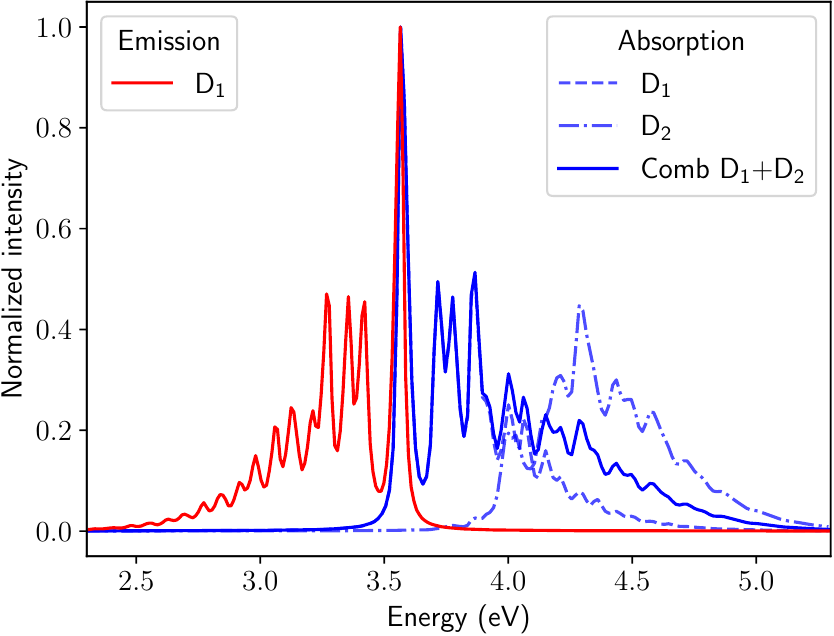}
    \includegraphics[width=0.31\textwidth]{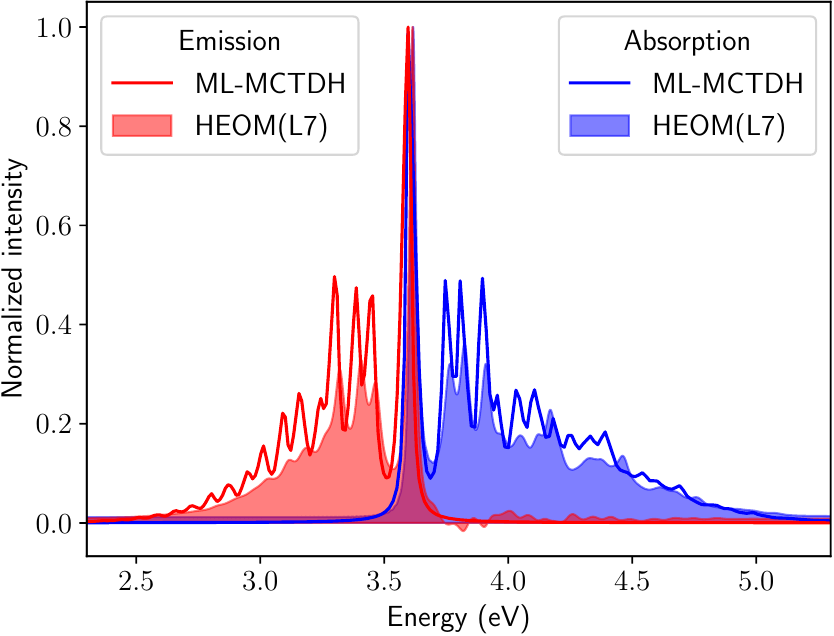}
    \includegraphics[width=0.31\textwidth]{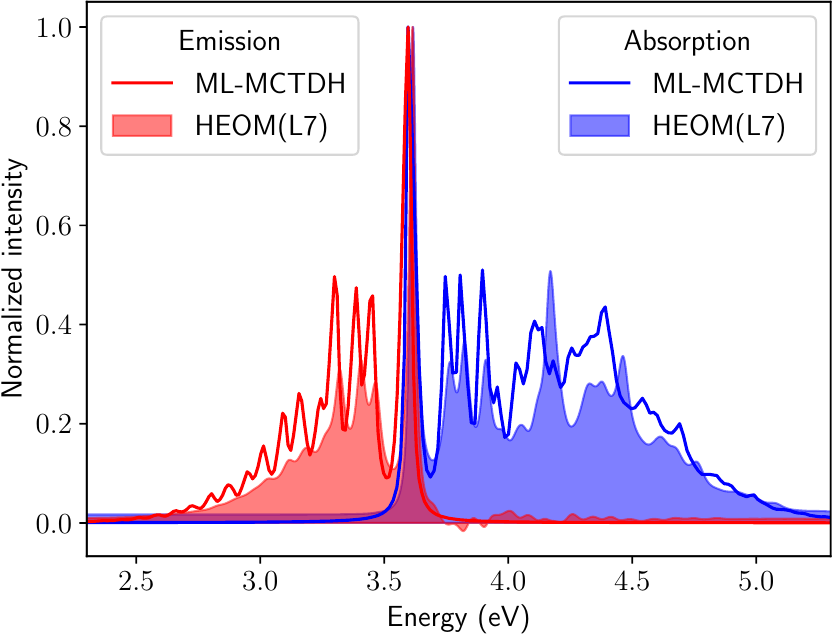}
    \caption{
    Individual or combined contributions to the absorption spectra from the ground state to D\textsubscript{1}, D\textsubscript{2}, or both: dashed blue, dotted blue, and plain blue, respectively.
    Emission spectra from the vibronic ground state within the excited (D\textsubscript{1}, D\textsubscript{2}) manifold: plain red line.
    The ML-MCTDH spectra are broadened by using a damped autocorrelation function with damping time $\tau=\qty{60}{\femto\second}$.
    The HEOM spectra are displayed as filled spectra for the sake of visibility.
    Left: ML-MCTDH spectra for the LVC$+\gamma$ model.
    Center: ML-MCTDH and HEOM spectra for the LVC model with dipole strength ponderation,  \cref{eq:absorption_ponderation_dipstr}.
    Right: ML-MCTDH and HEOM spectra for the LVC model with a ponderation corresponding to an electric field along an oblique polarisation (case (X+Y)(X+Y)),  \cref{eq:absorption_ponderation_xypola}.
    }
    \label{fig:linear_spectroscopy}
\end{figure*}

For ML-MCTDH, the exact same methodology as in a previous work for a reduced model is used to compute power spectra within the realm of steady-state (or linear) spectroscopy.\cite{galiana_excitation_2024}
Two contributions, $\sigma^{(1)}(\omega)$ and $\sigma^{(2)}(\omega)$, are computed for the absorption spectrum (toward the D\textsubscript{1} and D\textsubscript{2} states, respectively) and one for the emission spectrum toward the ground state.
In addition, we estimate the total absorption spectrum as
\begin{equation}
\label{eq:absorption_ponderation_dipstr}
\begin{aligned}
    \sigma^{\text{Total}}(\omega)
    &=
    \left(\mu_{01,x}^2+\mu_{01,y}^2\right)\sigma^{(1)}(\omega)\\
    &+
    \left(\mu_{02,x}^2+\mu_{02,y}^2\right)\sigma^{(2)}(\omega)
\end{aligned}
\end{equation}
which is further normalized.
The weighting is isotropic as regards the polarization direction.
It is also consistent with a sudden approximation for the excitation of both electronic excited states with electronic  transition dipole strengths at the Franck-Condon point as effective weights for each contribution.
They are all given in \cref{fig:linear_spectroscopy} (left panel) for the LVC$+\gamma$ model.

Note that the emission spectrum is approximated as the power spectrum initialized with the most important vibrational contribution in the ground vibronic eigenstate of the excited state manifold (with more than 99\% population in the D\textsubscript{1} electronic state) and projected to the electronic ground state. 

The same is done in \cref{fig:linear_spectroscopy} (center panel), obtained with the LVC model in order to compare with the HEOM linear-response spectrum obtained for the 3L model with the approximate level L7 without Matsubara terms. Convergence tests and comparison with the 2L model are presented in SI, section SI-V.C.
Let us recall that the total absorption spectrum to both excited states with HEOM is obtained \textit{a posteriori} by summing the spectra of the XX and YY cases so as to reproduce the weighting given in \cref{eq:absorption_ponderation_dipstr}.
Both absorption and emission spectra compare well for the two methods.
We observe a larger discrepancy in the spectral region after \qty{4.0}{\electronvolt}, corresponding to the transitions toward the D\textsubscript{2} electronic excited state.

In \cref{fig:linear_spectroscopy} (right panel), we now compare the results obtained by applying some specific oblique polarization (case (X+Y)(X+Y)) for checking its effect,
\begin{equation}
\label{eq:absorption_ponderation_xypola}
\begin{aligned}
    \sigma^{\text{Oblique}}(\omega)
    &=
    \left(\mu_{01,x}+\mu_{01,y}\right)^2\sigma^{(1)}(\omega)\\
    &+
    \left(\mu_{02,x}+\mu_{02,y}\right)^2\sigma^{(2)}(\omega) \quad.
\end{aligned}
\end{equation}
The oblique absorption spectrum, $\sigma^{\text{Oblique}}(\omega)$ is obtained with ML-MCTDH from the isolated spectra with weights in the sum that are different from the previous case, $\sigma^{\text{Total}}(\omega)$.
With HEOM, the oblique spectra are computed from \cref{eq:sigmaabs} and \cref{eq:sigmaem} with
${{\rho}_{{{\mu }_{+}}}}(0)=\sum\nolimits_{p}{\mu _{+}^{(p)}}{{\rho}_{S}}(0)$ and ${{\rho}_{{{\mu }_{-}}}}(\text{eq})=\sum\nolimits_{p}{\mu _{-}^{(p)}}{{\rho}_{S}}(\text{eq})$, respectively, and the dipole operators given by \cref{eq:mumoins} and \cref{eq:muplus}.

The previously mentioned difference as regards absorption toward D\textsubscript{2} is even more remarkable for this numerical experiment.
This is consistent with the fact that the $y$-polarized spectrum gives more importance to the absorption toward D\textsubscript{2}, which amplifies the effect of the discrepancy within this higher-energy spectral range.

In both approaches, we observe that the emission spectrum exhibits no Stokes shift.
Both absorption and emission spectra overlap within a marked Lorentzian-like peak at the band origin 0--0.
This is typical of very fast nuclear dynamics, which is to be expected when Mukamel’s $\kappa =\Lambda /\Delta $ parameter \cite{Mukamel1995} is larger than 1, where ${{\Lambda }^{-1}}$  and  $\Delta $ are estimates of the bath fluctuation timescale and of the amplitude of the fluctuations, respectively.
With HEOM, $\Lambda $ is taken as the spectral density cutoff and $\Delta ={{C}^{1/2}}(t=0)$ is given by the squareroot of the initial value of the bath correlation function.\cite{Mangaud2023}

A strict comparison of the spectra generated by ML-MCTDH or HEOM may seem somewhat disputable \textit{per se}.
Our objective is not to view such approaches as capable of competing for providing the same signal, and we did not aim at doing so from the onset. 
Our intention is rather to check that all such approaches, subject to their own built-in hypotheses, approximations, and limitations, can provide similar enough qualitative information as regards the overall dynamics of the system. Nevertheless, while the spectra generated with ML-MCTDH or HEOM match only qualitatively, their comparison remains much satisfactory in view of the great differences between both approaches.
Each discrete mode of the LVC model is enlarged according to a uniform and quite arbitrary manner (\cref{eq:delta}).
A more relevant simulation of the spectral densities would take the dynamics of the environment into account, as done for example in refs.\cite{Mangaud2015,Geva2020,Dunnett2021,Zuehlsdorff2024}. It is possible that the interaction with a condensed phase leads to a broadening of the LVC peaks that is less pronounced than the value that we used here.
On the other hand, one should be aware that the spectral density may be drastically modified by the solvent, anyhow, as illustrated in Ref. \cite{HunterChin2024}.
The fit of the spectral density obtained from the LVC model with three enlarged Tannor-Meier Lorentzian functions (\cref{eq:Jexpansion}) relies on some working hypotheses, together with assuming that the tuning baths are uncorrelated.
One may also invoke some explicit temperature effects since ML-MCTDH wavepackets are performed at \qty{0}{\kelvin} (further broadened conveniently, according to some well-assessed procedure) and HEOM densities are typically carried out at room temperature. 
In the latter case, computations at low temperature are much more demanding and require another strategy, for instance the free-pole method.\cite{Xu2022,LeDe2024}
Indeed, and as estimated in SI, section SI-V.C, table SI-II, the computational effort is expected to be unusually heavy in the present situation.

\subsection{\label{sec:nonlinear}Toward nonlinear spectroscopy}

\subsubsection{Time and frequency resolved spectra}
The computation of the responses (\cref{eq:R1}) or \cref{eq:R3}) requires a two-dimensional ($t_2,t_3$)-grid. 
The $t_2$ domain is chosen to probe the early nonadiabatic dynamics during \qty{150}{\femto\second}. 
The $t_3$-range spans \qty{300}{\femto\second}. 
It is fixed to ensure enough decay of the correlation function before the Fourier transform. 
The timestep is \qty{0.24}{\femto\second} (10~a.u.). 
A full time grid can be computed within a reasonable computational time for the 2L model only by choosing the HEOM level L7 and by discarding the Matsubara terms.
The required matrices for different approximations are discussed in SI, section SI-V.C, table SI-II. Only some cuts for selected delay times $t_2$ have been performed in the 3L model.
They are compared with the same cuts with the 2L model in \textcolor{black}{SI, fig. SI-10} to assess the validity of the approximation leading to the same qualitative information.

We first consider that the pump and the probe have the same polarization, either parallel to vector $e_x$ (case XX) or $e_y$ (case YY) (see \cref{fig:m23_est}). 
The polarization along an oblique direction $(e_x+e_y)/\sqrt{2}$ (case (X+Y)(X+Y)) is presented in \textcolor{black}{SI, fig. SI-13}. 
Due to the transition dipole strengths, the XX case mainly populates the $\text{D}_1$ state, implying excitation of the $\text{p}_3$ branch.
In contrast, in the YY case, both states are excited with nearly similar weights. 
The $\text{D}_2$ state is now significantly more populated than in the XX case and gives rise to EET toward the acceptor state $\text{D}_1$.

\begin{figure*}
    \centering
    \includegraphics[width=0.45\textwidth]{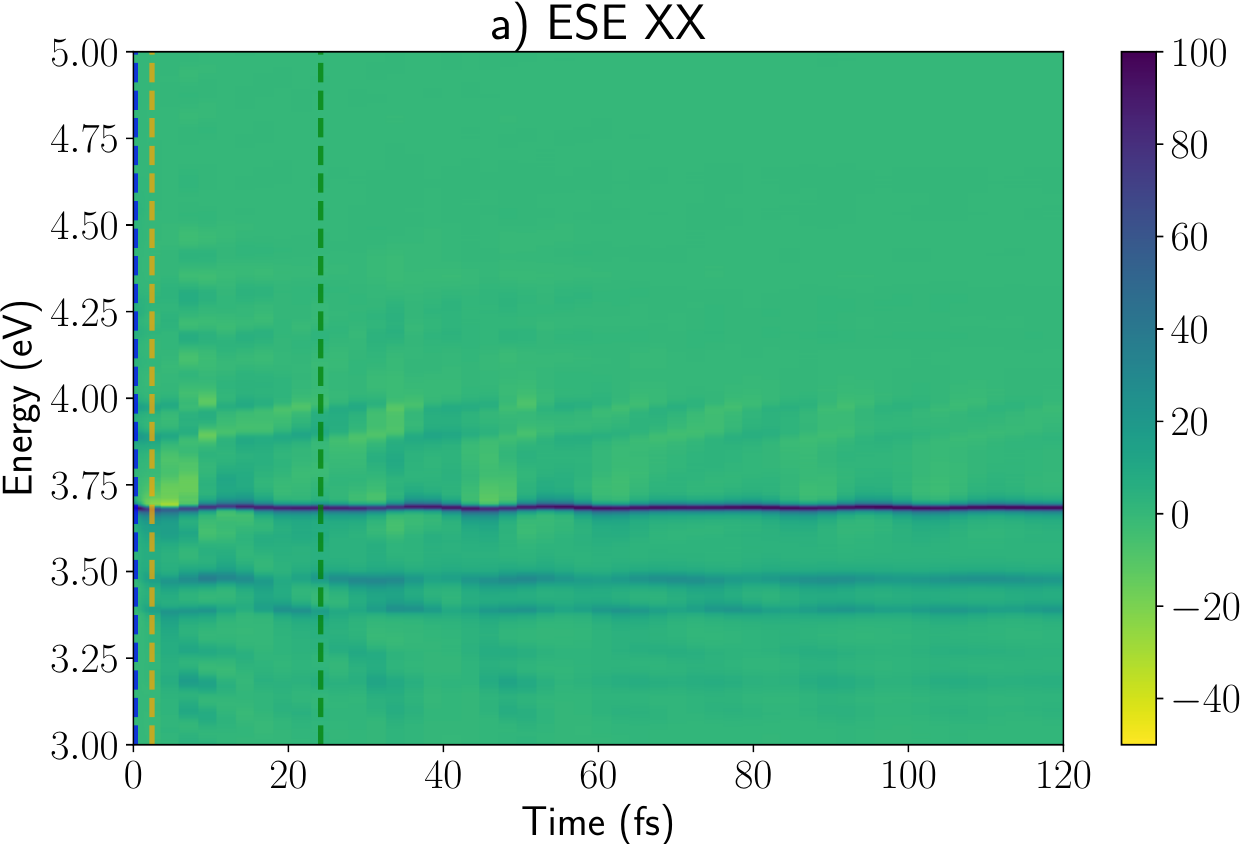}
    \includegraphics[width=0.45\textwidth]{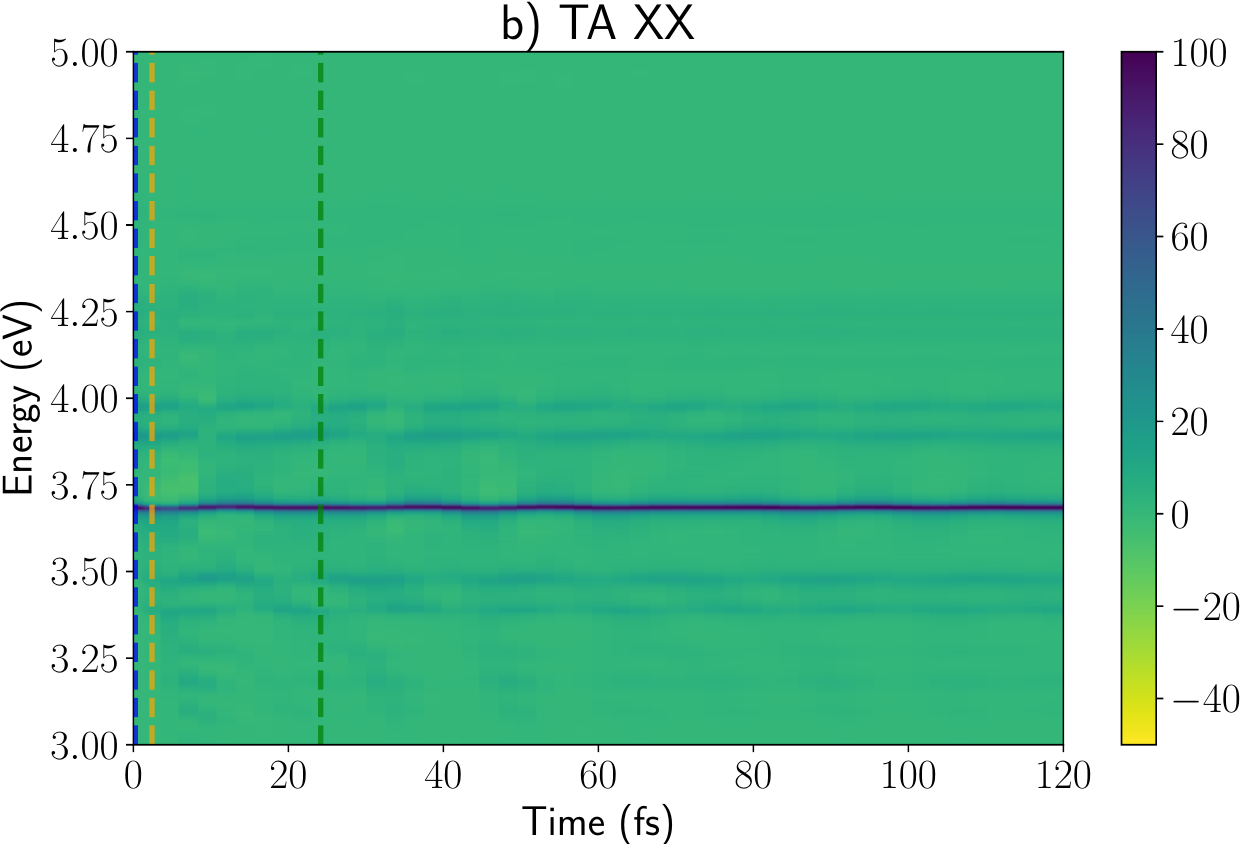}\\
    \includegraphics[width=0.45\textwidth]{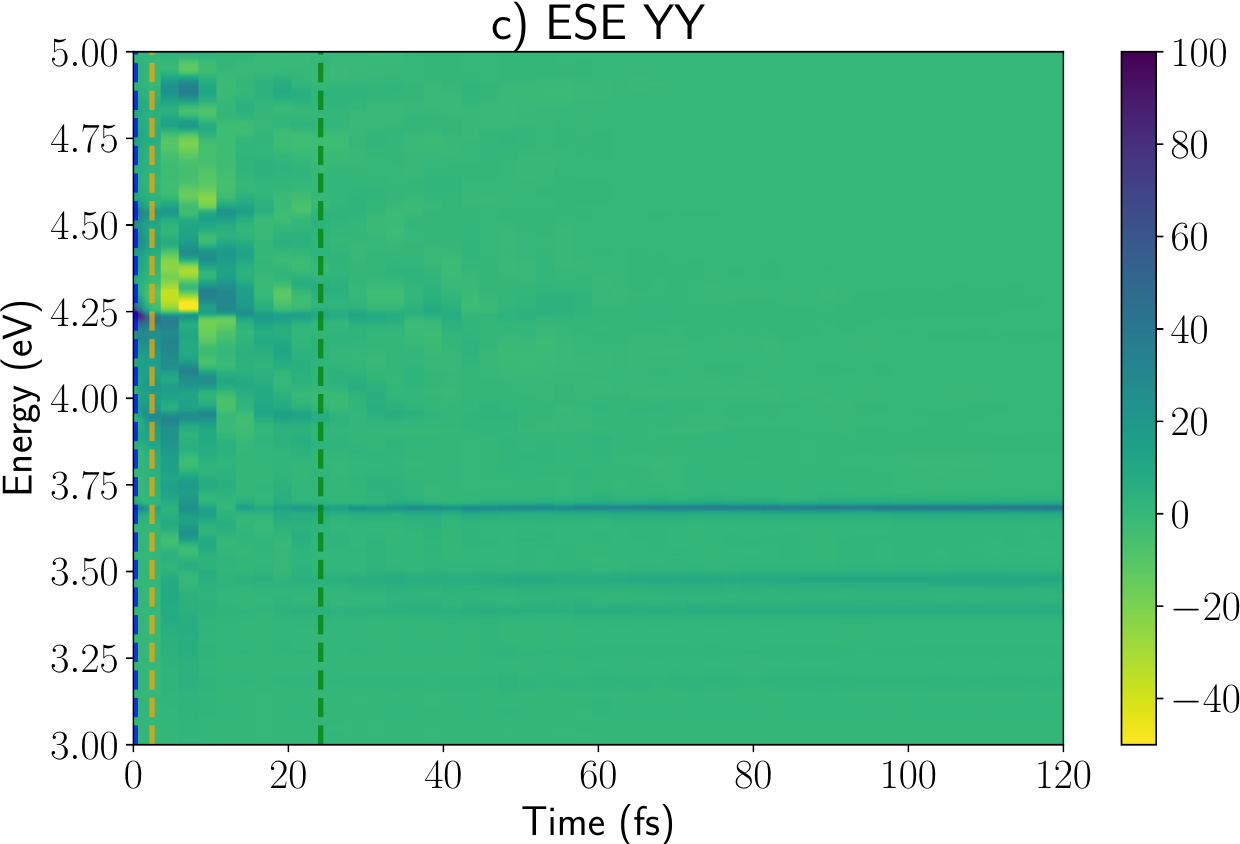}
    \includegraphics[width=0.45\textwidth]{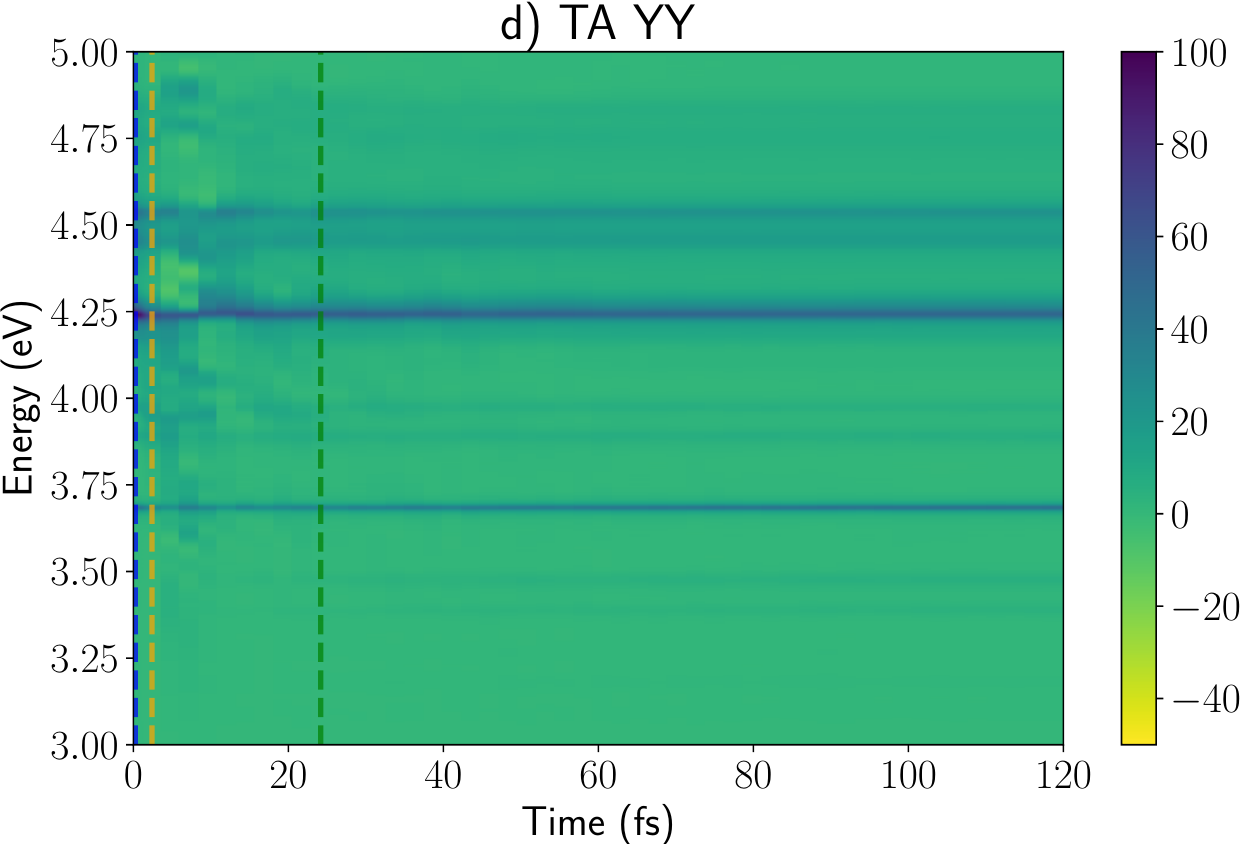}\\
    \caption{
    Time-and-frequency-resolved ESE (left panels)  and TA (right panels) spectrograms
    (\cref{eq:ESE,eq:TA})
    for the approximate 2L model with only the two high-frequency peaks in the spectral densities of \cref{fig:spectral_density}. The pump and probe are delta-like laser pulses. 
    The polarizations of the pump and the probe are parallel -- along $e_x$ (case XX, top panels) or $e_y$ (case YY, bottom panels).
    The vertical dashed lines indicate the cuts presented in \cref{fig:ESE_spectrogram_cuts} for the full 3L model, and in \textcolor{black}{SI, fig. SI-10} for the 2L model.
    }
    \label{fig:ESE_spectrograms}
\end{figure*}

\Cref{fig:ESE_spectrograms} presents time-and-energy-resolved spectrograms of ESE (\cref{eq:ESE}) (left panels) and TA (\cref{eq:TA}) (right panels) for the 2L model obtained with a sampling of 9 delay times $t_2$ with time step \qty{0.24}{\femto\second} (10~a.u.) up to \qty{2.4}{\femto\second} followed by 50 values with step \qty{2.4}{\femto\second}.
The TA spectrograms are the sum of the ESE and GSB spectra. As discussed in \cref{sec:timedptspectro}, the GSB contribution does not depend on $t_2$ and simply adds up the absorption spectrum for the chosen polarization at all times. 

\begin{figure*}
    \centering
    \includegraphics[width=0.9\textwidth]{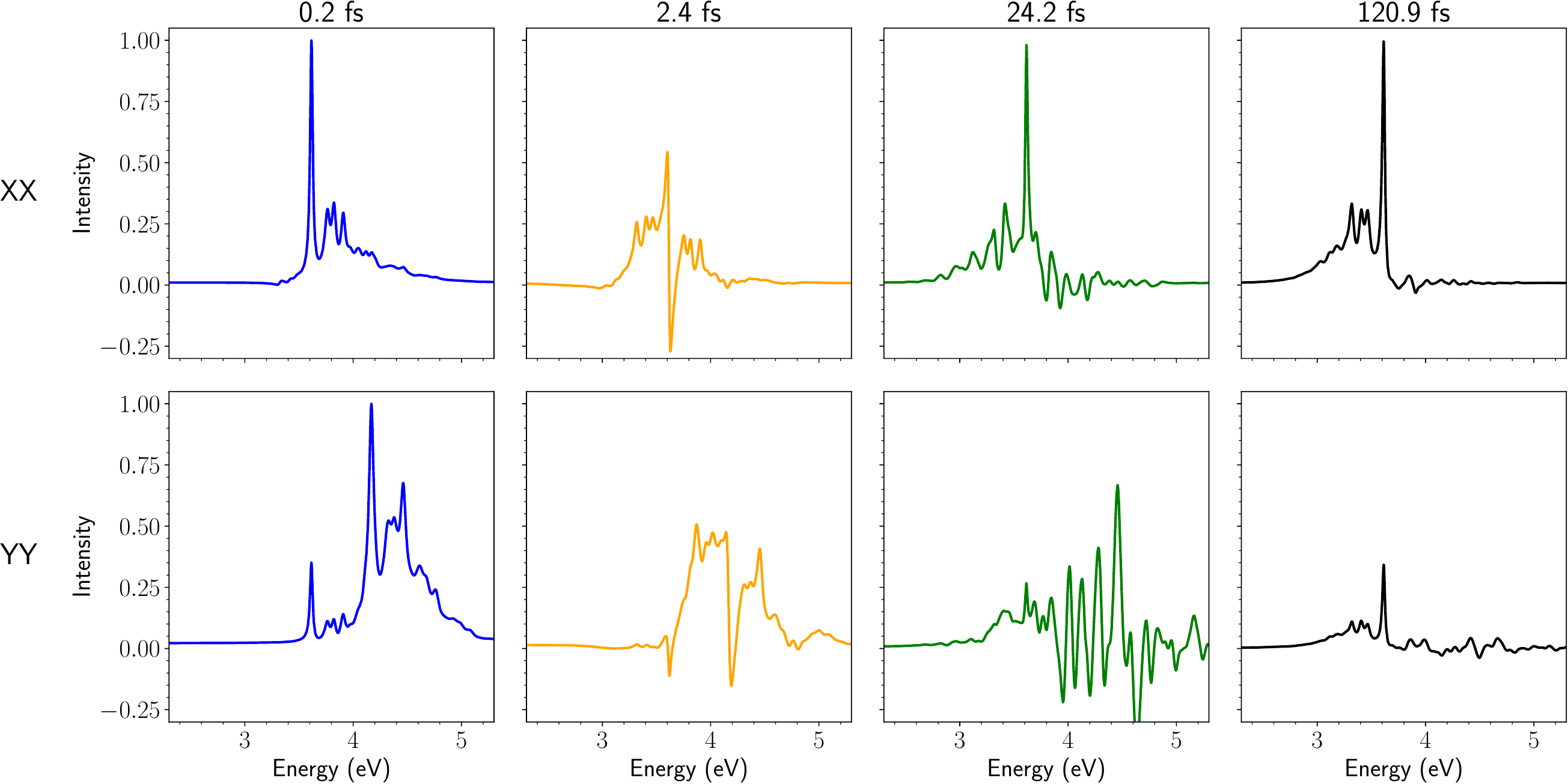}
    \caption{Cuts through the time-and frequency-resolved ESE spectrograms for different pump-probe delays shown in dashed lines in \cref{fig:ESE_spectrograms} obtained in the case of the full 3L model. 
    The corresponding cuts with the 2L model are given in \textcolor{black}{SI, fig. SI-10}. 
    The polarizations are along $e_x$ (top) and $e_y$ (bottom).
    The spectra are normalized with the maximum intensity among spectra along the same line (with the same polarization).}  
    \label{fig:ESE_spectrogram_cuts}
\end{figure*}

The ESE evolution during the early dynamics is more visible when comparing different cuts through the spectrograms for selected delays that are indicated by vertical dashed lines.
The time-resolved cuts obtained for the full 3L model are given in \cref{fig:ESE_spectrogram_cuts}.
They are given in \textcolor{black}{SI, fig. SI-10}, in the case of the 2L model for comparison, showing similar qualitative behavior. 
Cuts through the TA spectrograms are also given in \textcolor{black}{SI, figs. SI-11 and SI-12}, for both the 3L and 2L models.

For very short delays, ESE corresponds to the absorption spectrum. 
It is due to stimulated Rabi oscillations at the frequencies of the vertical transitions from the equilibrium ground state. 
As soon as the propagation begins, the energy gap varies and the ADOs representing the baths are gradually populated. 
The application of the transition dipole operators at $t_2$ affects all the ADOs.\cite{Tanimura2006} 
Around $t_2=\qty{2.5}{\femto\second}$, Fano profiles (abrupt changes from positive to negative intensities) appear due to interferences in the optical response between bath and system.\cite{ZhangYan2015,Finkelstein2016} 
These particular profiles are not a signature of the nonadiabatic interaction since they persist in the $\text{D}_1$ or $\text{D}_2$ states when the inter-state coupling is artificially suppressed (see \textcolor{black}{SI, fig. SI-18}). 
For a longer delay, one sees the turnover from the absorptive to the emissive profile that reaches its asymptotic value after about \qty{125}{\femto\second}.

\begin{figure*}
    \centering
    \includegraphics[width=0.90\textwidth]{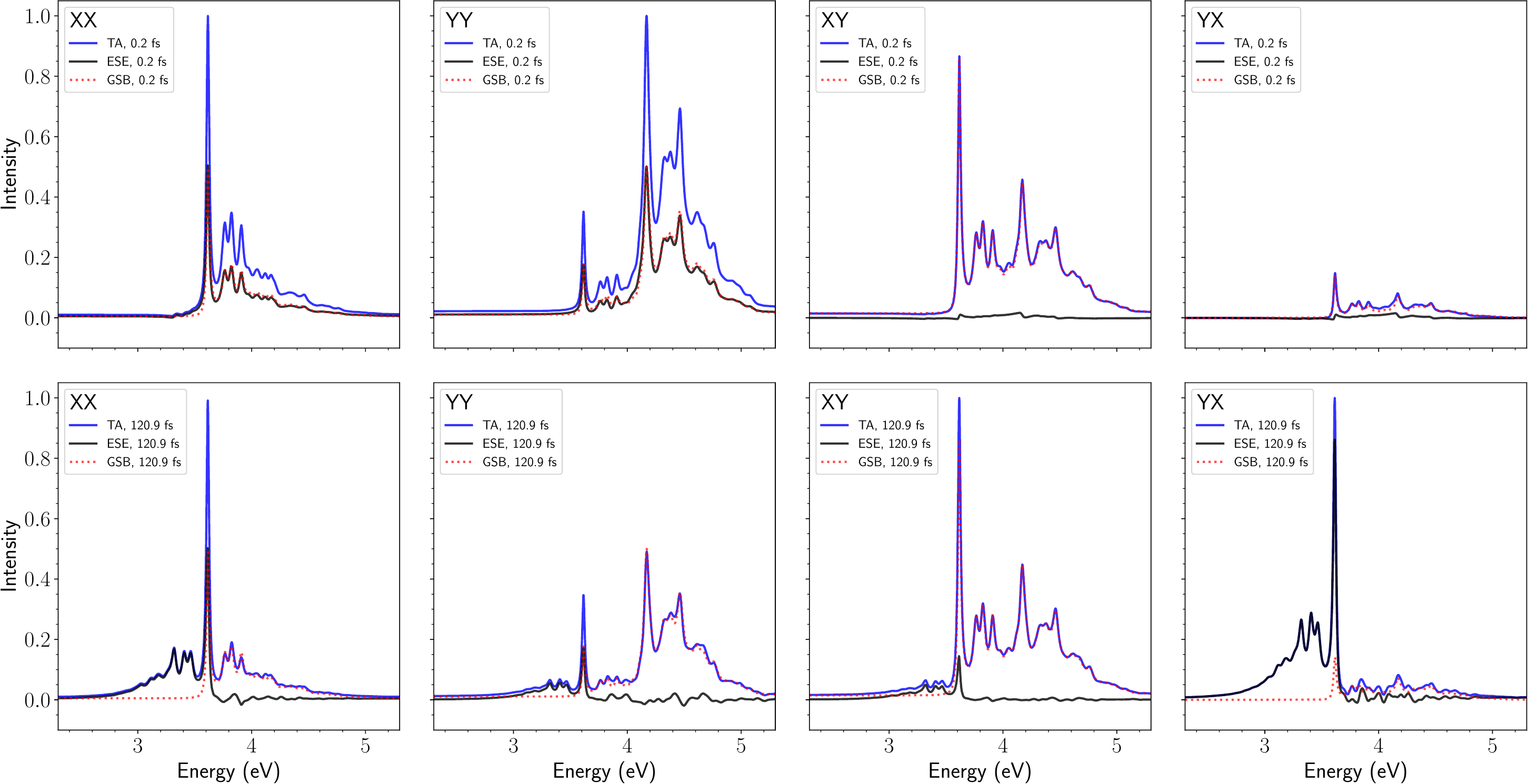}
    \caption{
    ESE and TA signals obtained within the 3L model at \qty{0.2}{\femto\second} and \qty{120.9}{\femto\second}, compared with GSB (constant at all times).
    From left to right, the polarizations of the pump and the probe are parallel -- along $e_x$ (case XX, top panels) or $e_y$ (case YY, bottom panels) -- and  perpendicular (crossed polarizations) -- along $e_x$ and $e_y$, respectively, in the case XY or \textit{vice} \textit{versa} in the case YX.
    The spectra are normalized with the maximum intensity among spectra in the same column (with the same polarization).
    }
    \label{fig:ESE_TA_GSB_spectrogram_cuts}
\end{figure*}

\Cref{fig:ESE_TA_GSB_spectrogram_cuts} illustrates the strong evolution of the ESE and TA profiles for the next-to-initial time (\qty{0.2}{\femto\second}) and the long time (\qty{121}{\femto\second}) close to the equilibrium. 
The GSB spectrum, which remains the same at all times, is drawn in dotted lines. 
For the very short delay, ESE and GSB are the same when the pump and probe pulses have the same polarization. 
The TA signal is then twice the GSB spectrum.
For the long delay, the TA signal becomes equal to the GSB one in the range of the absorption spectrum since ESE becomes negligible in this domain. 
On the contrary, TA coincides with ESE in the range of the stationary emission spectrum. 
This suggests that the ESE time-dependent signal can be obtained from the time-dependent TA signal upon subtracting from it half of the initial TA spectrum, which is equivalent to the GSB contribution, at least when the polarizations are identical.

\subsubsection{\label{sec:spectralfinger} S(t) signals as EET spectral fingerprints}
In this section, we illustrate how some features of the ideal impulsive pump-probe spectra may be related to EET dynamics. 
We first consider the evolution of some peak intensities (horizontal cuts through the spectrograms) and then we analyse the integrated signals over the range of frequencies. 

a) \textit{Peak evolution at relevant frequencies}

\begin{figure*}
    \centering
    \includegraphics[width=0.45\textwidth]{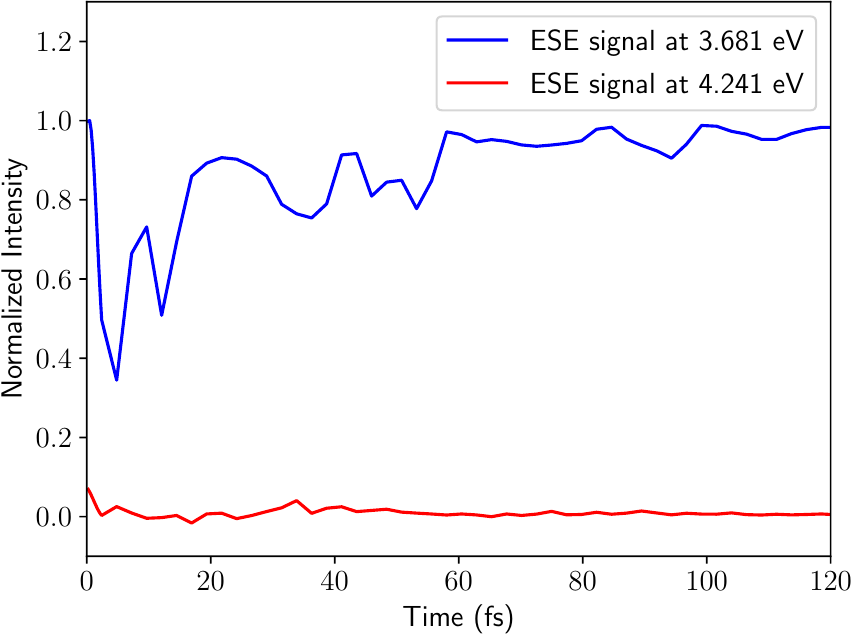}
    \includegraphics[width=0.45\textwidth]{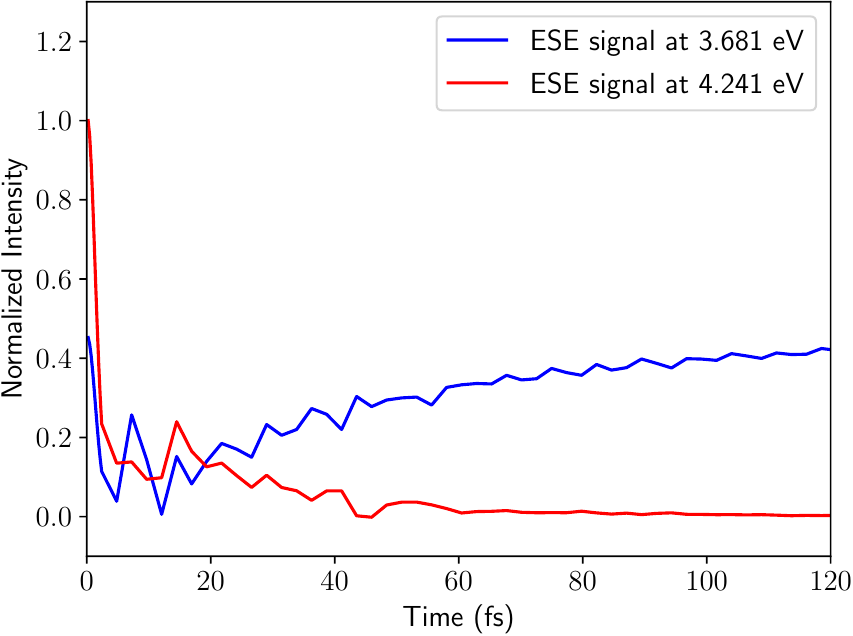}
    \caption{
    Time-resolved slices through the  ESE spectrogram (\cref{eq:ESE}) at two different frequencies corresponding to maxima in the steady-state absorption spectrum. 
    Left panel: polarization along $e_x$, right panel: polarization along $e_y$.
    }
    \label{fig:ESE_spectrogram_TRP}
\end{figure*}

\Cref{fig:ESE_spectrogram_TRP} gives the ESE intensity at two frequencies corresponding to the main transitions to state D\textsubscript{1} (p3 branch) or D\textsubscript{2} (p2 branch) at \qty{3.681}{\electronvolt} and \qty{4.241}{\electronvolt}, respectively. 
They are horizontal slices through the ESE spectrogram for the 2L model but the behavior is similar to that of the 3L model as may be seen by following the maximum of the peaks in \cref{fig:ESE_spectrogram_cuts}.
When the polarization is along $e_x$ (\cref{fig:ESE_spectrogram_TRP}, left panel), ESE concerns the evolution in D\textsubscript{1} that gives rise to a negligible inter-state transition. 
On the contrary, a signature of EET may be seen when following the ESE for the transition at \qty{4.241}{\electronvolt} with the polarization along $e_y$ (\cref{fig:ESE_spectrogram_TRP}, right panel). Indeed, the initial signal at \qty{4.241}{\electronvolt} completely disappears in about \qty{20}{\femto\second}.

b) \textit{Integrated signals with parallel polarizations}

We now examine how some fingerprint of the coherence and population evolution during the EET dynamics may be found in the integrated signal $S^\text{TA}(t)$ obtained upon integrating $\sigma^\text{TA}(\omega,t)$ (\cref{eq:TA}) over the frequency domain. When these signals with parallel polarizations are normalized, the information is given by $S^\text{ESE}(t)$ (\cref{eq:signal(t)}) since the contribution of GSB is only a constant. $S^\text{ESE}(t)$ gives the time evolution of the emission spectrum area. We consider four situations, some of which are fictitious cases without inter-state coupling, in order to dissect the various influences on the signal $S^\text{ESE}(t)$. 

(i) Only one state, D\textsubscript{1} (case XX) or D\textsubscript{2} (case YY), is excited by approximating the transition dipoles to their largest components only ($\mu_{01}\approx(-4.77,0,0)$ a.u., $\mu_{02}\approx(0,2.01,0)$ a.u.) and assuming that there is no inter-state coupling.
The initial electronic state is then stationary.
Hence, the initial value $R_1(t_3=0,t,0)$ does not depend on $t$ even if the response for $t_3 \neq 0$ depends on $t$.
Thus, $S^\text{ESE}(t)$ remains constant, and the shape of the ESE spectrum evolves with preserving its area.
Indeed, for any delay $t_2$, the density matrix remains constant with a single population $P_j = 1$ ($j=1$ or $2$).
The application of $\mu_{+}$ after the delay $t_2$ creates an optical coherence $\rho_{\text{S},j0}$ (between the electronic system states $j$ and $0$) that is also independent of $t_2$.
At the end of $t_3$, $\mu_{-}$ projects this coherent superposition into the ground state. 

\begin{figure*}
    \centering
    \includegraphics[width=0.31\textwidth]{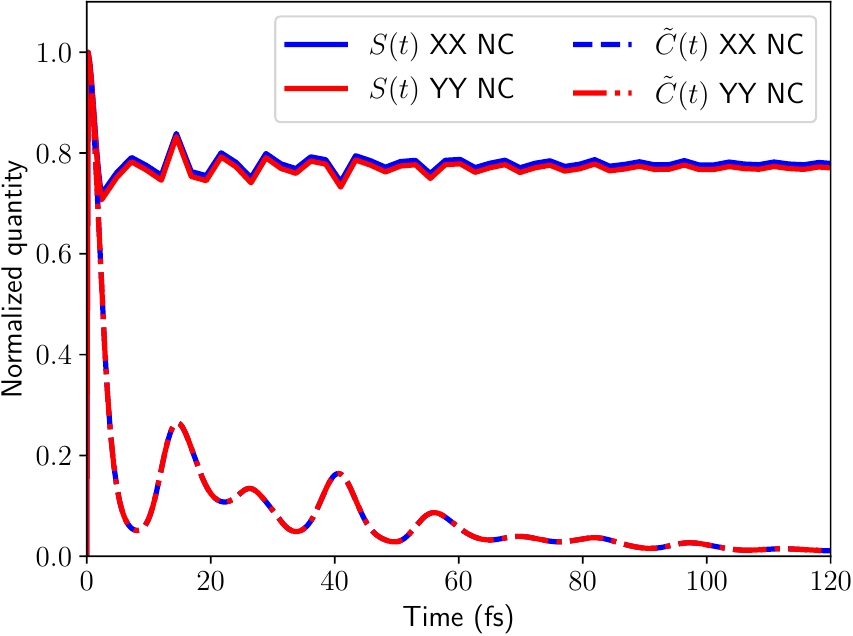}
    \includegraphics[width=0.31\textwidth]{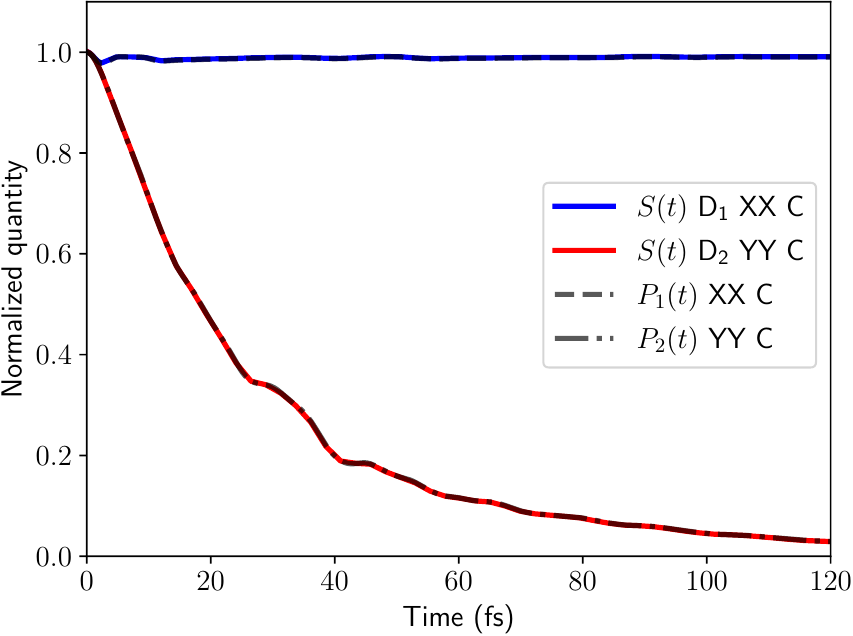}
    \includegraphics[width=0.31\textwidth]{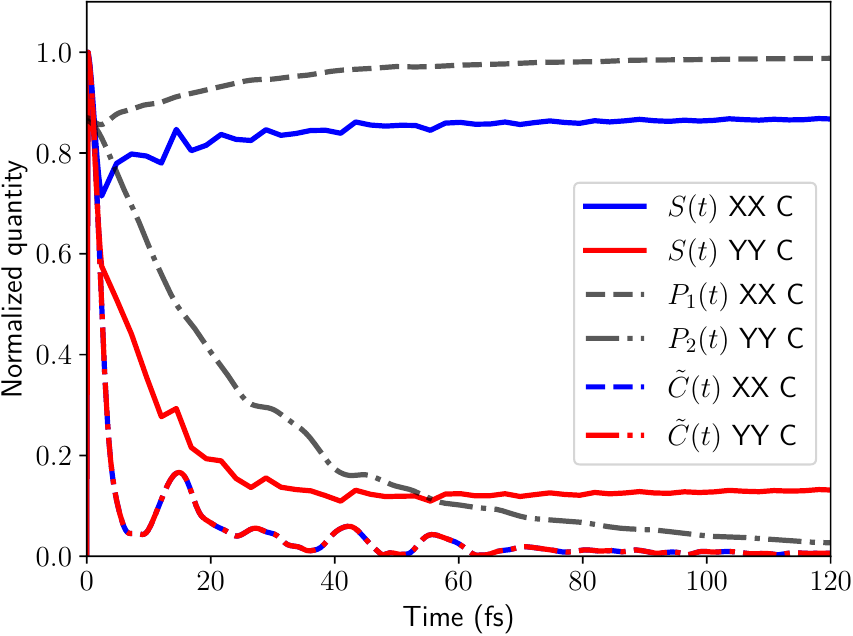}
       \caption{
       Normalized integrated ESE signal, $S(t)=S^\text{ESE}(t)$, relative modulus of the electronic coherence, $\tilde{C}(t)=|\rho_{\text{S},12}(t)|/|\rho_{\text{S},12}(0)|$, and electronic populations, $P_{1}(t)$, $P_2(t)$, in a field-free situation (when assuming the initial preparation).
       From left to right: cases (ii), (iii), and (iv), such as discussed in \cref{sec:spectralfinger}.
       In the case (ii): coherently superposed electronic state without inter-state coupling (NC), case (iii): single initial electronic state with inter-state coupling (C), and case (iv): coherently superposed electronic state with inter-state coupling (C).
       Note that (1,2)-populations are constant with time in case (ii), and the (1,2)-coherence in case (iii), which is why they are not plotted.
       }
    \label{fig:StCohePop}
\end{figure*}

(ii) The two states D\textsubscript{1} and D\textsubscript{2} are excited in the XX or YY cases ($\mu_{01}=(-4.77,0.80,0)$ a.u., $\mu_{02}=(1.84,2.01,0)$ a.u.), and the inter-state coupling is nil.
The initial electronic state is then a superposition with an initial electronic coherence but the populations remain constant.
This is called a pure dephasing case of the electronic system, induced by the vibrational bath in each state. 
The left panel of \cref{fig:StCohePop} compares the evolution of $S^\text{ESE}(t)$ for the two polarizations XX and YY with the decoherence function of a field-free case given by the relative modulus of the electronic coherence $\tilde{C}(t)=|\rho_{\text{S},12}(t)|/|\rho_{\text{S},12}(0)|$.
One observes that $S^\text{ESE}(t)$ now exhibits a fast initial decay with a rate similar to that of the (de)coherence function.
$S^\text{ESE}(t)$ then becomes nearly constant as in the previous case when the coherence becomes negligible.
In this situation, even if the populations are not varying, the electronic coherence strongly depends on the delay $t_2$ due to the effect of the tuning baths.
The action of $\mu_{+}$ at that time generates different optical coherences $\rho_{\text{S},j0}$ ($j=1,2$), and the projection into the ground state after $t_3$ is different for each delay as long as the coherence evolves.  

(iii) Similar to case (i) but the inter-state coupling now operates.
The middle panel of \cref{fig:StCohePop} illustrates the remarkable behavior of $S^\text{ESE}(t)$ that follows the evolution of the populations, in particular the decay of the most excited state.
The evolution of the reduced-density matrix before the application of $\mu_{+}$ is now due to the variation of the populations that are transformed to optical coherences and projected by $\mu_{-}$ after $t_3$ with a yield that evolves with the same rate as the population decay. 
After normalizing to disregard the value of the transition dipole, the $S^\text{ESE}(t)$ signal perfectly follows the population decay due to the inter-state coupling.

(iv) Similar to case (ii) but accounting for the inter-state coupling.
This is the most realistic situation. It is shown in \cref{fig:StCohePop}, right panel.
$S^\text{ESE}(t)$ now contains fingerprints of both the decoherence and the EET population transfer.
The (de)coherence function presents a similar initial decay as in case (ii) but is damped by the population transfer.
In the YY case, the most populated state is D\textsubscript{2}. 
Unlike the case without coupling displayed in the left panel, $S^\text{ESE}(t)$ now continues to decrease and the rate of decay is very similar to that of the D\textsubscript{2} state shown in dashed-dotted line (the initial populations, $P_1=0.13$ and $P_2=0.87$, are fixed by the normalized superposed state prepared at $t_1=0$).
In the XX case, the behavior of $S^\text{ESE}(t)$ is quite similar to that without coupling but it stabilizes to a higher value.
Again the slight increase follows the evolution of the D\textsubscript{1} population, shown in dashed line (the initial populations are $P_1=0.86$ and $P_2=0.14$).
Hence, case (iv) is a mixture of cases (ii) and (iii) as regards the time evolution of the $S^\text{ESE}(t)$ signal.
While it may be difficult to disentangle the two effects in practice, our numerical experiments thus provide a guideline for deciphering the relative roles played by decoherence and EET when probing the time-resolved spectroscopy of an optically active manifold of two vibronically coupled excited electronic states that can be excited coherently by the pump.
For the sake of completion, non-normalized signals are provided in \textcolor{black}{SI, fig. SI-17}.

c) \textit{Anisotropy decay signal}

\begin{figure*}
    \centering
    \includegraphics[width=0.45\textwidth]{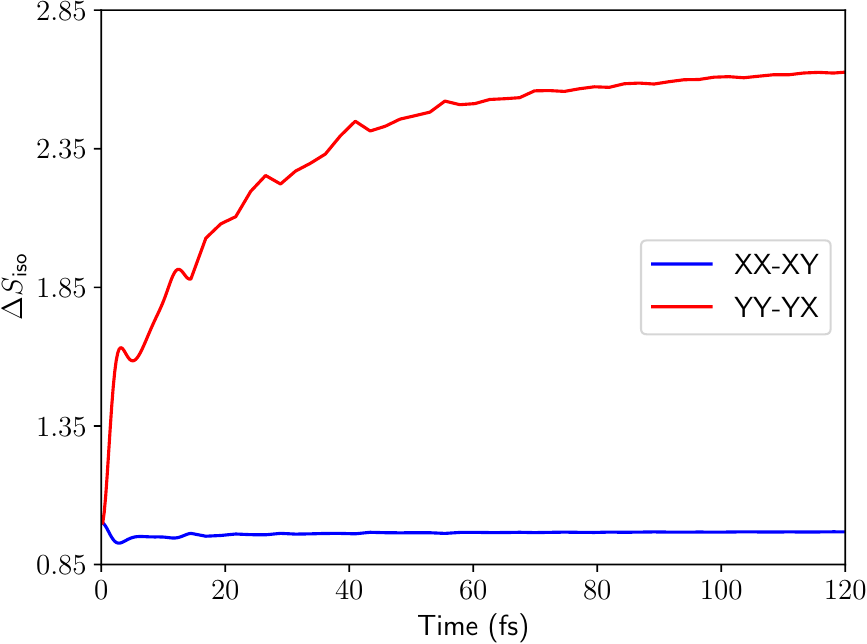}
    \includegraphics[width=0.45\textwidth]{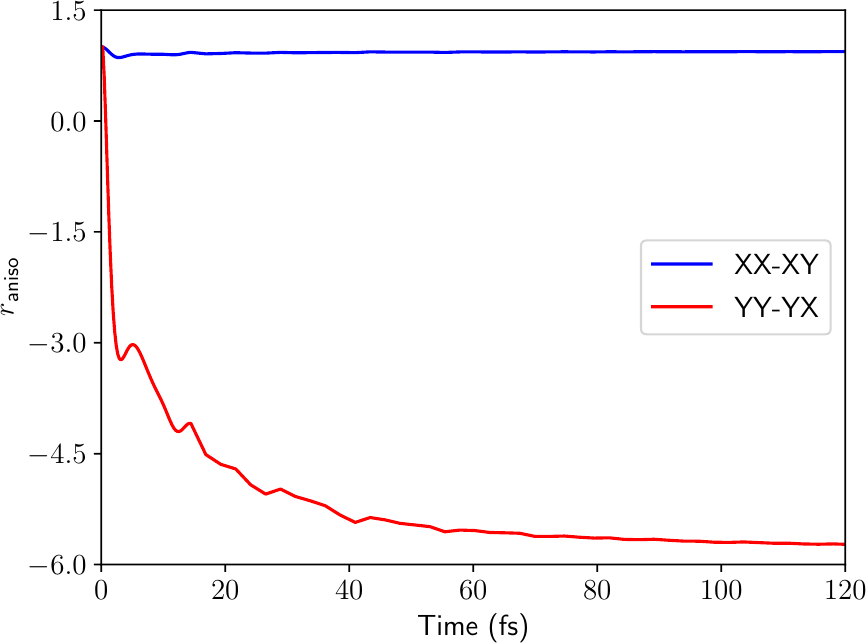}
    \caption{
    Signals with parallel and perpendicular polarizations when exciting with polarization along $e_x$ (case XX-XY) or polarization along $e_y$ (case YY-YX).    
    Left panel: $\Delta S_{\text{iso}}(t)$ (\cref{eq:deltaiso}), right panel: $r(t)$ (\cref{eq:raniso}).
    }
    \label{fig:Anisotropy}
\end{figure*}

Up to now, we extensively discussed the integrated signals of ESE for parallel polarizations of the pump and probe pulses, ${S}^{\text{ESE}}_{\text{XX}}(t)$ or ${S}^{\text{ESE}}_{\text{YY}}(t)$. 
Similar integrated signals can be obtained for perpendicular polarizations,  ${S}^{\text{ESE}}_{\text{XY}}(t)$ or ${S}^{\text{ESE}}_{\text{YX}}(t)$.
More importantly, these integrated signals can be evaluated for TA upon considering ${S}^{\text{TA}}_{\text{XX}}(t)$ and ${S}^{\text{TA}}_{\text{XY}}(t)$, or ${S}^{\text{TA}}_{\text{YY}}(t)$ and ${S}^{\text{TA}}_{\text{YX}}(t)$, which allows for the calculation of the isotropic signal, $\Delta S_{\text{iso}}(t)$ (\cref{eq:deltaiso}), and the anisotropy decay signal, $r(t)$ (\cref{eq:raniso}). 
Such signals are shown in \cref{fig:Anisotropy} and compared for the pump-probe setups with parallel XX and perpendicular (crossed) polarizations XY (blue lines) or with YY and YX (red lines).

Exciting with polarization along $e_x$ mainly populates the $\text{D}_1$ state while both excited states are reached with exciting polarization along $e_y$. The very fast variation of $r(t)$ in that case reveals the EET towards the acceptor state leading to a fast change of the polarization.
The change of polarization is completed in about \qty{40}{\femto\second}, as expected from the different intermediate signals discussed above. Indeed, when the pump is polarized along $e_y$, the spectrogram $\sigma^{\text{ESE}}(\omega,t)$, the integrated signal $S^{\text{ESE}}(t)$, and finally the anisotropic decay signal $r(t)$ contain all three the same signature of the ultrafast transfer from the D\textsubscript{2} to the D\textsubscript{1} states. 
In this context, we expect these experimentally observable signals to characterize the timescale and the yield of the electronic population transfer.
We note that the occurrence of weak oscillations in both $\Delta S_{\text{iso}}(t)$ and $r(t)$ is a fingerprint of the vibronic coherence as already observed in \cref{fig:StCohePop} (right panel).

\subsubsection{2D spectra}

\begin{figure*}
    \centering
    \includegraphics[width=0.90\textwidth]{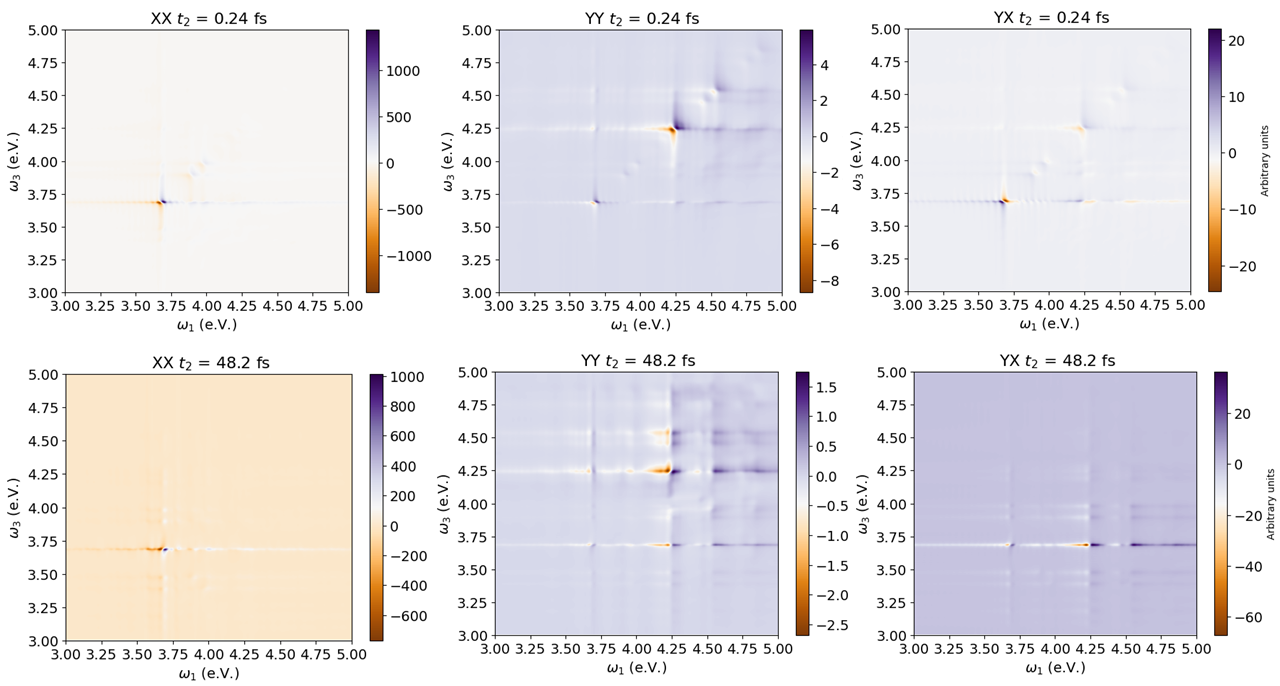}
    \caption{
    2D spectra for two population times $t_2$ obtained with parallel polarizations along $e_x$ (case XX) or $e_y$ (case YY) and crossed polarizations (case YX).
    }
    \label{fig:spectre2DM23}
\end{figure*}

\Cref{fig:spectre2DM23} gives the 2D spectra for two population times $t_2$ with parallel or crossed polarizations computed with the 2L model. 
The grids of the coherence time $t_1$ and the detection time $t_3$ extend over \qty{100}{\femto\second} and \qty{300}{\femto\second}, respectively, to ensure the decay of the corresponding correlation functions. 
For the ultra-short waiting time $t_2 = \qty{0.24}{\femto\second}$, the position of the diagonal peaks in the cases XX and YY matches the absorption or the early emission spectra at the Franck-Condon geometry (see \textcolor{black}{SI, fig. SI-10}).

Indeed, for very short population times, the GSB and ESE contributions are the same. 
In the XX case, the intense peak at 3.69 eV and the two peaks at \qtylist{3.9;3.99}{\electronvolt} are transitions toward the $\text{D}_1$ state reached by the high $\mu_x$ component. 
With the polarization along $e_y$ (case YY), the optical coherences toward the $\text{D}_1$ and $\text{D}_2$ states are of the same order of magnitude. 
The first three diagonal peaks coincide with those obtained in the XX case with a lower intensity due to the lower value of the transition dipole $\mu_y$ toward this $\text{D}_1$ state (see \cref{fig:m23_est}). 
The following diagonal peaks give the vibronic progression corresponding to the $\text{D}_2$ state at \qtylist{4.43;4.53;4.75;4.80}{\electronvolt}.
The off-diagonal peaks confirm the possible EET from the $\text{D}_2$ state toward the acceptor state, for instance an excitation at \qty{4.25}{\electronvolt} is likely to give a signal at \qty{3.69}{\electronvolt}.  
This effect of probing EET is more obvious when using the crossed-polarization XY case.
The lower off-diagonal peak linking excitation at \qty{4.25}{\electronvolt} and detection at \qty{3.69}{\electronvolt} from the $\text{D}_1$ state is very intense.
The other following peaks resulting from excitation at \qty{4.25}{\electronvolt} correspond to the detection of the states at \qtylist{3.90;3.99}{\electronvolt} also belonging to the acceptor state.

For a longer waiting time  $t_2= \qty{48.2}{\femto\second}$, even if the GSB remains the same, the ESE contribution does not match the Franck-Condon region anymore and nearly coincide with the asymptotic emission of the relaxed system. 
In the XX case, at excitation energy \qty{3.69}{\electronvolt}, one observes the two weak emission peaks at \qtylist{3.4;3.49}{\electronvolt} (see \textcolor{black}{SI, fig. SI-10}). 
The intensity of all the peaks in the YY case is lower than for the short waiting time, confirming the fast population decay from the $\text{D}_2$ state. 
The cross-peak at \qtylist{4.25;3.69}{\electronvolt} reveals the excitation in $\text{D}_2$ and detection from the acceptor state. 
In a similar way the two weak detection peaks at \qtylist{4.25;3.4}{\electronvolt} and \qtylist{4.25;3.49}{\electronvolt} are also a signature of the transfer toward the $\text{D}_1$ state. 
With the other crossed polarization YX, the diagonal peaks related to the Franck-Condon excitation disappear as in the previous case. 
The main peak at \qtylist{4.25;3.69}{\electronvolt} and the satellite ones (\qtylist{4.25;3.4}{\electronvolt} and \qtylist{4.25;3.49}{\electronvolt}) confirm the presence of EET. 
The difference of the intensity when detecting with polarization Y or X comes from the difference of the components of the transition dipoles, which occur at power four in 2D spectra.

\section{Conclusions and outlook}

The first focus of the present work was to extend to high dimensionality and/or dissipative approaches our previous reduced-dimensionality study of ultrafast EET occurring in the smallest asymmetrical \textit{meta}-substituted PPE oligomer, m23, made of 2-ring (p2) and 3-ring (p3) \textit{para}-substituted pseudo-fragments.\cite{galiana_excitation_2024}

As regards this prototypical system, the donor state is mainly localized on p2 and the acceptor state on p3.
For the isolated m23 system, we have constructed an \textit{ab initio} high-dimensional vibronic coupling Hamiltonian (VCH) model for the first two singlet electronic excited states accounting for all the 93 in-plane vibrational degrees of freedom in terms of a linear vibronic coupling (LVC) model and various types of quadratic vibronic coupling (QVC) extensions.

Wavepacket simulations using a 93-dimensional ML-MCTDH approach confirmed a typical timescale of about \qty{25}{\femto\second} for intramolecular EET, which was confirmed with the dissipative reduced-density approach known as HEOM.
The primary role of the high-frequency acetylenic and quinoidal vibrations was confirmed. 

The second focus of our work was to investigate the nonlinear spectroscopy of m23 with particular attention paid to time-dependent TA spectra as regards their GSB and ESE components and 2D spectroscopy, upon decephering the respective roles of the different polarizations.
Our primary intention here was not to improve an already-present
method with new features, but rather to show how to take benefit of approaches developed by different communities to get a pragmatical recipe/strategy aimed at addressing
a realistic situation: a molecule of decent size, of much experimental 2D-spectroscopic interest for EET, which also occurs to cause challenging numerical problems as opposed
to better-behaved toy models, and that will eventually be subject to some approximations. 
The time-dependent spectra and the 2D spectra of the m23 dimer have recently
raised interest and been computed with a semi-classical approach. 
Hence, we believe that it remains interesting to probe the possibilities and the limits of the HEOM
approach for such a complex and interesting system for EET.

Highlighting the role of the high-frequency modes allowed us to retain only the essential parts of the spectral densities and generate spectrograms, which deserve some care so as not to become too computationally expensive.
However, we also showed that low-cost simulations of integrated signals is possible in the delta-like pulse appoximation upon using the response function at $t_3 = 0$ insteead of integrating the spectra.

All the nonlinear spectra presented here correspond to ideal impulsive delta-like laser pulses, and some fingerprints of EET have been characterized in this context.
This illustrates that a minimal model deserves the essential description of the EET process as a complement to different approaches with more realistic pulses and more electronic states. \cite{hu_spectral_2021,zhang_what_2025}
Further investigations,  both on the theoretical and experimental fronts, are awaited, for instance to take into account the solvent effect.

Even if nonlinear signals were only computed with HEOM in this work, there is no \textit{a priori} limitation in getting the same signals with (ML-)MCTDH but we purposefully did not aim at following such avenues in the present work, since we wanted to focus on the HEOM approach.
Such wavepacket protocols have been recently proposed by other authors according to the time-dependent Davydov method \cite{Zhang2019} or in Refs. \cite{segatta_nonlinear_2023,segatta_time-resolved_2024}, where the time-resolved signals directly depend on the overlaps of the nuclear wavepackets propagated with standard MCTDH. 
Some practical complications may arise from the use of ML-MCTDH wavepackets for such a large-dimensional system, as opposed to standard MCTDH with a dimensionally-reduced model.
Along the same line, we expect 2DES lineshapes to be accessible within the same framework, only with more extensive computational resources and clever-crafted input/output data management.
This is left for further work.

\section*{Acknowledgments}
J. G. acknowledges the French MESR (Ministère chargé de l’Enseignement Supérieur et de la Recherche) and the ENS (Ecole Normale Supérieure) of Lyon for funding his PhD grant, hosted at the University of Montpellier. We warmly thank J.M. Teuler of the University Paris-Saclay for the upgrade of his JMFFT-7.2 library (CNRS-IDRIS France). 

\section*{Supporting information}
Supporting information is provided and includes: a discussion on bilinear terms in the VCH models, a representation of the ML-MCTDH wavefunction, additional information on nuclear displacements, and on spectral densities, operational equations for HEOM, examples of bath dynamics in HEOM, comparisons of the population dynamics in the 2L- or 3L models, expressions of the nonlinear responses, and additional figures for the discussion of nonlinear spectroscopy.

\section*{Data availabilty statement}
The data that support the findings of this study are available from the corresponding author upon reasonable request. 

\section*{References}


\end{document}


\title{\underline{Supplementary Information to}:\\
Wavepacket and Reduced-Density Approaches for High-Dimensional Quantum Dynamics: 
Application to the Nonlinear Spectroscopy of Asymmetrical Light-Harvesting Building Blocks}
\author{Joachim Galiana}
\altaffiliation{Current affiliation: Departamento de Química, Universidad Autónoma de Madrid, Madrid, Spain}
\affiliation{ICGM, Univ Montpellier, CNRS, ENSCM, Montpellier, France}
\author{Michèle Desouter-Lecomte}%
\affiliation{Institut de Chimie Physique, Université Paris-Saclay-CNRS, UMR8000, F-91400 Orsay, France
}%

\author{Benjamin Lasorne}
\email{benjamin.lasorne@umontpellier.fr}
\affiliation{%
ICGM, Univ Montpellier, CNRS, ENSCM, Montpellier, France
}%

\date{\today}

\maketitle
\tableofcontents

\section{A note on bilinear terms and VCH reduced models}

\begin{figure*}[!ht]
\centering
\includegraphics[width=1.0\textwidth]{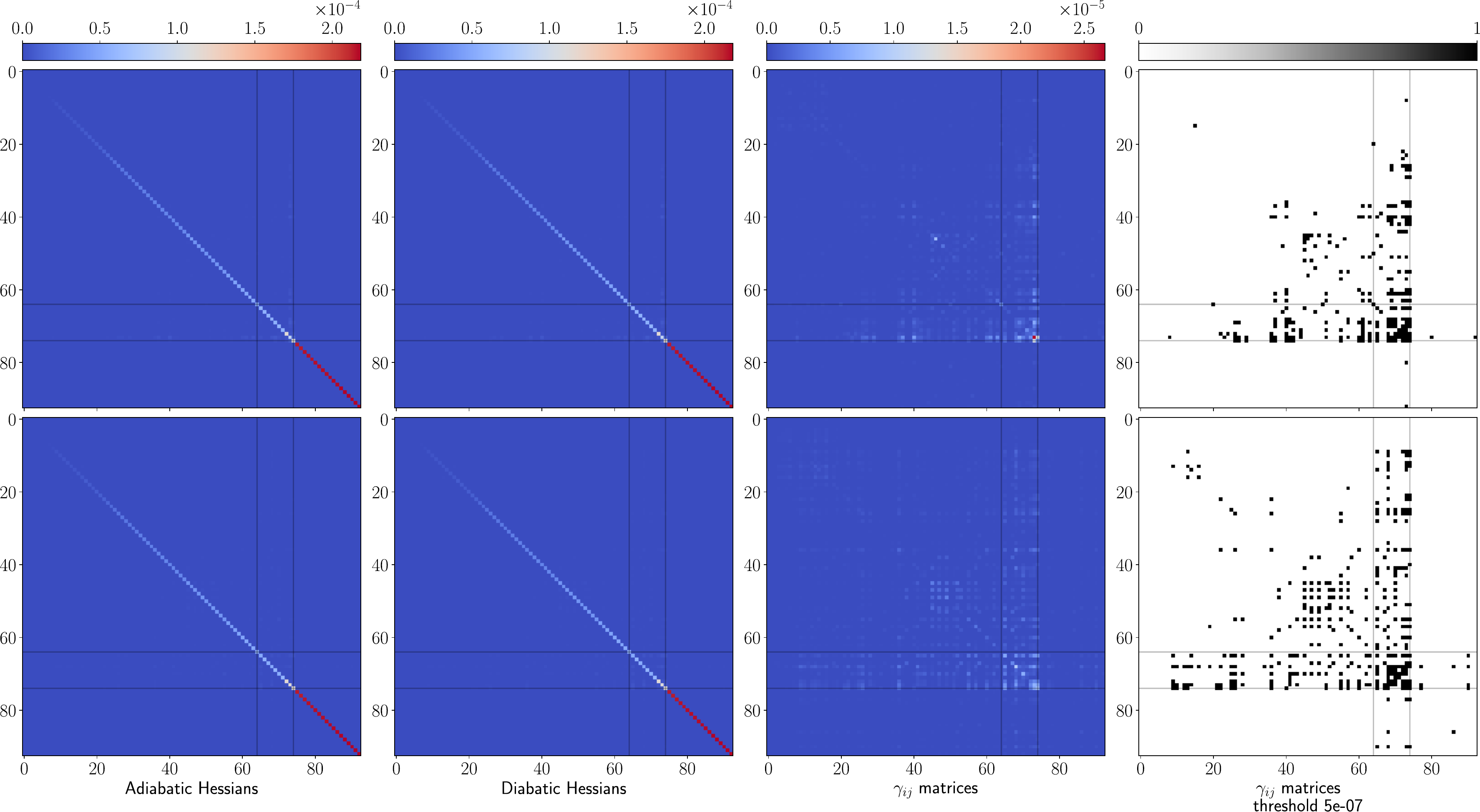}
\caption{
  From left to right, absolute values of the:
  i) adiabatic Hessians $\mathbf{K}_{\text{S}}$ of S\textsubscript{1} and S\textsubscript{2} at the the FC geometry (FCP);
  ii) post-processed diabatic Hessians $\mathbf{K}_{\text{D}}$ of D\textsubscript{1} and D\textsubscript{2} at FCP;
  iii) mode-mixing matrices $\bm{\gamma}^{(s)}=\mathbf{K}_{\text{D}_s}-\mathbf{K}_{\text{S}_0}$ for the two electronic states;
  iv) same with blanks where values are $\leq \qty{5e-7}{\frac{\hartree}{\bohr^2\electronmass}}$ (unselected values).
  All Hessians are projected onto the normal modes computed at FCP.
  The top and bottom panels correspond to the first and second excited states, respectively.
  The grey horizontal and vertical lines define the region of quinoidal, anti-quinoidal, and acetylenic normal modes.
}
\label{fig:m23_HDQD_HESSIANS}
\end{figure*}

The adiabatic Hessians, diabatic Hessians, and $\mathbf{\gamma}$-matrices are given in the first three columns of \cref{fig:m23_HDQD_HESSIANS}.
The terms selected in the LVC+$\gamma$ model are shown in the fourth column of \cref{fig:m23_HDQD_HESSIANS}.

The relevance of the $\gamma$-matrices for a dimensionally-reduced model is detailed hereafter.
We compare the EET dynamics simulated for the LVC model parametrized \emph{via} a global fit (fitting of \emph{ab initio} PESs, see Ref.\cite{galiana_excitation_2024}), to the LVC-parametrized model \emph{via} a local fit (identification of energy derivatives; this work).
For such a comparison, we extract the 8-dimensional \emph{child} model from the 93-dimensional \emph{parent} model, upon freezing all the unselected normal modes in quantum-dynamics calculations.

\begin{figure*}[!hb]
\centering
\includegraphics[width=0.75\textwidth]{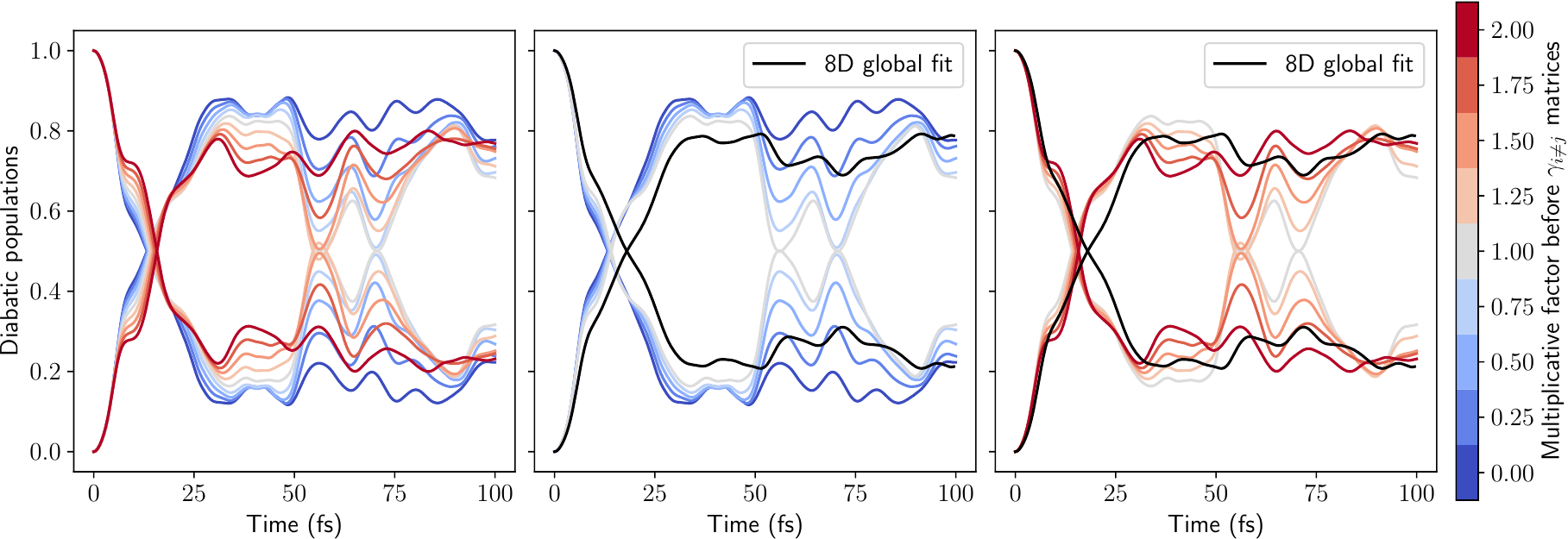}
\includegraphics[width=0.75\textwidth]{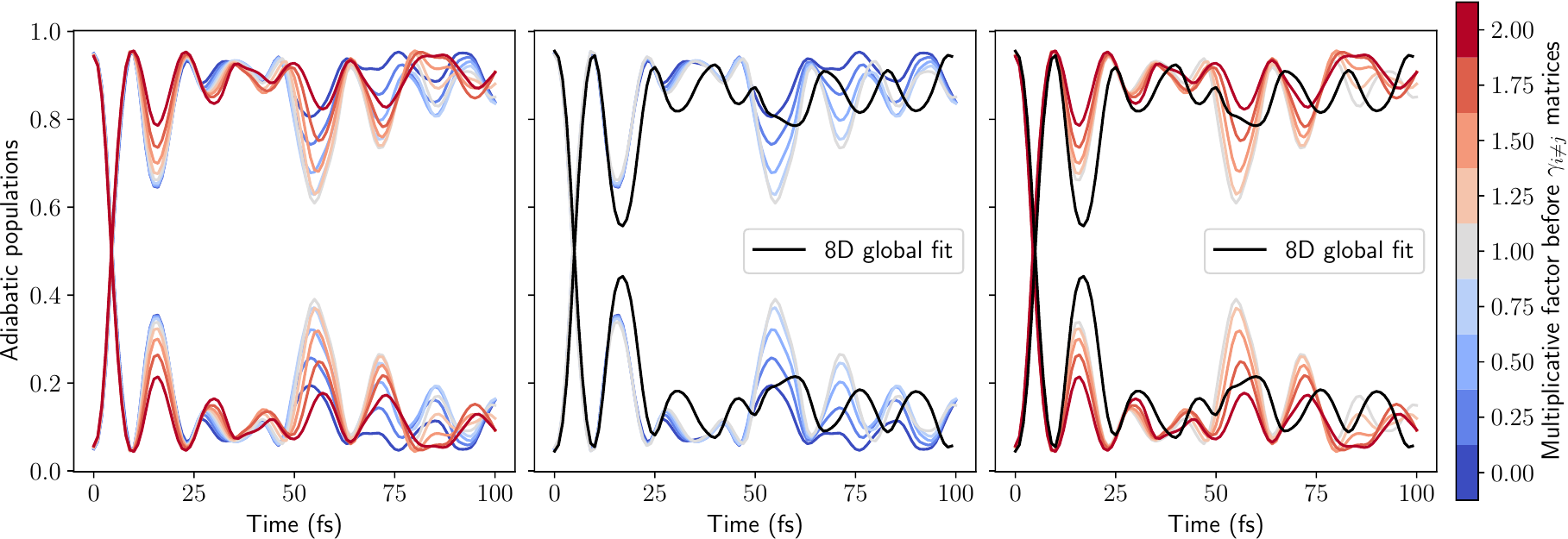}
\caption{
  Diabatic and adiabatic populations (top and bottom panels, respectively) of the first two excited states for various LVC models.
  The results obtained when varying the off-diagonal bilinear matrices from 0 (LVC model with different curvatures for the excited states but no mode mixing) to $\gamma_{i\neq j}$ (same but with actual mode mixing) [left and center panels], and from $\gamma_{i\neq j}$ to $2\gamma_{i\neq j}$ (articifially enhanced mode mixing) [left and right panels] are shown with a colormap from blue to red as regards the multiplicative factor.
  The results obtained with the 8D global fit are shown in black lines for comparison [center and right panels].
}
\label{fig:m23_varyingGammas}
\end{figure*}

We show in \cref{fig:m23_varyingGammas} the population transfer dynamics of EET for simulations for which the matrices $\gamma_{ij,i\neq j}$ are smoothly switched on from 0 (blue lines) to 1 (gray lines), and to 2 (red lines).`

For both diabatic and adiabatic populations, the early EET population transfer (up to \qty{50}{\femto\second}) seems to be only slightly affected by the variations of the $\gamma_{ij,i\neq j}$-matrices.
In particular, we note that the diabatic population transfer is slightly faster when switching on, and enhancing, the mode mixing.

This is likely due to the contribution of the $\gamma_{ij,i\neq j}$-parameters to the diabatic gradient difference, which really drives the early dynamics.
Quite counter-intuitively, the diabatic quantum yield (again, before \qty{50}{\femto\second}) is higher for simulations with no mode mixing, while we would expect bilinear intra-state couplings to allow for more efficient relaxations in the S\textsubscript{1} electronic state.
For longer times, we observe that the diabatic states mix again with a 50:50 population ratio for the case of $1\times\gamma_{ij}$, at both \qtylist{55;70}{\femto\second}.

Quite counter-intuitively again, the transfer is more monotonic in the extreme case of $0\times\gamma_{ij}$ than in the (supposedly more realistic) case of $1\times\gamma_{ij}$.
A plausible explanation would be that accounting for the mode mixing \emph{via} non-zero bilinear intra-state couplings is not accurate enough in low-dimensional systems.
Indeed, as we have observed above, with the same model ($1\times\gamma_{ij}$), but without freezing any in-plane mode, there is no diabatic recrossing.
The same is found for the extreme case of $2\times\gamma_{ij}$, which is an artificial enhancement of mode-mixing parameters.

Our interpretation is that accounting for mode mixing (that is, intra-state bilinear couplings), \emph{via} a local fit of the excited-state Hessians, might not be adapted to low-dimensional models.
Indeed, with the \emph{a priori} more realistic local fit parametrization ($1\times\gamma_{ij}$), the results are significantly different from the global fit parametrization.
This can be explained by the fact that the $\gamma_{ij}$-parameters are obtained for the full-dimensional system, so that they might take too much importance when used in a reduced model.

We note that these effects are not found to the same extent in the adiabatic population transfer, although the simulation of EET with $1\times\gamma_{ij}$ also exhibits significant recrossings at \qtylist{55;70}{\femto\second}.

\clearpage

\section{ML-MCTDH wavefunction}
The primitive basis set was defined with 15 harmonic-oscillator (Gauss-Hermite) basis functions per mode, with the equilibrium geometry set to zero (the FCP origin), effective masses set to one (reduced masses included in the mass-weighted coordinates), and ground-state frequencies for each normal mode of the model.
The multilayer tree was built according to a strategy meant to reflect the different types of molecular vibrations among the 93 normal modes.
Four groups are defining the upper layer: low- and high-frequency modes with small reduced masses (frequencies: \qtyrange{0}{1000}{\per\centi\meter} and \qtyrange{2400}{3500}{\per\centi\meter}, for soft modes and rigid C-H elongations, respectively), and two sets of high-frequency modes with large reduced masses (frequencies: \qtyrange{1000}{1500}{\per\centi\meter} and \qtyrange{1500}{2400}{\per\centi\meter}, which gather triangular, quinoidal, and acetylenic modes).

\begin{figure*}[!ht]
    \centering
    \includegraphics[width=0.99\textwidth]{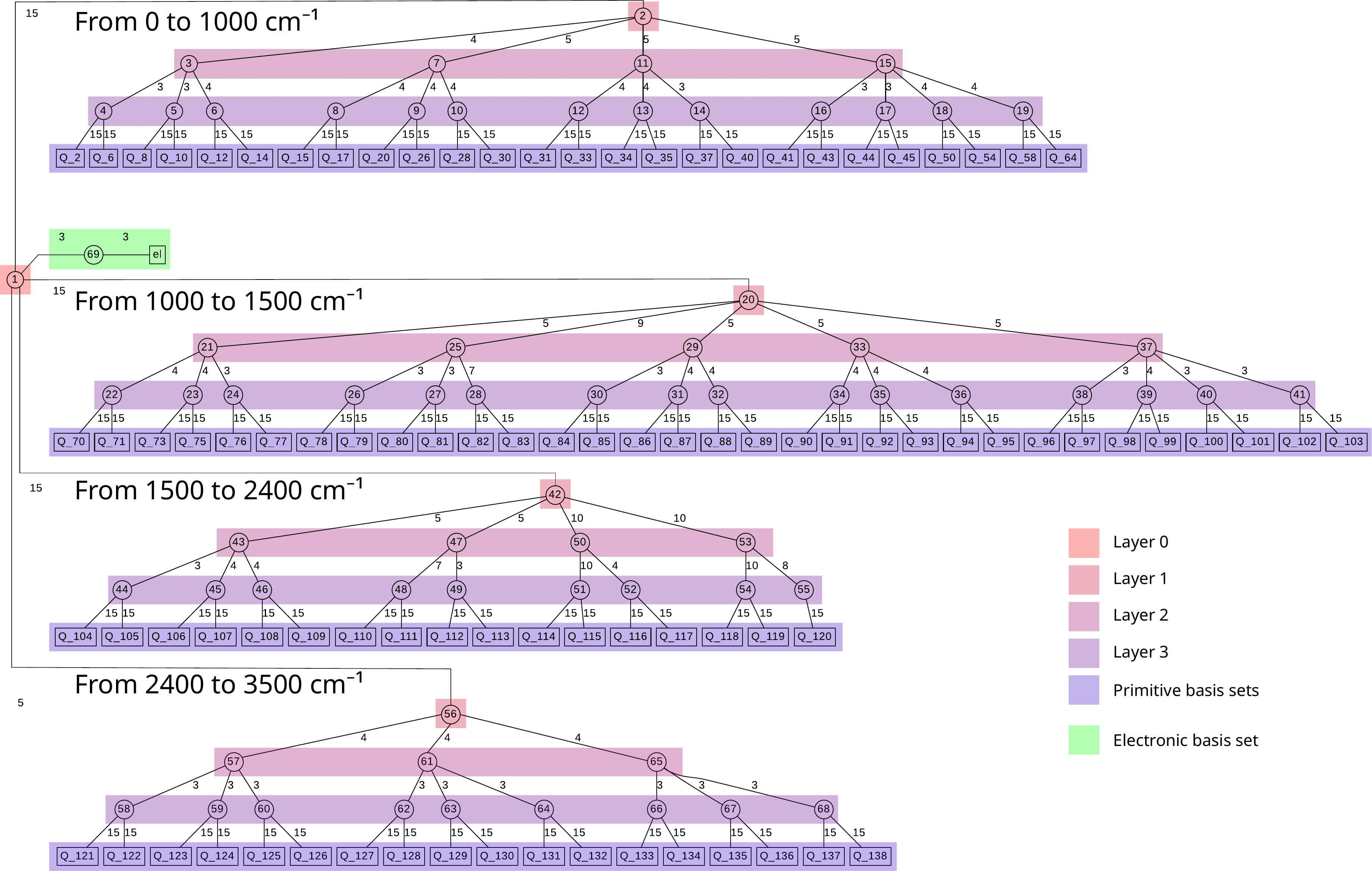}
    \caption{
    Representation of the ML-tree used for the quantum-dynamics simulations with the (1+2)-state 93-dimensional LVC model.
    The layer for the electronic states is given in green.
    The layers for the vibrational degrees of freedom are given in red to blue shading, from the top layer to the primitive basis, respectively.
    Nodes are in circles, primitive modes are in squares.
    The size of the basis set for each expansion (nodes or modes) is given above the corresponding node.
}
    \label{fig:mltree}
\end{figure*}

\clearpage

\section{Nuclear displacements: discrete modes}

\begin{figure*}[!ht]
    \centering
    \includegraphics[width=0.9\textwidth]{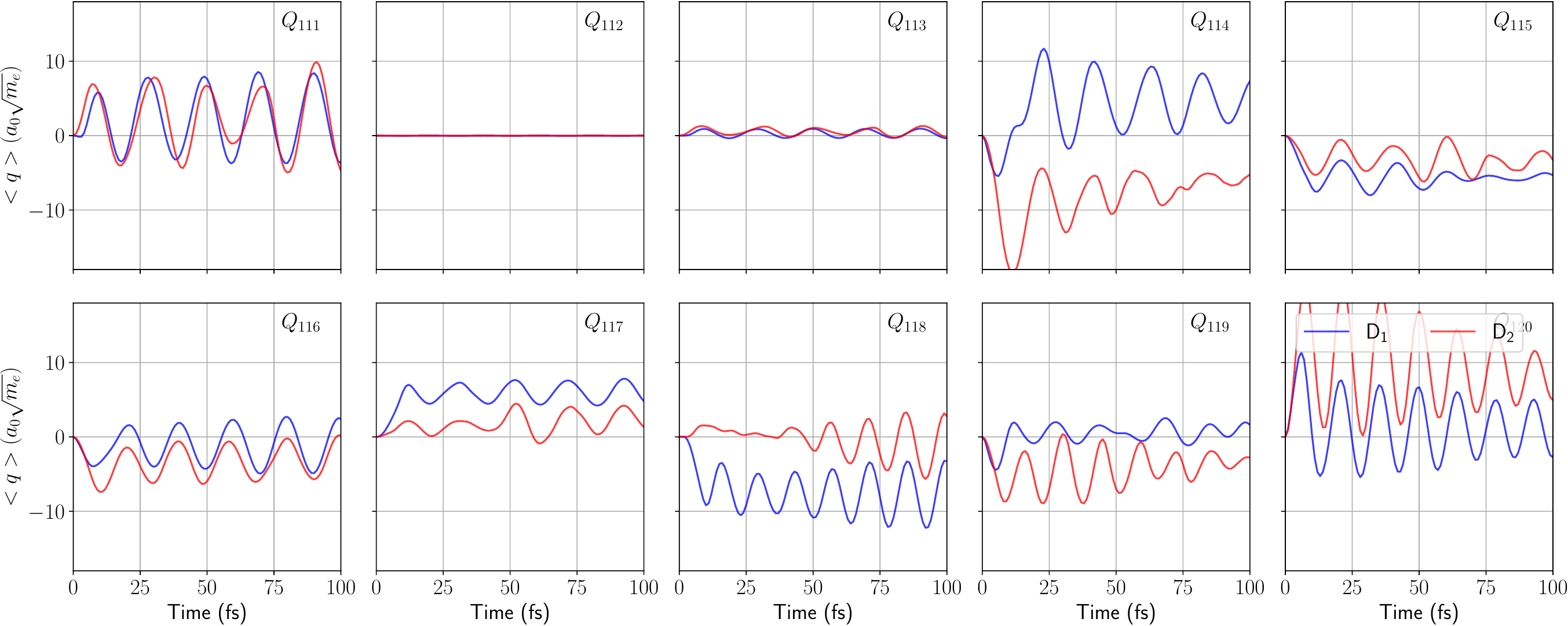}
    \caption{
    Time evolution of the state-specific expectation values of the position operator in diabatic states D\textsubscript{1} and D\textsubscript{2} for a selection of normal modes of vibration among the 93-dimensional model.
    The expectation values are computed with the propagated ML-MCTDH wavepacket for the LVC$+\gamma$ model after initial excitation to D\textsubscript{2}. 
    }
    \label{fig:QExp}
\end{figure*}

\begin{figure*}[!ht]
    \centering
    \includegraphics[width=0.9\textwidth]{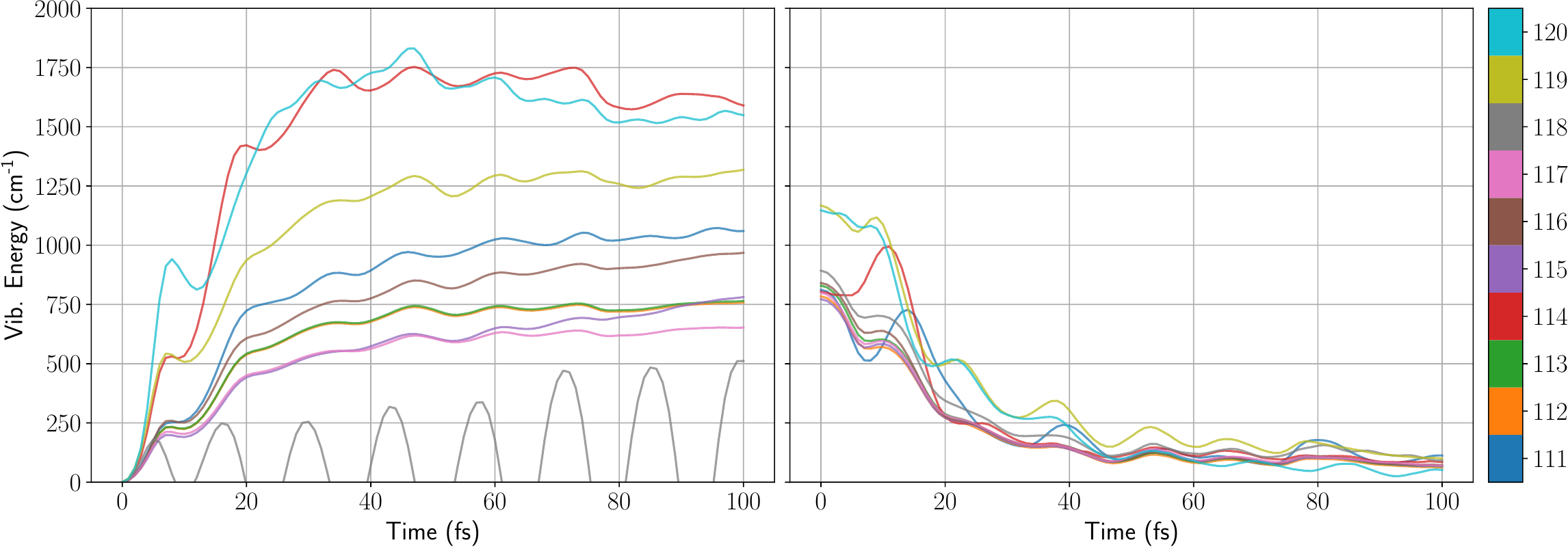}
    \caption{
    Time evolution of the vibrational energy per mode (for a selection of modes among the 93-dimensional model) and per state, (D\textsubscript{1} or D\textsubscript{2}, left and right, respectively), calculated with the propagated ML-MCTDH wavepacket for the LVC$+\gamma$ model after initial excitation to D\textsubscript{2}.
    }
    \label{fig:VibE}
\end{figure*}

As studied in previous work,\cite{galiana_excitation_2024} intramolecular EET can be characterized by following the geometry of the molecule (here, the expectation values of positional operators, see \cref{fig:QExp}), or the vibronic relaxations (vibrational energy per mode and per state, see \cref{fig:VibE}, relative to the energy within the electronic system at FCP).

\section{Spectral densities}
\label{sec:Specdens}

\begin{figure}[!ht]
    \centering
    \includegraphics[width=0.5\textwidth]{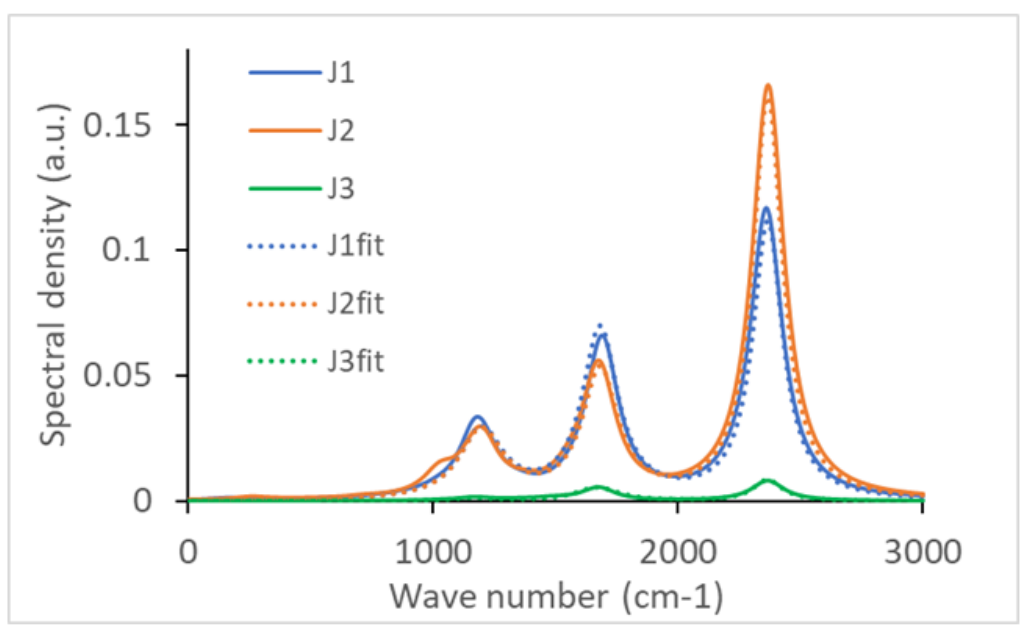}
    \caption{Spectral densities associated to the correlation functions for the three baths: $J_1$ tuning for D\textsubscript{1}, $J_2$ tuning for D\textsubscript{2}, and $J_3$ inter-state coupling. Solid lines: continuous functions from the LVC model obtained with the smoothing function,  \cref{eq:delta}; dotted lines: fit with the three Tannor-Meier Lorentzian functions.}
    \label{fig:spectral_density}
\end{figure}

The spectral densities of the baths $n$ ($n=1,3$) using the LVC parameters are given as, 
\begin{equation}
 {{J}_{n }}(\omega )=\frac{\pi }{2}\sum\nolimits_{k}{\frac{f_{k}^{(n )2}}{{{\omega }_{k}}}}\,\delta (\omega -{{\omega }_{k}}) \quad.
\end{equation}
They only involve the intramolecular vibrations. In order to account for the existence of an environment, the delta distribution is broadened by a Lorentzian smoothing function,
\begin{equation}
\delta (\omega -{{\omega }_{j}})\sim\frac{1}{\pi }\frac{\Gamma}{{{\left( \omega -{{\omega }_{j}} \right)}^{2}}+{{\Gamma }^{2}}} \quad,
\label{eq:delta}
\end{equation}
leading to a continuous spectral density shown in \cref{fig:spectral_density}.
We take $\Gamma =\qty{160}{\per\centi\meter}$ (HWHM). 

The spectral density of each tuning or coupling bath is fitted by a sum of ${n}_\text{lor}$ Tannor-Meier Lozentzian functions,\cite{Meier1999,Pomyalov2010}
\begin{equation}
 J^\text{TM}\left( \omega  \right)  \approx\sum\limits_{l=1}^{{{n}_\text{lor}}}{\frac{{{p}_{l}}\omega }{\hbar^3\left[ {{\left( \omega +{{\Omega }_{l}} \right)}^{2}}+\Gamma _{l}^{2} \right]\left[ {{\left( \omega -{{\Omega }_{l}} \right)}^{2}}+\Gamma _{l}^{2} \right]}} \quad.
\label{eq:JTM}
\end{equation}
The parameters $p_{l}$, $\Omega_{l}$, and $\Gamma_{l}$ are given in Table \ref{tab:table_threebaths}. 

\begin{table}[ht]
    \begin{tabular}{cccc}
    \toprule
    
    & $p_{l}$ (Ha$^4$) & $\hbar \Omega_{l}$ (Ha) & $\hbar \Gamma_{l}$ (Ha) \\ 
\midrule
$J_{1}$ & $5.5 \times 10^{-10}$ & $1.078687 \times 10^{-2}$ & $3.40 \times 10^{-4}$ \\
~      &  $3.3 \times 10^{-10}$ & $7.6634\times 10^{-3}$ & $4.0 \times 10^{-4}$ \\
 ~      & $1.5\times 10^{-10}$ & $5.4517 \times 10^{-3}$ & $5.0 \times 10^{-4}$ \\
 $J_{2}$ & $8.0 \times 10^{-10}$ & $1.078687 \times 10^{-2}$ & $3.40 \times 10^{-4}$ \\
~      &  $2.5 \times 10^{-10}$ & $7.6634\times 10^{-3}$ & $4.0 \times 10^{-4}$ \\
 ~      & $1.5\times 10^{-10}$ & $5.4517 \times 10^{-3}$ & $5.0 \times 10^{-4}$ \\
$J_{3}$ & $4.1 \times 10^{-11}$ & $1.078275 \times 10^{-2}$ & $3.40 \times 10^{-4}$ \\
 ~       & $4.1 \times 10^{-11}$ & $7.61 \times 10^{-3}$ & $5.0 \times 10^{-4}$  \\
 ~       & $5.0 \times 10^{-12}$ & $5.286 \times 10^{-3}$ & $4.5 \times 10^{-4}$ \\
\bottomrule
    \end{tabular}
\caption{\label{tab:table_threebaths} Parameters of the Tannor-Meier Lorentzian functions fitting the spectral densities of the three-bath model. $J_{1}(\omega)$ and $J_{2}(\omega)$ correspond to the tuning bath that makes the relative energies of states D$_1$ and D$_2$ fluctuate to first order (tuning). $J_{3}(\omega)$ refers to the coupling bath inducing first-order variations of the interstate coupling.}
\end{table}

\clearpage
\section{\label{sec:HEOM}HEOM}
\subsection{\label{subsec:Operationalequ}Operational equations}
The HEOM formalism aims at solving the non-Markovian master equation of an open quantum system \cite{Kubo1989,Tanimura2006,Tanimura2020,Yan2007,Yan2009,Mangaud2023}. It consists in a local-in-time system of coupled equations among auxiliary operators (ADOs). Each ADO has the dimension of the system reduced-density matrix (3 $\times$ 3 in our application), and is labelled by a global index-vector denoted $\mathbf{m} = (m_1,...,m_k,...,m_K)$, which gives the occupation number of each artificial mode of the expansion fitting the correlation funtion of bath $n$
\begin{equation}
{{C}_{n}}(t)=\sum\nolimits_{k=1}^{{{K}_{n}}}{\alpha _{k}^{(n)}}{{e}^{i\gamma _{k}^{(n)}t}} \quad,
\label{eq:Cdetexpan}
\end{equation}
with complex amplitudes and phases, $\alpha_k$ and $\gamma_{k} $. The system reduced-density matrix is ${{\rho}_\mathrm{S}}={\rho}_{\mathbf{m=0}}={{\rho}_{0,...,0}}$, with all the $m_j$ indices in $\mathbf{m}$ set to zero.
The number $K=\sum\nolimits_{n}^{{{N}_\text{bath}}}{{{K}_{n}}}$ represents the total number of artificial modes belonging to the $N_{\text{bath}}$ baths (size of the $\mathbf{m}$-vector).

The hierarchical master equations for the ADOs thus take the following form,
\begin{align}
  & {{{\dot{\rho}}}_{\mathbf{m}}}(t)={\mathbf{L}_\text{S}}{{\rho}_{\mathbf{m}}}(t)+i\sum\limits_{n}^{{{N}_{\text{bath}}}}{\sum\limits_{k=1}^{{{K}_{n}}}{m_{k}^{(n)}\gamma _{k}^{(n)}{{\rho}_{\mathbf{m}}}(t)}} \nonumber \\ 
 & -i\sum\limits_{n}^{{{N}_{\text{bath}}}}{\left[ {{\mathbf{S}}_{n}},\sum\limits_{k=1}^{{{K}_{n}}}{{{\rho}_{\mathbf{m}_{k}^{+(n)}}}(t)} \right]} \nonumber \\ 
 & -(i/\hbar^2)\sum\limits_{j}^{{{N}_{\text{bath}}}}{\sum\limits_{k=1}^{{{K}_{n}}}{m_{k}^{(n)}\left( \alpha _{k}^{(n)}{{\mathbf{S}}_{n}}{{\rho}_{\mathbf{m}_{k}^{-(n)}}}(t)-\tilde{\alpha }_{k}^{(n)}{{\rho}_{\mathbf{m}_{k}^{-(n)}}}(t){{\mathbf{S}}_{n}} \right)}} \quad,
 \label{eq:eqscouHEOM}
\end{align}
where ${\mathbf{L}}_{\text{S}}\bullet =-(i/\hbar)\left[ {\mathbf{H}_{\text{S}}},\bullet \right]$ is the Liouvillian `superoperator matrix representation' of the system Hamiltonian matrix, ${\mathbf{S}}_{n}$ are the matrices of the system operators of the $n^{\textrm{th}}$ bath in the system-bath Hamiltonian $\mathbf{H}_{\text{{SB}}}$, and $\mathbf{m}_{k}^{\pm(n) }=\{{{m}_{1,}},...,{{m}_{k}}\pm 1,...,{{m}_{K}}\}$ are the indices of the ADOs for which the occupation number of one artificial mode of the $n^{\textrm{th}}$ bath has been changed by one unit.
Note that due to their definition in term of multi-integrals over time the ADOs have a dimension $[\text{time}]^{-L}$ where $L$ is the hierarchy level \cite{Meier1999,Ishizaki_Fleming2009,Fleming2010,Mangaud2023} given by the total occupation number of vector $\mathbf{m}$. The hierarchy is thus structured in layers.
Each level interacts only with the two neighboring layers.

When the spectral density is fitted by a sum of Tannor-Meier (TM) Lorentzian functions,\cite{Meier1999}
\begin{equation}
 J\left( \omega  \right)=\sum\limits_{l=1}^{{{n}_\text{lor}}}{\frac{{{p}_{l}}\omega }{\hbar^3\left[ {{\left( \omega +{{\Omega }_{l}} \right)}^{2}}+\Gamma _{l}^{2} \right]\left[ {{\left( \omega -{{\Omega }_{l}} \right)}^{2}}+\Gamma _{l}^{2} \right]}} \quad.
 \label{eq:Jexpansion}
\end{equation}
each Lorentzian leads to two artificial decay modes with opposite frequencies, ${{\Omega }_{l}}+i{{\Gamma }_{l}}$ and $-{{\Omega }_{l}}+i{{\Gamma }_{l}}$. They correspond to poles in the upper half-plane.
The corresponding weights, $\alpha_k$, take into account the Bose function and satisfy the Boltzmann relation between system-bath absorption and emission. The two $\alpha_k$ complex coefficients for each function $l$ are
\begin{align}  \label{eq:alphaTM}
  & {{\alpha }_{l1}}=\frac{{{p}_{l}}}{8{\hbar^2{\Omega }_{l}}{{\Gamma }_{l}}}\left[ \coth \left( \frac{\hbar\beta }{2}({{\Omega }_{l}}+i{{\Gamma }_{l}}) \right)-1 \right] \nonumber \quad, \\  
 & {{\alpha }_{l2}}=\frac{{{p}_{l}}}{8{\hbar^2{\Omega }_{l}}{{\Gamma }_{l}}}\left[ \coth \left( \frac{\hbar\beta }{2}({{\Omega }_{l}}-i{{\Gamma }_{l}}) \right)+1 \right]\quad. 
\end{align} 
The corresponding rates are
\begin{align}  \label{eq:gammaTM}
 {{\gamma_{l1}=\Omega }_{l}} +i{{\Gamma }_{l}} \nonumber \quad, \\ 
 {{\gamma_{l2}=-\Omega }_{l}}+i{{\Gamma }_{l}}\quad.
 \end{align} 
The expansion in \cref{eq:Cdetexpan} also involves the poles of the Bose function on the imaginary axis ($\omega=0$) also known as the Matsubara modes $M$ with parameters 
 \begin{equation}
 \alpha_{Mk}=2iJ(\gamma_{Mk})/\beta \quad,
 \end{equation}
and
\begin{equation}
\gamma_{Mk}=i2\pi k/(\hbar \beta) \quad.    
\end{equation}
With this TM parametrization, the $\tilde{\alpha}_k$ of  \cref{eq:eqscouHEOM} are defined as follows: $\tilde{\alpha}_{l1}=\alpha_{l2}^*$, $ \tilde{\alpha}_{l2}=\alpha_{l1}^*$ and $\tilde{\alpha}_{Mk}=\alpha_{Mk}^*$ where $*$ designates the complex conjugate.

\subsection{\label{subsec:Bathdynamics}Bath dynamics in HEOM}
Some information about the bath dynamics during the relaxation can be extracted from the ADOs.\cite{ZhuShi2012,LiuShi2014,Desouter2017,Desouter2019}
For instance, the average value of each tuning collective mode, ${{\bar{X}}_{n}}={{\bar{B}}_{n}}/{{D}^{(n)}}$ (with $n=1,2$), illustrates the trajectory toward the new equilibrium position.
${{\bar{B}}_{n}}(t)$ is obtained upon summing the corresponding diagonal elements of each ADO (referred to as with index $j$) of the first hierarchy level, $L = 1$, with a single-occupation number in the artificial modes of this bath $n$,
\begin{equation}
 {{\bar{B}}_{n}}(t)=-\hbar \sum\nolimits_{j}{\rho _{j,nn}^{L=1}}(t) \quad. \label{eq:Bn(t)}  
\end{equation}
The normalisation factor is computed in the continuous case by \begin{equation}
 {{D}^{(n)2}}=\frac{2}{\pi }\int_{0}^{\infty }{\omega {{J}_{n}}(\omega )\mathrm{d}\omega } \quad.
 \label{eq:D(n)}
\end{equation}

\begin{figure}[!ht]
    \centering
    \includegraphics[width=0.5\textwidth]{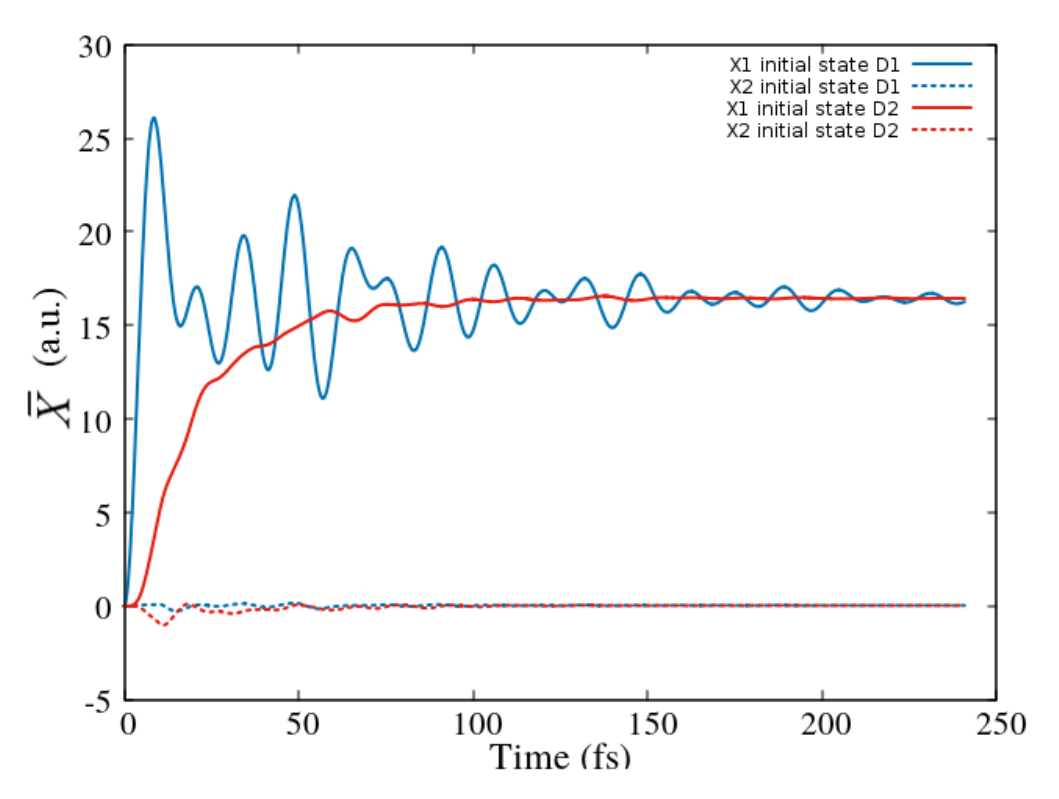}
    \caption{Time evolution of the expectation value for the collective mode ${\bar{X}}_{n}$ (with $n=1,2$) for bath 1 (full lines) and 2 (dashed lines) when the initial state is D$_1$ or D$_2$ (blue and red, respectively).}
    \label{fig:AverageX}
\end{figure}

Figure \ref{fig:AverageX} gives the average position of the collective mode of bath 1 or 2 when the initial state is in D$_1$ or D$_2$, respectively.

When the system is prepared in D$_1$, one observes a fast motion far from the Franck-Condon region, followed by an oscillatory relaxation toward the equilibrium position of D$_1$.
Very little population reaches the D$_2$ state, so that the average remains quasi zero.
A similar result would be obtained with a wavepacket having a vanishing norm in an excited state.
When the preparation is in state D$_2$, the transfer is ultrafast and the relaxation is overdamped and directly reaches the same equilibrium position in D$_1$.
Again, the norm of the remaining wavepacket in D$_2$ is negligible.

\subsection{Convergence tests for the 2L and 3L model}

The simulation of the non-linear spectra in the m23 dimer is computationally very demanding since it involves three vibrational baths strongly coupled to the electronic system requiring high HEOM levels. 
Even if a single propagation may be driven with a converged HEOM level, the computation of 2D grids become prohibitive and we choose a non-converged but reasonable level to ensure a good balance between qualitative accuracy and computational resources. 
The first approximation concerns the representation of the spectral densities (see \cref{fig:spectral_density}) by retaining the three main peaks (3L model) or only the two high-frequency peaks (2L model). 
The second approximation is related to the number of effective artificial modes retained in the bath correlation functions (\cref{eq:Cdetexpan}). 
As we adopt the Tannor-Meier method, each Lorentzian function generates two artificial modes (i.e., 18 in the 3L model and 12 in the 2L model). Rigorously, some Matsubara terms corresponding to the poles of the Bose function must be taken into account mainly at very low temperature. 
We first give in \cref{tab:ADOnumber} the number of auxiliary matrices (dimension 3 $\times$ 3 here) necessary for different approximations at room temperature. 
To justify our choice of the TM strategy in our application, we also mention the number of matrices required in the Free-Pole methodology (FP) \cite{Xu2022} or in a discrete undamped mode approach (DM) \cite{LiuShi2014} that have not been used in our simulation but have been checked recently for the m22 dimer in Ref. \cite{LeDe2024}. In the m23 case, at least six poles are necessary for each bath in the FP method leading to 36 artificial modes. To illustrate the  discrete mode case, one assumes a discretization of the spectral density with only 30 or 50 modes per bath at HEOM level
L = 5. The effective mode number for each bath is then 2 × 30 or 2 × 50 because one has to consider positive and negative
frequencies.

It is worthy to note that the approximation of discarding the Matsubara modes in a TM approach is conceivable only at high temperature since at very low temperature, their number may be a few hundreds and alternative strategies are required.

\begin{table}[ht]
    \begin{tabular}{cccc}
    \toprule
    
     Modes & Matsubara & HEOM level & number of matrices \\ 
\midrule
12 TM & 0 & 7 & 50388 \\
~     & 0 & 8 & 125970 \\
 ~    & 0 & 9 & 293930 \\
 ~    & 3 & 7 & 1184040 \\
 ~    & 5 & 7 & 5379616 \\ 
18 TM & 0 & 7 & 480700 \\
 ~    & 0 & 8 & 1562275 \\
~     & 0 & 9 & 4686825 \\
 ~    & 3 & 7 & 5379616  \\
36 FP & $-$ & 6 & 5245786 \\
~    & $-$ & 7 & 32224114  \\
30 DM & $-$ & 5 & 8259888 \\
50 DM & $-$ & 5 & 96560646 \\
\bottomrule
    \end{tabular}
\caption{\label{tab:ADOnumber} Number of HEOM matrices at different levels for a parametrization using Tannor-Meier (TM) fit of the spectral density described in \cref{sec:Specdens},  free poles (FP) used in references \cite{Xu2022,LeDe2024} or discrete undamped modes (DM) \cite{LiuShi2014,LeDe2024} (not used here). The 2L and 3L models involve 12 and 18 TM modes respectively. The ``Matsubara'' column gives the number of terms coming from the Bose function in the TM parametrization (see \cref{subsec:Operationalequ}).  }
\end{table}

\begin{figure}[!ht]
    \centering
    \includegraphics[width=0.5\textwidth]{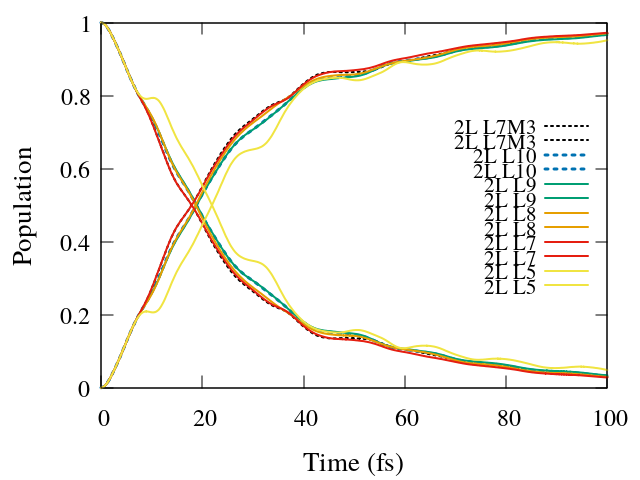}
    \caption{Time evolution of the populations when the initial state is D$_2$ in the 2L model for different hierarchy levels (denoted LX) and with three Matsubara terms for the level 7 (L7M3).}
    \label{fig:Pop2L}
\end{figure}

\begin{figure}[!ht]
    \centering
    \includegraphics[width=0.5\textwidth]{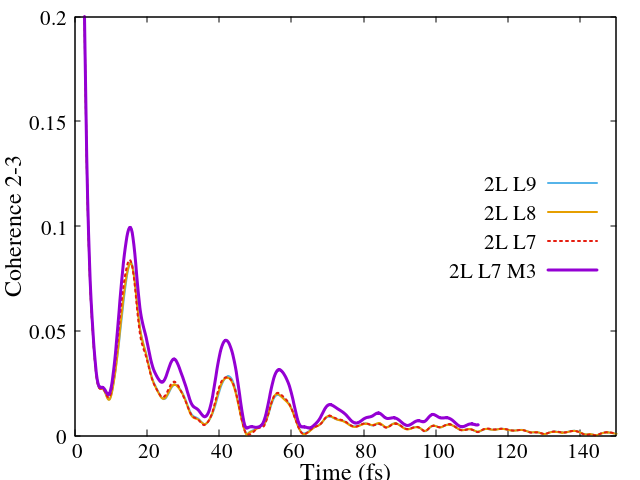}
    \caption{Time evolution of the coherence modulus of an initial superposed state with equal weights in the two excited states in the 2L model for different hierarchy levels (denoted LX) and with three Matsubara terms for the level 7 (L7M3).}
    \label{fig:Coh2L}
\end{figure}

We now illustrate the impact of different approximations in the 2L model on the population in \cref{fig:Pop2L} and on the coherence in \cref{fig:Coh2L}. 
Figure \ref{fig:Pop2L} shows that the populations in the 2L model are converged at level L9 and L10 and that they remain acceptable at level L7. 
One also sees that adding three Matsubara (L7M3) does not modify the populations. 
However, coherences are more sensitive to these terms and we will illustrate that their neglect to keep reasonable computational time leads to approximate results retaining the main lines of the processes.
As seen in \cref{fig:Coh2L}, coherences converge more slowly than the populations. 
We compare the evolution of the modulus of the coherence of an initial superposed state with equal weights in each excited state. 
In this case, adding three Matsubara terms does not modify the general oscillatory pattern but slighy increases the amplitude revealing a slower dephasing. 
We will illustrate in \cref{subsec:ESEuncoupledcase} the impact of the different approximations on a specific spectrum in a purely dephasing case for which computations may be compared to an analytical result. 

 To estimate how the 2L model may provide the main characteristics of the EET, we compare the HEOM populations calculated using both the 2L and 3L models at level 7 for the 2L model and levels 7, 8 and 9 for the 3L model. 
 The populations are not converged in the 3L model at level L $= 7$ but the behavior remains similar to that predicted by upper levels. 
 For the populations, this approximation agrees in a satisfactory manner with the level L $= 7$ used for our simulations in the 2L model.

\begin{figure}[!ht]
    \centering
    \includegraphics[width=0.5\textwidth]{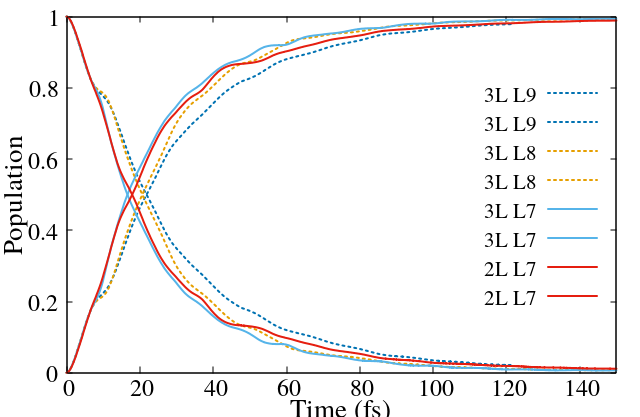}
    \caption{Time evolution of the populations when the initial state is D$_2$ in the 2L or 3L model for different hierarchy levels without Matsubara terms.}
    \label{fig:Pop2L-3Lmodel}
\end{figure}

\begin{figure*}[!ht]
    \centering
    \includegraphics[width=0.45\textwidth]{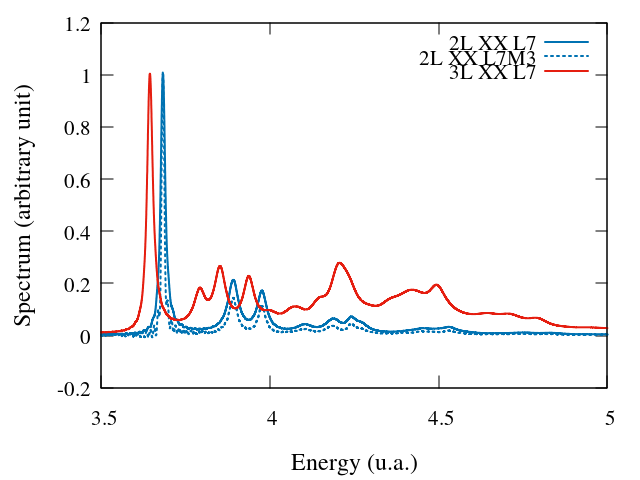}
    \includegraphics[width=0.45\textwidth]{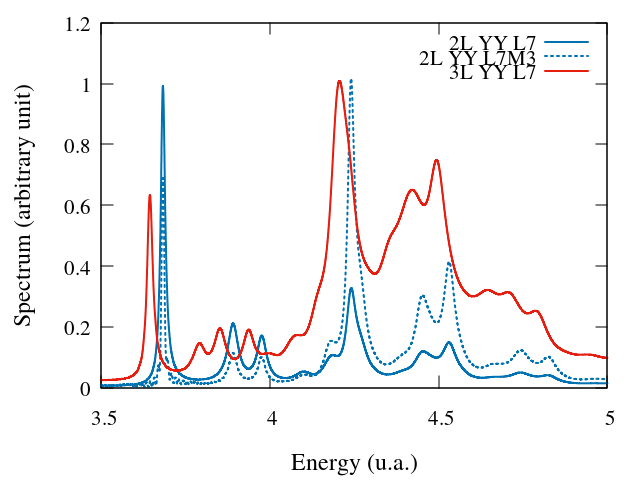}\\
    \includegraphics[width=0.45\textwidth]{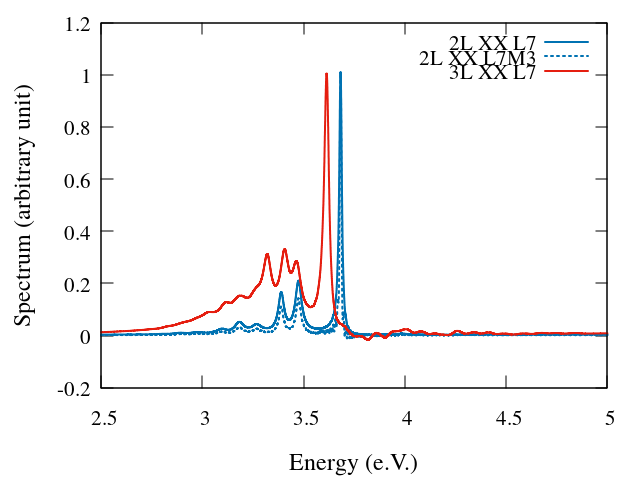}
    \includegraphics[width=0.45\textwidth]{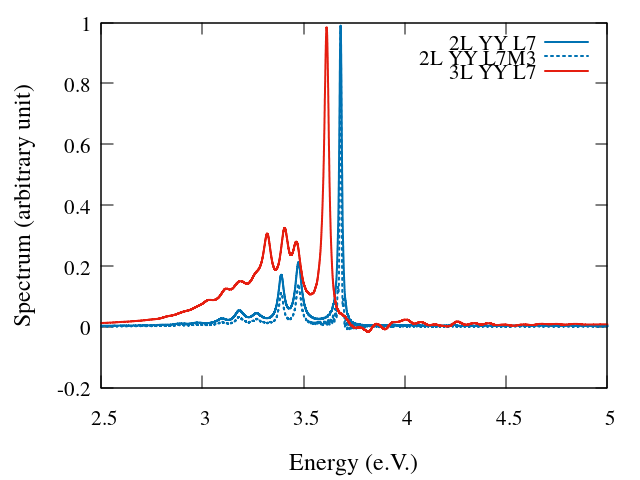}\\
    \caption{Normalized absorption (upper panels) and emission (lower panels) spectra for the 2L and 3L models with L$=7$ HEOM level and three Matsubara (L7M3) in the 2L case. The polarization is along  $e_x$ (left) (case XX) or $e_y$ (right) (caseYY).}
    \label{fig:spectre2L3L}
\end{figure*}

Finally, in \cref{fig:spectre2L3L}, we compare the spectra obtained with polarisation along $e_x$ (left) or $e_y$ (right). 
In the 2L model, taking into account three Matsubara terms does not modify the peaks position but their intensity.
There is a notable intensity difference for the peak at 4.2 eV with the polarization along $e_y$. 
The computation in the 3L model at level L $=7$ leads to an enlargment of the peaks and to a red shift of about 0.05 eV of the main peaks. 
The peak at 4.25 eV with $e_y$ polarization has an intensity similar that that obtained in the 2L model with the Matsubara terms. 
These examples suggest that even if it is not satisfactory to work with a non-converged level, this provides a reasonable qualitative representation of the spectra at a reasonable computational cost.
The ratio of the computational times on the same system with a similar parallelization scheme is 10 for the case L$=7$ in the 2L or 3L model. 
It is 25 for the case L$=7$ in the 2L model without or with three Matsubara terms.

Another example is given in \cref{subsec:ESEuncoupledcase}. 
\clearpage

\section{Nonlinear spectroscopy}

\subsection{Third-order response funtion}
The third-order optical response function  ${{R}^{(3)}}({{t}_{3}},{{t}_{2}},{{t}_{1}})$ that describes the system response to the light-induced perturbations is abundantly documented.\cite{Mukamel1992,Mukamel1995,Khalil2003,Mukamel2009,Domcke2015,Mukamel2017,Zhang2019,Collini2021,Krich2023,segatta_time-resolved_2024,Bittner2023}. 
We summarize here the expressions used with HEOM propagation.

By assuming a factorisation, the initial total density matrix is the product of the system density times the thermally equilibrated density of the bath, ${{\rho}_{\text{S}}}(0){{\rho}_{\text{eq}}}(T)$. 
In the HEOM simulation, the propagation involves an ensemble of matrices $\bar{\rho} (t) = \left\{ \rho_{\mathbf{m}=0}(t),\rho_{\mathbf{m\ne 0}}(t) \right\}$   containing the system density matrix $\rho_S=\rho_{\mathbf{m}=0}$ and all the ADOs $\rho_{\mathbf{m\ne 0}}$ where $\mathbf{m}$ is the global index vector defined in \cref{subsec:Operationalequ}. 
As extensively reviewed, the computation of ${{R}^{(3)}}({{t}_{3}},{{t}_{2}},{{t}_{1}})$ may be split into different responses that correspond to multi-time correlation functions and are associated to different processes: ESE (excited-state stimulated emission), GSB (ground-state bleaching), and ESA (excited-state absorption).
The 2D signal involves six responses (for the sake of simplicity, the polarization $(p)$ is not specified in the lowering or raising operators but it must be specified when using $\mu_{-}$ or $\mu_{+}$).
The operational relations are summarized as follows and explained below,
\begin{equation}
{{R}_{1}}({{t}_{3}},{{t}_{2}},{{t}_{1}})=\left\langle {{\bar{\mu} }_{-}}G({{t}_{3}})\left\{ G({{t}_{2}})\left[ G({{t}_{1}})\left( {{\bar{\mu} }_{+}}{{\bar{\rho}}_{t=0}} \right){{\bar{\mu} }_{-}} \right]{{\bar{\mu} }_{+}} \right\} \right\rangle \quad,  
\label{eq:R1}
\end{equation}
\begin{equation}
 {{R}_{2}}({{t}_{3}},{{t}_{2}},{{t}_{1}})=\left\langle {{\bar{\mu} }_{-}}G({{t}_{3}})\left\{ G({{t}_{2}})\left[ {{\bar{\mu} }_{+}}G({{t}_{1}})\left( {{\bar{\rho}}_{t=0}}{{\bar{\mu} }_{-}} \right) \right]{{\bar{\mu} }_{+}} \right\} \right\rangle \quad,
 \label{eq:R2}
\end{equation} 
\begin{equation}
 {{R}_{3}}({{t}_{3}},{{t}_{2}},{{t}_{1}})=\left\langle {{\bar{\mu} }_{-}}G({{t}_{3}})\left\{ {{\bar{\mu} }_{+}}G({{t}_{2}})\left[ G({{t}_{1}})\left( {{\bar{\rho}}_{t=0}}{{\bar{\mu} }_{-}} \right){{\bar{\mu} }_{+}} \right] \right\} \right\rangle \quad, 
 \label{eq:R3}
\end{equation}
\begin{equation}
 {{R}_{4}}({{t}_{3}},{{t}_{2}},{{t}_{1}})=\left\langle {{\bar{\mu} }_{-}}G({{t}_{3}})\left\{ {{\bar{\mu} }_{+}}G({{t}_{2}})\left[ {{\bar{\mu} }_{-}}G({{t}_{1}})\left( {{\bar{\mu} }_{+}}{{\bar{\rho}}_{t=0}} \right) \right] \right\} \right\rangle \quad,
 \label{eq:R4}
\end{equation}   
\begin{equation}
 {{R}_{5}}({{t}_{3}},{{t}_{2}},{{t}_{1}})=\left\langle {{\bar{\mu} }_{-}}G({{t}_{3}})\left\{ {{\bar{\mu} }_{+}}G({{t}_{2}})\left[ {{\bar{\mu} }_{+}}G({{t}_{1}})\left( {{\bar{\rho}}_{t=0}}{{\bar{\mu} }_{-}} \right) \right] \right\} \right\rangle \quad,
\end{equation}
\begin{equation}
{{R}_{6}}({{t}_{3}},{{t}_{2}},{{t}_{1}})=\left\langle {{\bar{\mu} }_{-}}G({{t}_{3}})\left\{ {{\bar{\mu} }_{+}}G({{t}_{2}})\left[ G({{t}_{1}})\left( {{\bar{\mu} }_{+}}{{\bar{\rho}}_{t=0}} \right){{\bar{\mu} }_{-}} \right] \right\} \right\rangle \quad.  
\end{equation} 
$G(t)$ is the HEOM propagator driving the ensemble of matrices during a time $t_1$, $t_2$ or $t_3$. $\bar{\mu }={{I}_{N}}\otimes \mu $ where $I$ is the identity matrix,  $N$ is the number of HEOM matrices and $\mu$ is the dipole transition matrix of the system. This means that this $\mu$ matrix must be applied to the system density matrix and to every ADO \cite{Tanimura2006}. In the initial ${\bar{\rho}}_{t=0}$, all the ADOs are nil and the system is in the ground state. The first application of the dipole matrix is done on the left or on the right as given in the central parenthesis. The HEOM propagator is then driven during $t_1$. The second application of the relevant $\mu_+$ or $\mu_-$ to all the matrices is performed to prepare the right bracket [ ]. The HEOM propagator then operates during $t_2$. The third application of the transition operator on the left or on the right of all the matrices prepares the brace bracket $\{ \}$ that is propagated during $t_3$. Finally, the fourth application on the left of the lowering dipole operator mainly concerns the system density matrix and the relevant trace is over this system matrix. The trace over the bath is implicitly done by the HEOM algorithm.

The first two responses, ${{R}_{1}}$ and ${{R}_{2}}$, are associated to the ESE process, ${{R}_{3}}$ and ${{R}_{4}}$ describe GSB, while ESA is treated by ${{R}_{5}}$ and ${{R}_{6}}$. The latter process is not considered here because the electronic basis set is truncated to the ground and the two D\textsubscript{1} and D\textsubscript{2} excited states.

\begin{figure}[!ht]
    \centering
    \includegraphics[width=0.5\textwidth]{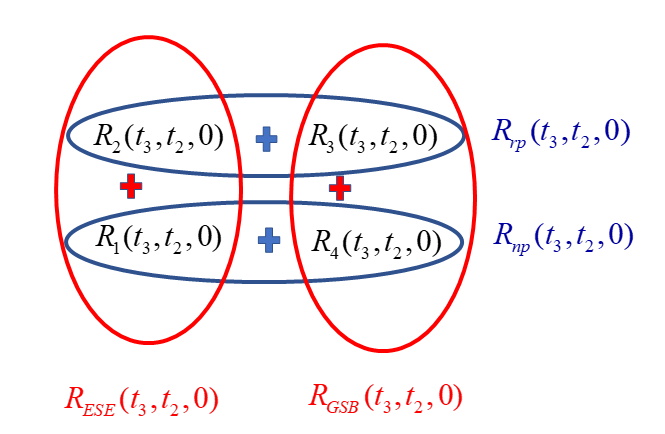}
    \caption{Scheme of the responses when ESA is neglected.}
    \label{fig:Response}
\end{figure}

The response function ${{R}^{(3)}}({{t}_{3}},{{t}_{2}},{{t}_{1}})={{R}_\text{rp}}({{t}_{3}},{{t}_{2}},{{t}_{1}})+{{R}_\text{nr}}({{t}_{3}},{{t}_{2}},{{t}_{1}})$ may be splitted in two contributions, the rephasing one, ${R}_\text{rp}$ (with detected wave vector $k=-{{k}_{1}}+{{k}_{2}}+{{k}_{3}}$), and the nonrephasing one, ${R}_\text{nr}$ (with $k={{k}_{1}}-{{k}_{2}}+{{k}_{3}}$).
They are given as
\begin{align}
  & {{R}_\text{rp}}({{t}_{3}},{{t}_{2}},{{t}_{1}})={{R}_{2}}({{t}_{3}},{{t}_{2}},{{t}_{1}})+{{R}_{3}}({{t}_{3}},{{t}_{2}},{{t}_{1}})-{{R}_{5}}({{t}_{3}},{{t}_{2}},{{t}_{1}}) \quad, \nonumber \\ 
 & {{R}_\text{nr}}({{t}_{3}},{{t}_{2}},{{t}_{1}})={{R}_{1}}({{t}_{3}},{{t}_{2}},{{t}_{1}})+{{R}_{4}}({{t}_{3}},{{t}_{2}},{{t}_{1}})-{{R}_{6}}({{t}_{3}},{{t}_{2}},{{t}_{1}}) \quad.  \nonumber\\ 
 \label{eq:RrpRnr}
\end{align}
Each rephasing or non-rephasing term contains two positive contributions related to excited stimulated emission (ESE) and ground state bleaching (GSB) and a negative part due to excited state absorption (ESA). When this absorption towards super excited states is neglected (the corresponding transition dipoles are assumed to vanish), both ${{R}_{5}}({{t}_{3}},{{t}_{2}},{{t}_{1}})$ and ${{R}_{6}}({{t}_{3}},{{t}_{2}},{{t}_{1}})$ are discarded. 
Computing the time-dependent spectra  
\begin{equation}
 {{\sigma }^{\text{ESE}}}({{\omega }},{{t}})=\operatorname{Re}\int_{0}^{\infty }{\mathrm{d}{{t}_{3}}{{e}^{i{{\omega }}{{t}_{3}}}}({R}_{1}(t_3,t,0) + {{R}_{2}}({{t}_{3}},{{t}},0)})\quad,
 \label{eq:ESE1}
\end{equation}
and
\begin{equation}
 {{\sigma }^{\text{GSB}}}({{\omega }},{{t}})=\operatorname{Re}\int_{0}^{\infty }{\mathrm{d}{{t}_{3}}{{e}^{i{{\omega }}{{t}_{3}}}}({R}_{3}(t_3,t,0) + {{R}_{4}}({{t}_{3}},{{t}},0)}) \quad,
 \label{eq:GSB1}
\end{equation}
involves the responses ${{R}_{1}}({{t}_{3}},{{t}_{2}},t_1)$, ${{R}_{2}}({{t}_{3}},{{t}_{2}},t_1)$ for ESE and ${{R}_{3}}({{t}_{3}},{{t}_{2}},t_1)$, ${{R}_{4}}({{t}_{3}},{{t}_{2}},t_1)$ for GSB as schematized in \cref{fig:Response}. 

When  ${{t}_{1}}=0$, one has $G({{t}_{1}=0})=1$ so that ${{R}_{1}}({{t}_{3}},{{t}_{2}},0)={{R}_{2}}({{t}_{3}},{{t}_{2}},0)$ and ${{R}_{3}}({{t}_{3}},{{t}_{2}},0)={{R}_{4}}({{t}_{3}},{{t}_{2}},0)$. A single response is necessary in each case and the two time-dependent spectra are finally given as
\begin{equation}
 {{\sigma }^{\text{ESE}}}({{\omega }},{{t}})=\operatorname{Re}\int_{0}^{\infty }{\mathrm{d}{{t}_{3}}{{e}^{i{{\omega }_{3}}{{t}_{3}}}}{{R}_{1}}({{t}_{3}},{{t}},0)} \quad,
 \label{eq:ESE2}
\end{equation}
\begin{equation}
 {{\sigma }^{\text{GSB}}}({{\omega }},{{t}})=\operatorname{Re}\int_{0}^{\infty }{\mathrm{d}{{t}_{3}}{{e}^{i{{\omega }_{3}}{{t}_{3}}}}{{R}_{3}}({{t}_{3}},{{t}},0)} \quad.
 \label{eq:GSB2}
\end{equation}

In all the expressions given in \cref{eq:R1,eq:R2,eq:R3,eq:R4} with ${{t}_{1}}=0$, one has $G({{t}_{1}=0})=1$, and the right part of the bracket describes the initial state that is assumed to be prepared by impulsive delta-like pulse associated to the lowering and raising operators.
For ESE, this initial state, ${{\mu }_{+}}{{\rho}_{0}}{{\mu }_{-}}$, is the system reduced-density matrix of a superposition of the excited states weighted by the transition dipoles and leading to an unnormalized state, which may be normalized. All the ADOs are zero in the intial equilibrium baths. For GSB, ${{\rho}_{0}}{{\mu }_{-}}{{\mu }_{+}}$ or ${{\mu }_{-}}{{\mu }_{+}}{{\rho}_{0}}$ corresponds to the system ground state populated with the norm of the previous superposition and all ADOs nil.

\subsection{Additional cuts and slices through the ESE and TA spectrograms}

\begin{figure*}[!ht]
    \centering
    \includegraphics[width=0.9\textwidth]{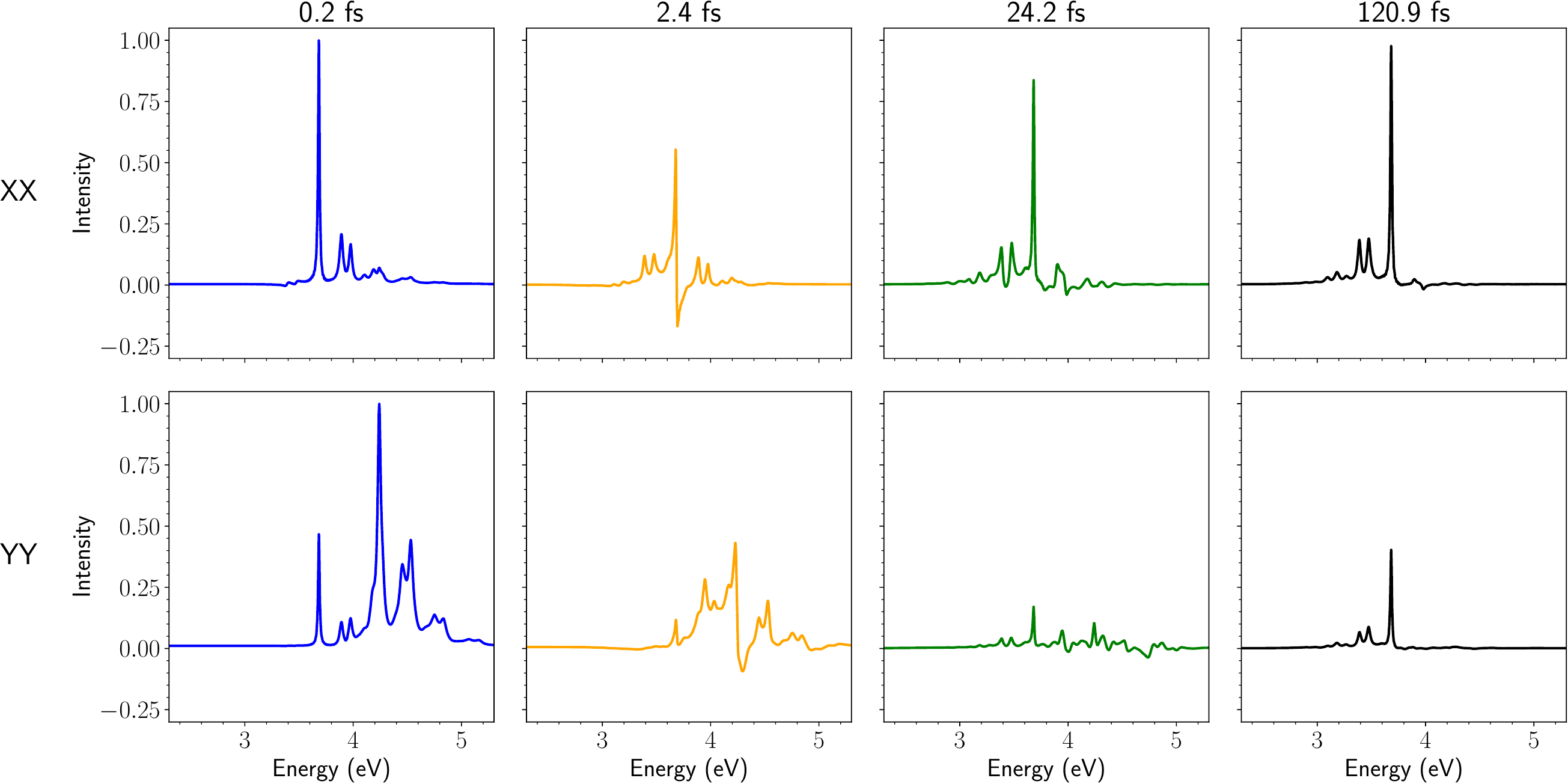}
    \caption{Energy-resolved cuts through the time-and-frequency-resolved ESE spectrograms for different pump-probe delays obtained in the full 2L model. 
    The polarizations are along $e_x$ (top) and $e_y$ (bottom). 
    The spectra are normalized with the maximum intensity among spectra along the same line (with the same polarization).}
    \label{fig:ESE_spectrogram_cuts_2L}
\end{figure*}

The energy-resolved cuts through the ESE spectrograms for the 2L model presented in fig.8 of the main text are given in \cref{fig:ESE_spectrogram_cuts_2L} here.
The corresponding cuts for the full 3L model are shown in fig. 9 in the main text.
The spectra are less structured but the general behavior is the same, passing indeed from the absorption profile toward the emissive one.

\begin{figure*}[!ht]
    \centering
    \includegraphics[width=0.90\textwidth]{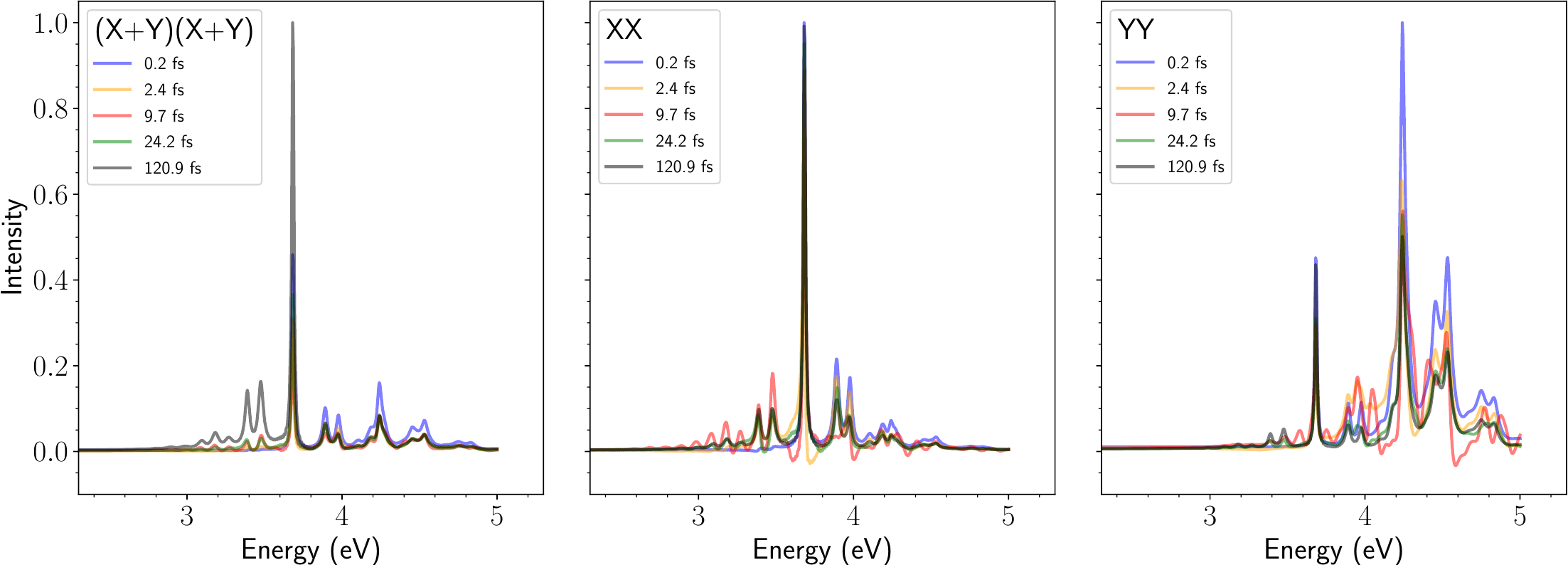}
    \caption{Energy-resolved cuts through the time-and-frequency-resolved TA spectrograms for different pump-probe delays obtained in the full 2L model. 
    The polarizations are along $(e_x + e_y)/\sqrt{2}$ (left panel), $e_x$ (center panel), and $e_y$ (right panel). 
    The spectra are normalized with the maximum intensity among spectra of the same panel (with the same polarization).}
    \label{fig:TA_spectrogram_cuts_2L}
\end{figure*}

\begin{figure*}[!ht]
    \centering
    \includegraphics[width=0.90\textwidth]{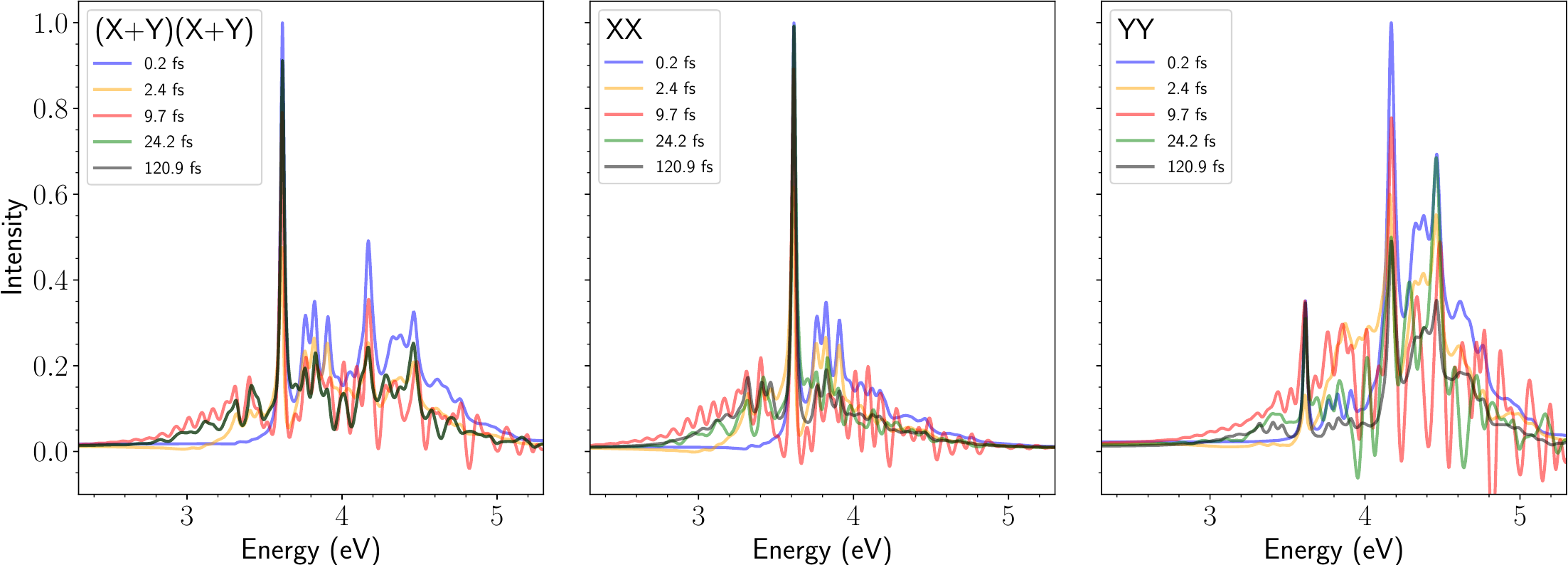}
    \caption{Energy-resolved cuts through the time-and-frequency-resolved TA spectrograms for different pump-probe delays obtained with the full 3L model. 
    The polarizations are along $(e_x + e_y)/\sqrt{2}$ (left panel), $e_x$ (center panel), and $e_y$ (right panel). 
    The spectra are normalized with the maximum intensity among spectra of the same panel (with the same polarization).}
    \label{fig:TA_spectrogram_cuts}
\end{figure*}

In \cref{fig:TA_spectrogram_cuts_2L} and \cref{fig:TA_spectrogram_cuts} here, we compare the evolution of the TA spectrum in the 2L and 3L models, respectively.
Since the initial ESE spectrum is similar to the GSB one, the TA spectrum evolves from nearly twice the GSB spectrum, \textit{i.e.}, twice the absorption spectrum, toward the sum of the absorption and the emission spectra.
The 3L model, also discussed in the maint text, accounts for the additional effect of the low frequencies at room temperature but shows similar trends.

\clearpage
\subsection{ESE and TA signals for an oblique direction of polarization}
\begin{figure*}[!ht]
    \centering
    \includegraphics[width=0.45\textwidth]{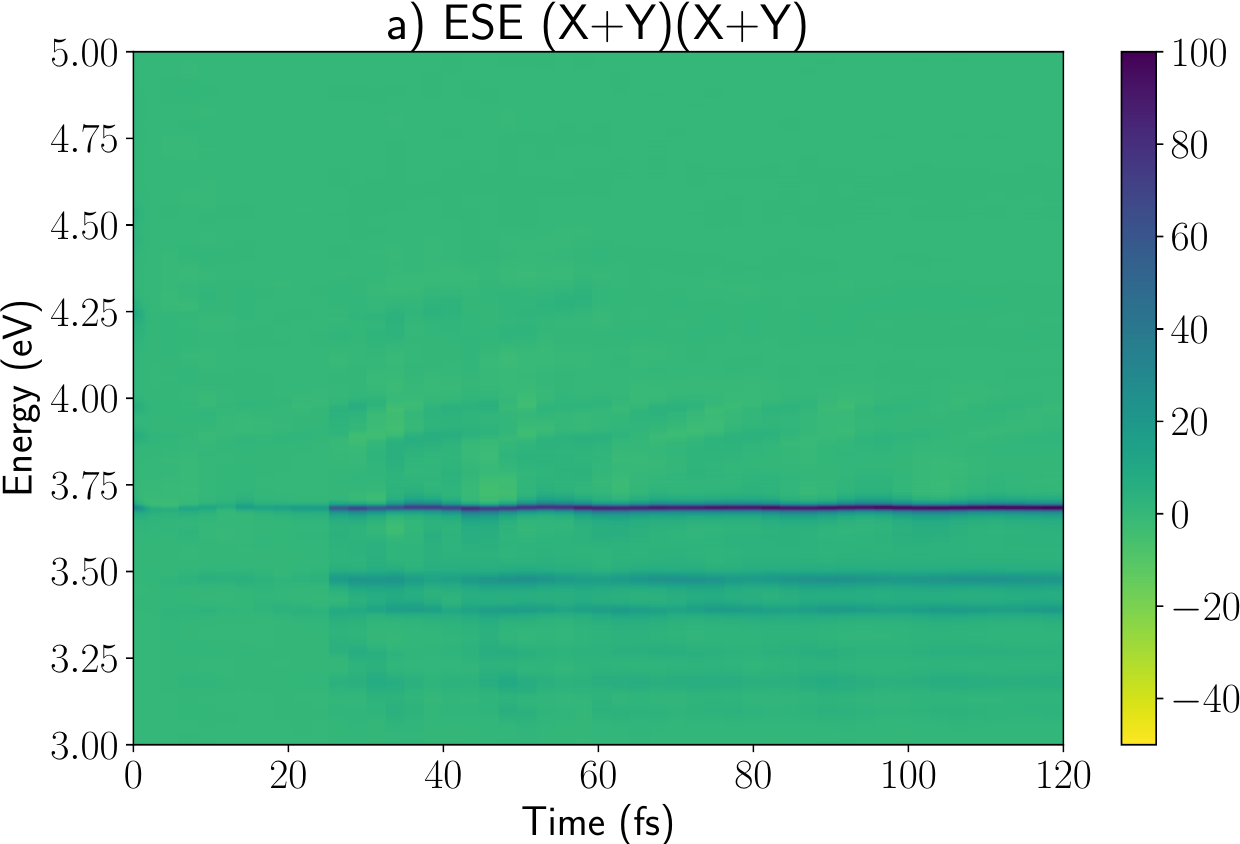}
    \includegraphics[width=0.45\textwidth]{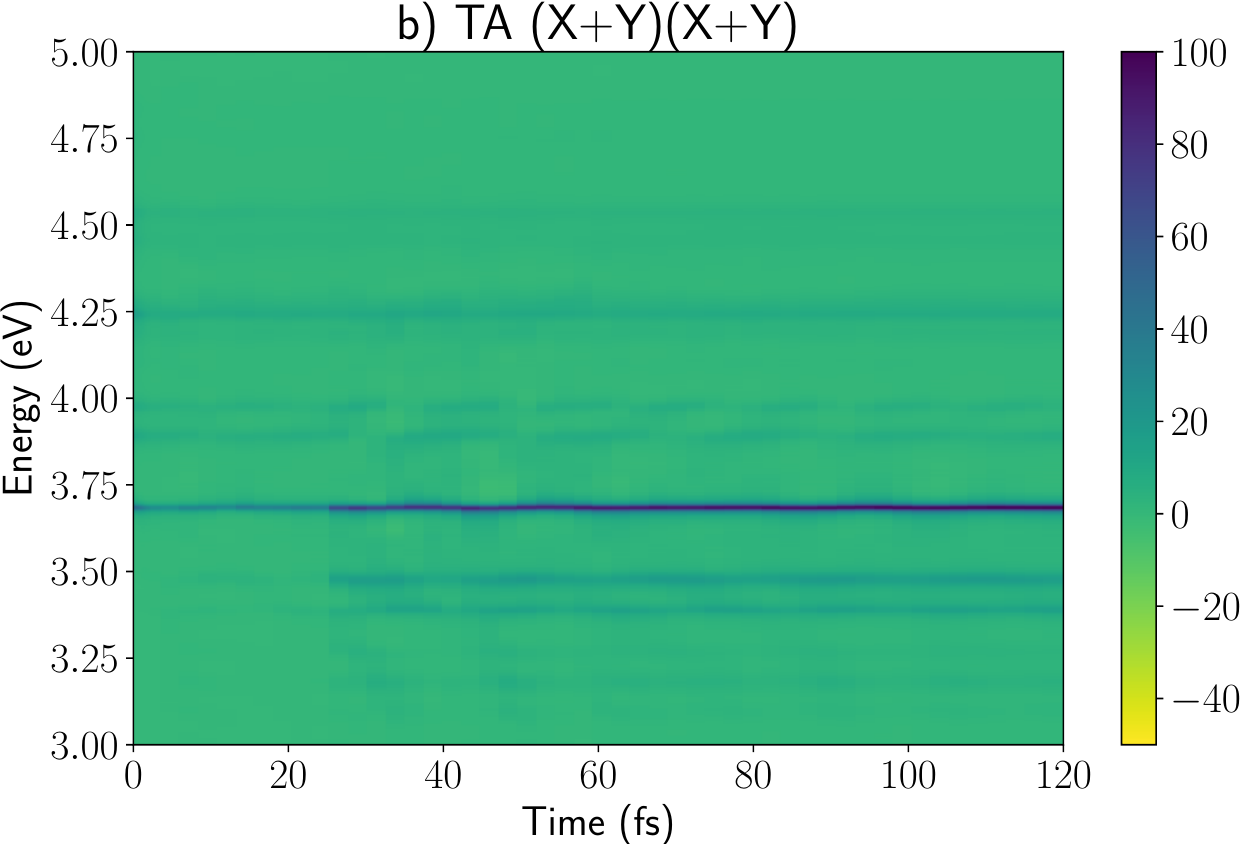}\\
    \caption{Time-and-frequency resolved ESE and TA spectrograms for the approximate 2L model with only the two high-frequency peaks in the spectral densities of \cref{fig:spectral_density}. The pump and probe are delta-like laser pulses. The polarization is along an oblique direction $(e_x+e_y)/\sqrt{2}$ (case (X+Y)(X+Y)).}
    \label{fig:ESE_TA_spectrograms_XpY}
\end{figure*}

Figure \ref{fig:ESE_TA_spectrograms_XpY} gives the ESE and TA spectrograms obtained for the 2L model when the polarization is oblique with respect to $e_x$ and $e_y$, that is $(e_x+e_y)/\sqrt{2}$ (case (X+Y)(X+Y)

\clearpage 
\subsection{ESE signals in the crossed-polarization cases}
\begin{figure*}[!ht]
    \centering
    \includegraphics[width=0.45\textwidth]{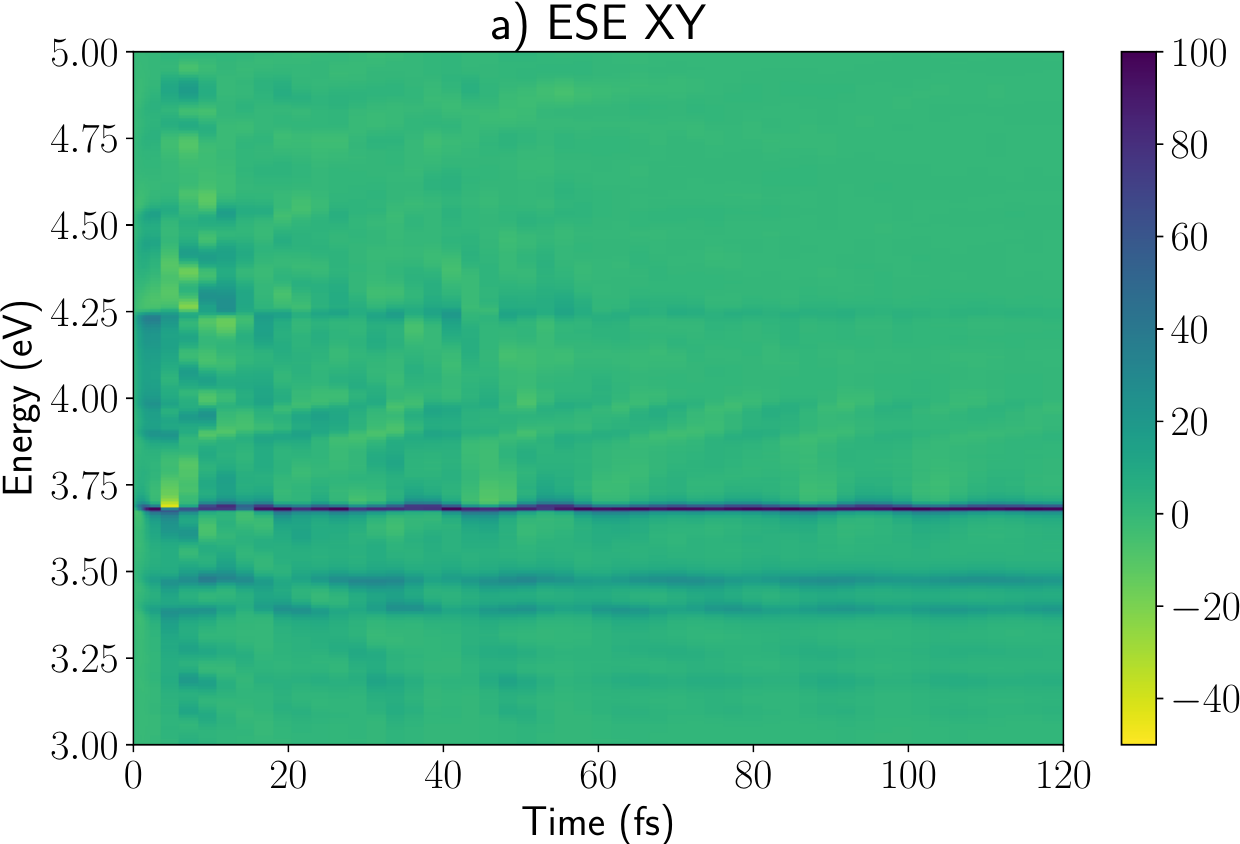}
    \includegraphics[width=0.45\textwidth]{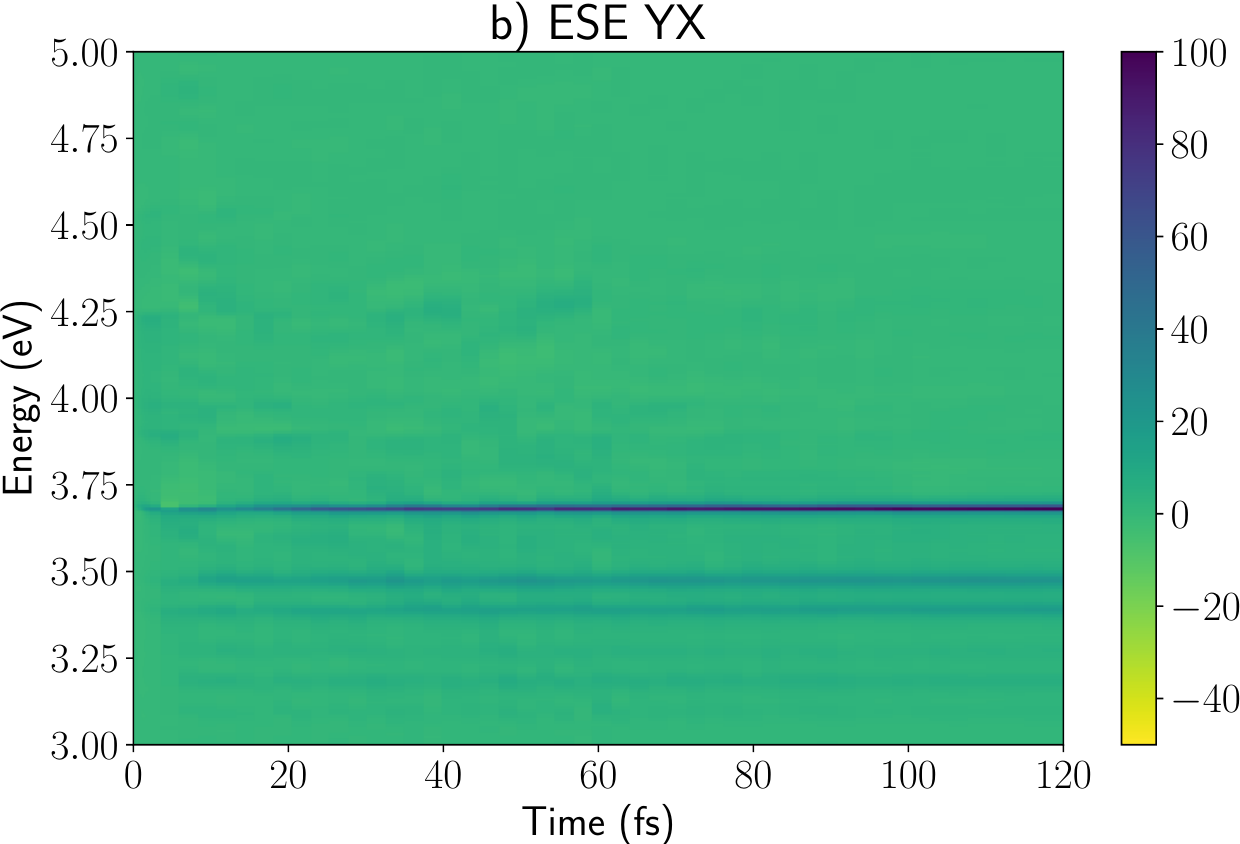}\\
    \caption{Time-and-frequency resolved ESE spectrograms for the approximate 2L model with only the two high-frequency peaks in the spectral densities of \cref{fig:spectral_density}. The pump and probe are delta-like laser pulses. The crossed polarizations for the pump and the probe are along $e_x$  and $e_y$, respectively, in the case XY (left panel) and \textit{vice} \textit{versa} in the case YX (right panel).}
    \label{fig:ESE_spectrograms_XY-YX}
\end{figure*}

Figure \ref{fig:ESE_spectrograms_XY-YX} gives the ESE spectrograms obtained for the 2L model with crossed-polarization cases XY and YX.

\subsection{ESE signals in the uncoupled case}
\label{subsec:ESEuncoupledcase}
\begin{figure*}[!ht]
    \centering
    \includegraphics[width=0.45\textwidth]{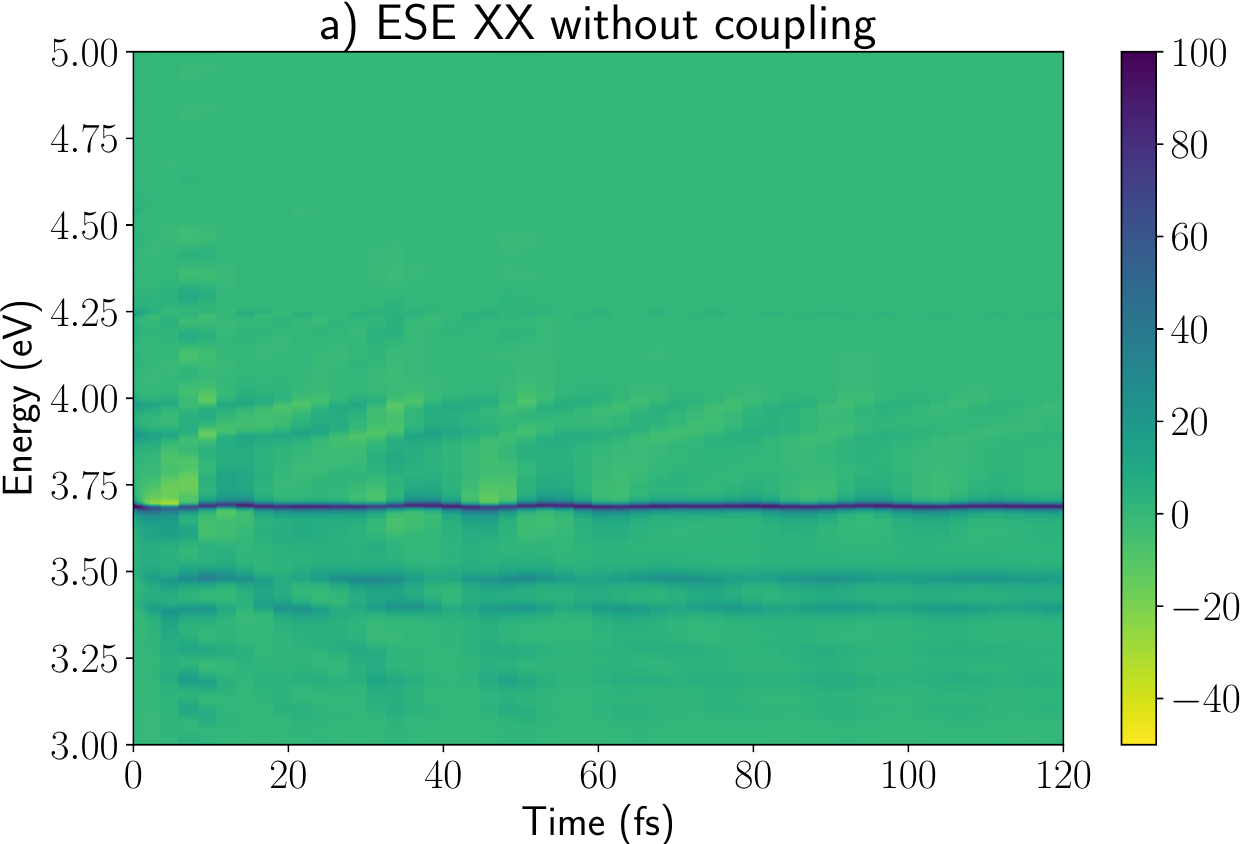}
    \includegraphics[width=0.45\textwidth]{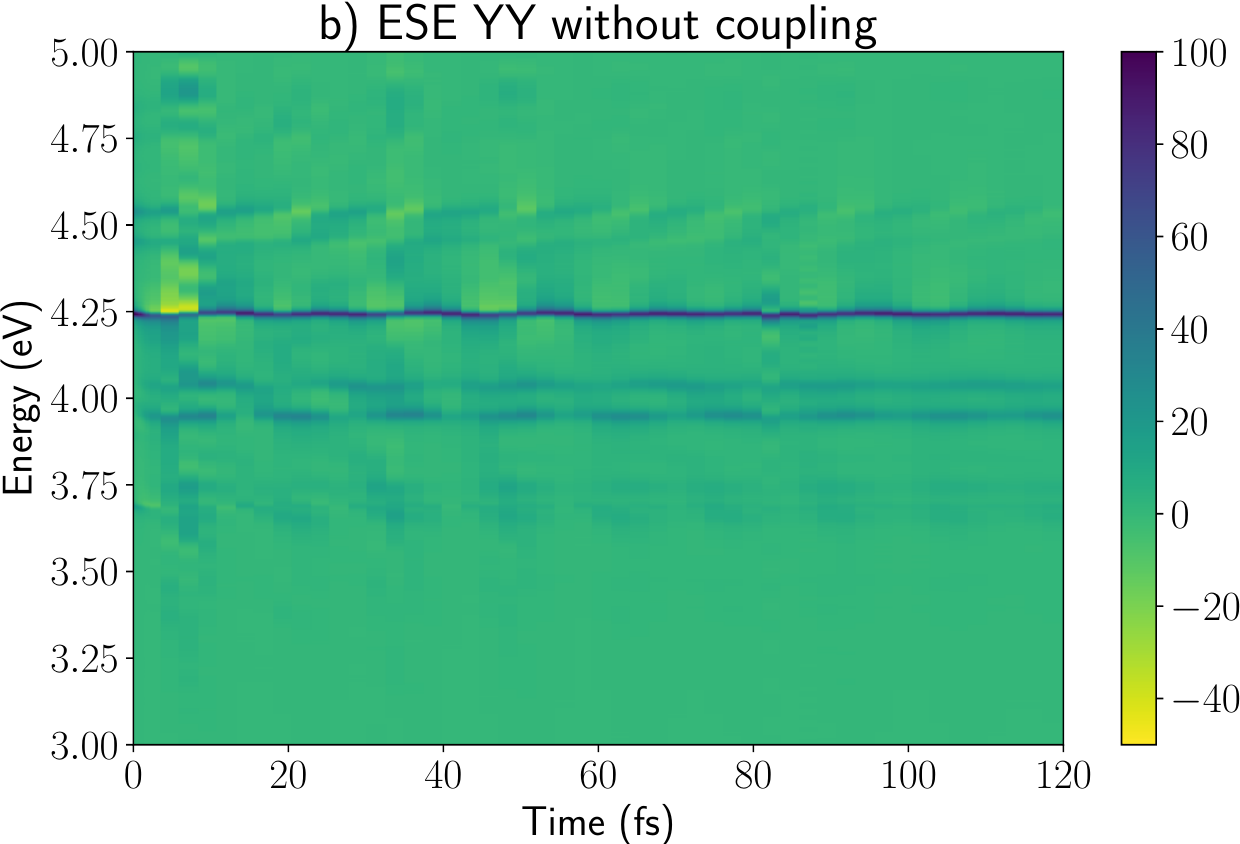}\\
    \caption{Time-and-frequency resolved ESE spectrograms when setting the inter-state coupling to zero, for the approximate 2L model with only the two high-frequency peaks in the spectral densities of \cref{fig:spectral_density}. The pump and probe are delta-like laser pulses. The polarizations are along  $e_x$ (left panel) or $e_y$ (right panel).}
    \label{fig:ESE_spectrograms}
\end{figure*}

Figure \ref{fig:ESE_spectrograms} gives the ESE spectrograms obtained for the 2L model with polarizations XX and YY in the case where the coupling bath is discarded so that there is no inter-state coupling.
This corresponds to a pure dephasing case where the populations remain constant but the initial electronic coherence is destroyed by the tuning baths.

\begin{figure*}
    \centering
    \includegraphics[width=0.45\textwidth]{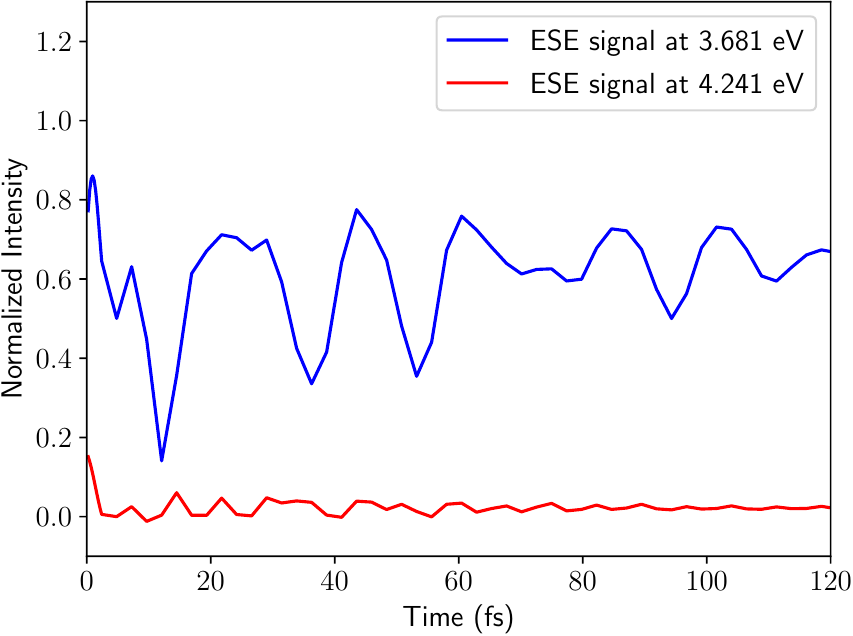}
    \includegraphics[width=0.45\textwidth]{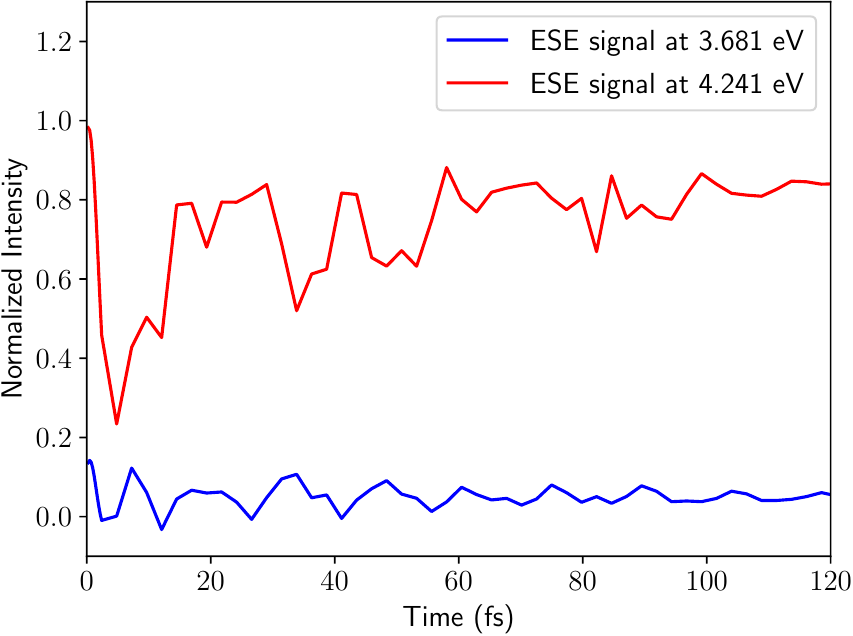}
    \caption{Time-resolved slices through the ESE spectrograms when setting the inter-state coupling to zero (\cref{fig:ESE_spectrograms}) at two different frequencies corresponding to distinct maxima in the absorption spectrum. The polarizations are along  $e_x$ (left panel) or $e_y$ (right panel).}
    \label{fig:ESE_spectrogram_TRP}
\end{figure*}

\begin{figure*}
    \centering
    \includegraphics[width=0.45\textwidth]{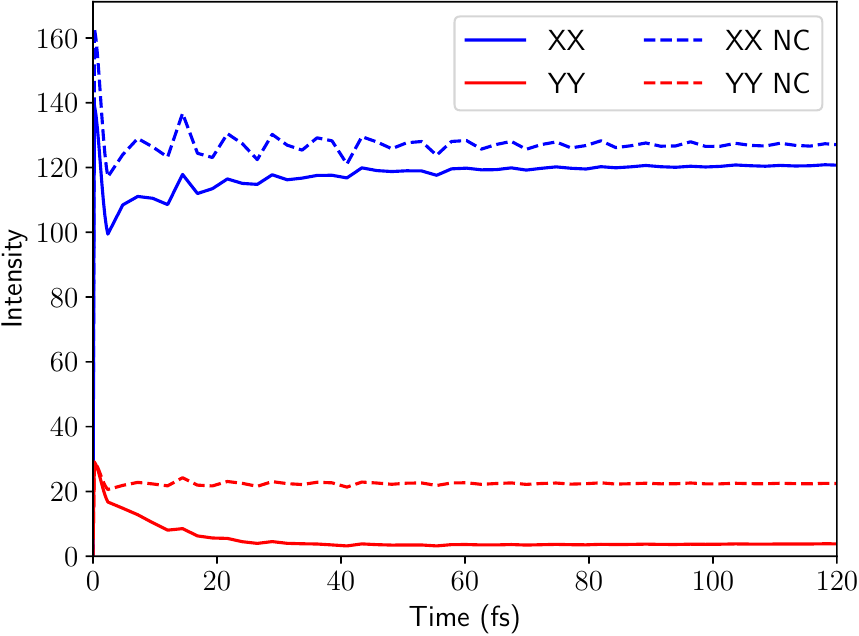}
    \includegraphics[width=0.45\textwidth]{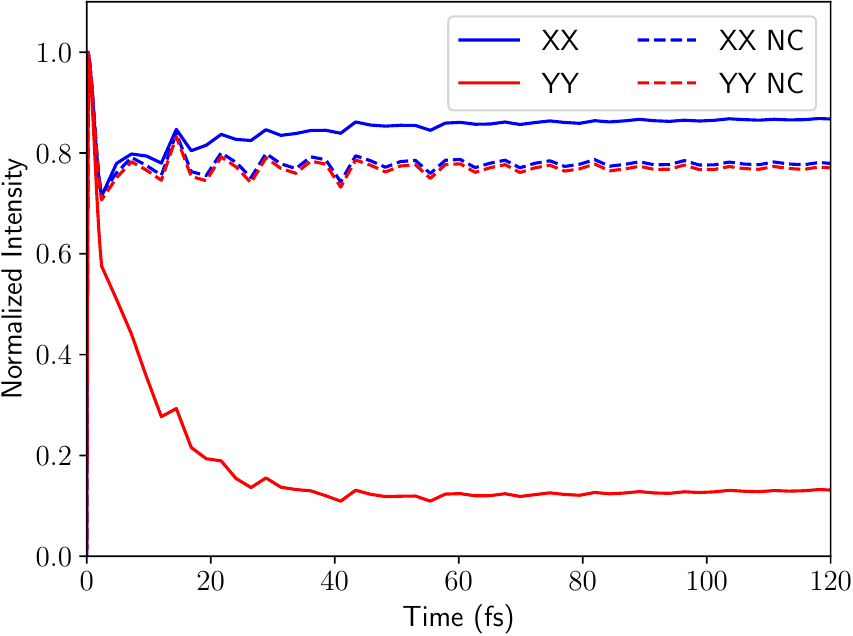}
    \caption{Integrated ESE signal, $S^{\text{ESE}}(t)$, for pump-probe polarizations $e_x$ or $e_y$ with and without inter-state coupling (NC: no coupling bath), as plain and dashed lines, respectively. The right panel corresponds to normalized signals for each polarization.}
    \label{fig:integratedsignal}
\end{figure*}

The time-resolved horizontal slices through the ESE spectrograms are shown in \cref{fig:ESE_spectrogram_TRP}. They are selected at two different frequencies corresponding to distinct maxima in the absorption spectrum.
The integrated ESE signals with and without the inter-state coupling are compared in \cref{fig:integratedsignal} with or without normalization.

\begin{figure*}
    \centering
        \includegraphics[width=0.9\textwidth]{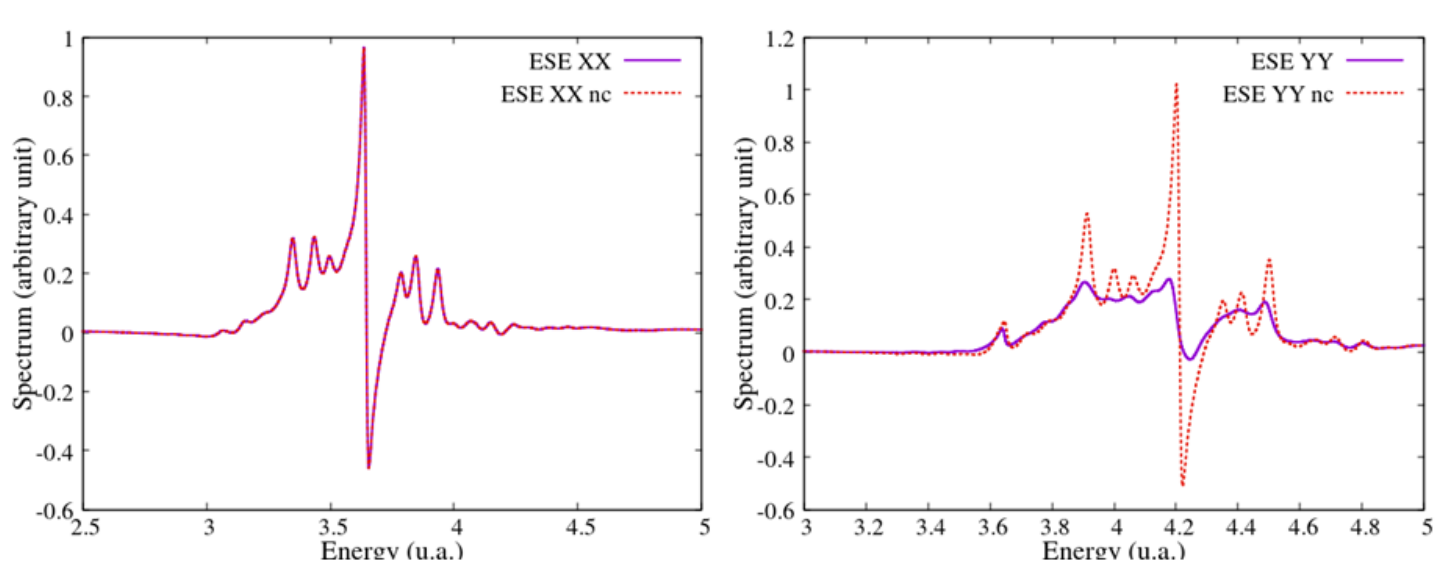}
    \caption{ESE spectra with and without inter-state coupling (nc: no coupling) for the 3L model at 2.4 fs. The polarizations are along  $e_x$ (left panel) or $e_y$ (right panel).}
    \label{fig:ESE-C-NC-100}
\end{figure*}

In \cref{fig:ESE-C-NC-100}, we illustrate that the Fano profile observed in the ESE spectra for a delay of about 2.5 fs is not due to the nonadiabatic transitions since it remains when the inter-state coupling is cancelled.  

\begin{figure*}
    \centering
        \includegraphics[width=0.5\textwidth]{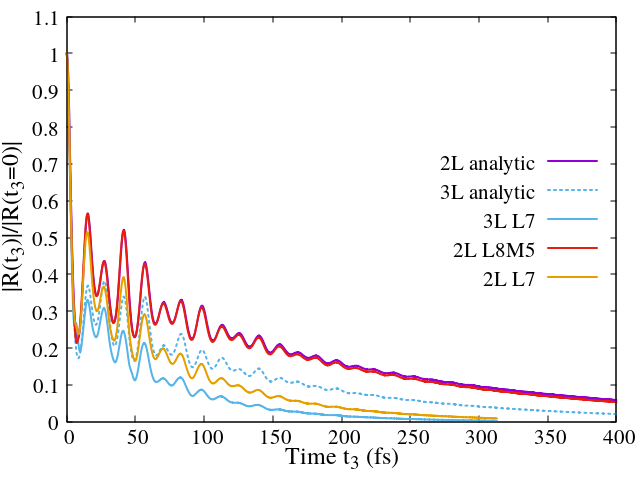}
    \caption{Modulus of the ESE response $R^{\text{ESE}}(0,t_2,t_3)$ for a delay $t_2 =$ 2.4 fs without inter-state coupling and optical excitation towards the D$_2$ state only for the 2L model or the 3L model for different HEOM levels. The polarizations are along $e_y$. The analytical expression is given by \cref{eq:responsecumulant}. }
    \label{fig:ESE-analytic}
\end{figure*}

\begin{figure*}
    \centering
        \includegraphics[width=0.5\textwidth]{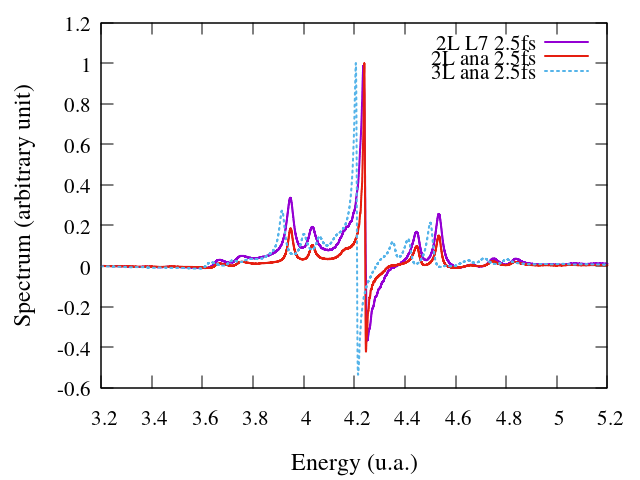}
    \caption{ESE spectrum  for a delay $t_2 =$ 2.4 fs obtained with the analytical responses displayed in \cref{fig:ESE-analytic} for the 2L model or the 3L model and by the simulation with level L $=7$ without Matsubara term.   }
    \label{fig:spectre-analytic}
\end{figure*}

Finally, we choose this example giving a particular Fano profile to illustrate the impact of the computational HEOM level on the simulation of the response and on the corresponding spectrum. By neglecting the inter state coupling and assuming an excitation towards the D$_2$ state only, the response of this pure dephasing case may be estimated by the cumulant expansion to second order \cite{Mukamel1995}. In this pure dephasing example, the tuning bath is associated to the  spectral density $J(\omega)=J_2(\omega)$ (\cref{fig:spectral_density}). The ESE response is expressed as a function of the so-called line shape function \cite{Mukamel1995}
\begin{align}
  & g(t)=\frac{1}{\pi }\int_{0}^{\infty }{d\omega J(\omega )\frac{1-\cos (\omega t)}{{{\omega }^{2}}}}\coth \left( \frac{\beta \hbar \omega }{2} \right) \nonumber \\ 
 & -\frac{i}{\pi }\int_{0}^{\infty }{d\omega }J(\omega )\frac{\sin (\omega t)-\omega t}{{{\omega }^{2}} }  \quad,
\end{align}
where $\beta$ is the Boltzmann factor. The response using the cumulant expansion at second order (cso) is given by:
\begin{equation}
R_{^{cso}}^{\text{ESE}}(0,{{t}_{2}},{{t}_{3}})={{e}^{i{{\omega }_{\text{FC}}}t}}{{e}^{-\left[ {{g}^{*}}({{t}_{3}})+f({{t}_{2,}}{{t}_{3}}) \right]}\quad,}
\label{eq:responsecumulant}
\end{equation}
where $\omega_{\text{FC}}$ is the Franck-Condon frequency and
\begin{equation}
f({{t}_{2}},{{t}_{3}})=2i\Im m\left[ \left( g({{t}_{2}}+{{t}_{3}} \right)-g({{t}_{2}})\right] \quad,   
\end{equation}
with $\Im m$ designating the imaginary part.
The modulus of the analytical expression $R_{^{cso}}^{\text{ESE}}(0,{{t}_{2}},{{t}_{3}})$ for $t_2= 2.4$ fs are compared in \cref{fig:ESE-analytic} for the 2L and 3L model at different HEOM levels. The oscillatory pattern is similar but the decay is faster in the 3L model as expected due to a stronger relaxation. In each case, the computation at level L$=7$ preserves the oscillatory scheme but leads to a faster decay than the analytical prediction. In the 2L model, the inclusion of three Matsubara terms allows the convergence with the analytical result. Finally, one sees in \cref{fig:spectre-analytic} that the approximate spectrum at the computationaly chosen level (2L model with level L$=7$ without Matsubara term) gives similar qualitative informations about the main peaks and the Fano profile.

\clearpage
\section{Cartesian coordinates}
Cartesian coordinates of the equilibrium geometry of the electronic ground state.
\addtolength{\tabcolsep}{10pt}
\renewcommand{\arraystretch}{1.0}
\begin{longtable}{ccrrr}
\toprule
 Center number &  Atomic number    &       $x$ &          $y$         & $z$           \\
\midrule
      1  &       6    &  -3.367756  & 1.150305  & 0.000000 \\
      2  &       6    &  -4.716618  & 1.519594  & 0.000000 \\
      3  &       6    &  -2.361587  & 2.121823  & 0.000000 \\
      4  &       6    &  -5.054623  & 2.880441  & 0.000000 \\
      5  &       6    &  -2.717211  & 3.478181  & 0.000000 \\
      6  &       6    &  -4.056541  & 3.846742  & 0.000000 \\
      7  &       6    &  -7.627865  &-1.326104  & 0.000000 \\
      8  &       6    &  -8.980862  &-0.958731  & 0.000000 \\
      9  &       6    &  -9.971511  &-1.932430  & 0.000000 \\
     10  &       6    &  -9.628748  &-3.282475  & 0.000000 \\
     11  &       6    &  -8.286733  &-3.655477  & 0.000000 \\
     12  &       6    &  -7.290982  &-2.687031  & 0.000000 \\
     13  &       6    &  -6.604726  &-0.326130  & 0.000000 \\
     14  &       6    &  -5.738314  & 0.518099  & 0.000000 \\
     15  &       6    &   1.555791  & 1.021874  & 0.000000 \\
     16  &       6    &   2.570019  & 1.990696  & 0.000000 \\
     17  &       6    &   3.903361  & 1.616019  & 0.000000 \\
     18  &       6    &   4.264804  & 0.260673  & 0.000000 \\
     19  &       6    &   3.250270  &-0.708028  & 0.000000 \\
     20  &       6    &   1.916931  &-0.333456  & 0.000000 \\
     21  &       6    &   0.180433  & 1.408047  & 0.000000 \\
     22  &       6    &  -0.985043  & 1.733223  & 0.000000 \\
     23  &       1    &  -9.244546  & 0.093989  & 0.000000 \\
     24  &       1    & -11.016198  &-1.635963  & 0.000000 \\
     25  &       1    & -10.405139  &-4.041640  & 0.000000 \\
     26  &       1    &  -8.013834  &-4.706564  & 0.000000 \\
     27  &       1    &  -6.244445  &-2.974219  & 0.000000 \\
     28  &       1    &  -3.099216  & 0.099747  & 0.000000 \\
     29  &       1    &  -4.325513  & 4.898619  & 0.000000 \\
     30  &       1    &  -6.100711  & 3.168152  & 0.000000 \\
     31  &       1    &  -1.937396  & 4.232475  & 0.000000 \\
     32  &       1    &   2.300306  & 3.041696  & 0.000000 \\
     33  &       1    &   4.681071  & 2.372639  & 0.000000 \\
     34  &       1    &   3.520044  &-1.758985  & 0.000000 \\
     35  &       1    &   1.139209  &-1.090078  & 0.000000 \\
     36  &       6    &   5.639941  &-0.125836  & 0.000000 \\
     37  &       6    &   6.804923  &-0.453676  & 0.000000 \\
     38  &       6    &   8.181560  &-0.841269  & 0.000000 \\
     39  &       6    &   9.193348  & 0.129460  & 0.000000 \\
     40  &       6    &   8.538124  &-2.197329  & 0.000000 \\
     41  &       6    &  10.529326  &-0.250569  & 0.000000 \\
     42  &       1    &   8.920374  & 1.179806  & 0.000000 \\
     43  &       6    &   9.876103  &-2.570264  & 0.000000 \\
     44  &       1    &   7.757187  &-2.950903  & 0.000000 \\
     45  &       6    &  10.875323  &-1.599824  & 0.000000 \\
     46  &       1    &  11.304269  & 0.510173  & 0.000000 \\
     47  &       1    &  10.140193  &-3.623604  & 0.000000 \\
     48  &       1    &  11.920542  &-1.894163  & 0.000000 \\
\bottomrule
\end{longtable}
\addtolength{\tabcolsep}{-10pt}
\renewcommand{\arraystretch}{1.0}
%

\clearpage